%% file: Arellano-Cordova-et-al.tex
\documentclass[fleqn,usenatbib]{mnras}

\usepackage{newtxtext,newtxmath}

\usepackage[T1]{fontenc}
\usepackage{ae,aecompl}

\usepackage[authoryear]{natbib}
\usepackage{psfrag,color}
\usepackage[latin1]{inputenc}
\usepackage{graphicx}	
\usepackage{amsmath}	
\usepackage{amssymb}	

\title[The abundance gradients in the Milky Way]{The Galactic radial abundance gradients of C, N, O, Ne, S, Cl and Ar from deep spectra of \ion{H}{ii} regions}

\author[K. Z. Arellano-C\'ordova et al.]{
K. Z. Arellano-C\'ordova$^{1,3}$\thanks{E-mail: karlaz@inaoep.mx and karlaz$\_$ext@iac.es (KZAC)}
C. Esteban,$^{1,2}$
J. Garc\'ia-Rojas$^{1,2}$
and J. E. M\'endez-Delgado$^{1,2}$
\\
$^{1}$Instituto de Astrof\'{\i}sica de Canarias, E-38200 La Laguna, Tenerife, Spain \\
$^{2}$Departamento de Astrof\'{\i}sica, Universidad de La Laguna, E-38206 La Laguna, Tenerife, Spain\\
$^3$Instituto Nacional de Astrof\'isica, \'Optica y Electr\'onica, Luis Enrique Erro $\# 1$,
Tonantzintla, Puebla 72840, Mexico\\
}

\date{Accepted XXX. Received YYY; in original form ZZZ}

\pubyear{2020}

\begin{document}
\label{firstpage}
\pagerange{\pageref{firstpage}--\pageref{lastpage}}
\maketitle

\begin{abstract}
We present a reassessment of the radial abundance gradients of C, N, O, Ne, S, Cl and Ar in the Milky Way using deep spectra of 33 \ion{H}{ii} regions gathered from the literature, covering Galactocentric distances from 6 to 17 kpc. The distances of the objects have been revised using {\it Gaia} parallaxes. We recalculate the physical conditions and ionic abundances in an homogeneous way using updated atomic data. All the objects have direct determination of the electron temperature, permitting to derive their precise ionic abundances. We analyze and compare different ICF schemes for each element in order to obtain the most confident total abundances. 
Due to the revised distances, our results do not support previous claims about a possible flattening of the O/H gradient in the inner Galactic disk. We find that the Galactic N/O gradient is rather flat, in contrast to what has been found in other spiral galaxies. The slope of the gradients of some elements is sensitive to the ICF scheme used, especially in the case of Ne. The dispersion around the fit for the gradients of C, N, O, S, Cl and Ar is of the order of the typical uncertainties in the determination of the abundances, implying the absence of significant inhomogeneities in the chemical composition of the ionized gas phase of the ISM. We find flat gradients of log(S/O) and log(Cl/O) and very shallow or flat ones for log(Ne/O) and log(Ar/O), consistent with a lockstep evolution of Ne, S, Cl and Ar with respect to O.   
\end{abstract}



\section{Introduction}
\label{sec:intro}
Stellar interiors are the factories where a large fraction of metals with Z$<$30 are formed in the Universe. \ion{H}{ii} regions, in particular Galactic regions, are excellent places to study the chemical abundances of metals such as C, N, O, Ne, S, Cl and Ar.

O is the third most abundant element in the Universe, and is usually taken as a proxy of the metallicity in photoionized nebulae. O is almost entirely produced by massive stars (MS, M$\gtrsim$8 M$_{\odot}$) and later ejected to the interstellar medium (ISM) by type II supernovae explosions in short time scales.  On the other hand, C and N present a much more complicated origin as they can be produced by different mechanisms in different stellar mass ranges \citep{Garnett:1990, Chiappini:2003, Clayton:2003, Carigi:2005}. For C, \citet{Carigi:2005} built  Galactic chemical evolution models to reproduce the observed behavior of negative C/O ratio gradients in the Galaxy; in the model that better reproduced the observations, the fraction of C produced by massive stars and low-to-intermediate mass stars (LIMS, M$\lesssim$8 M$_{\odot}$) is dependent  on time and on the Galactocentric distance. This model also predicted that, at the present time in the solar vicinity, half of the C in the ISM has been produced by MS and half by LIMS. For N, the behaviour is even more complicated as this element can have different origins: the main channel of N production is as a product of the CNO cycle, being formed at the expense of C and O already present in the star (secondary origin). However, it can also have a primary origin, both in asymptotic giant branch (AGB) stars through efficient nuclear burning at the base of the convective envelope (hot bottom burning) in intermediate-mass stars \citep[3.5 M$_{\odot}$ $\lesssim$ M $\lesssim$ 8 M$_{\odot}$, see][and references therein]{Karakas:2016}, or in MS through stellar rotation \citep{Meynet:2002}. 
Therefore, the study of N/O and C/O ratios can provide precious constraints to the chemical evolution  of the Galaxy. 

Ne, S and Ar, are produced via the alpha process in MS and ejected to the ISM along with O in core-collapse supernovae explosions. Therefore their abundance ratios with respect to O are, in principle, expected to be constant \citep[e.g.][]{Croxall:2016a, Esteban:2020, Berg:2020}. A similar behaviour is expected for Cl, which is produced by single particle -- proton or neutron -- capture by a S or Ar isotope \citep[see][]{Esteban:2015}. 

%
%



Among the different sources of uncertainty in chemical abundance determinations,  ionization correction factors (ICFs) play a critical role for most of the elements observed in the spectra of photoionized regions \citep[e.~g.][]{Peimbert:2017, Delgado-Inglada:2019}. ICFs take into account the contribution of unseen ions, and can be calculated both by using ionization potential similarities between observed and unseen ions \citep{Peimbert:1969, Peimbert:1977} or making photoionization models grids of complete extragalactic \ion{H}{ii} regions, which in principle should be more reliable as they include the physics that drives the spatial distribution of ions in photoionized nebulae. Some remarkable examples of this last approach can be found in \citet{Stasinska:1978, Izotov:2006, Dors:2016} and \citet{Berg:2020}. Finally, our group has also made an empirical approach for specific elements in our Galaxy, such as Cl and N, based on observations of Galactic \ion{H}{ii} regions that do not need ICF correction \citep{Esteban:2015, Esteban:2018}. 
%

The local metallicity, parameterized by the O/H ratio, is directly related to the star formation rate that, in spiral galaxies, is correlated with the total surface density \citep[e.g.][]{Vila-Costas:1992,Henry:1999,Carigi:2019}. Moreover, the absolute value of the O/H ratio at a given point of the disc of a spiral galaxy may be an indicator of the upper mass limit of the initial mass function of the whole galaxy \citet{Carigi:2019}. Since the earlier works by \citet{Guesten:1982}, \citet{Lacey:1985} or \citet{Goetz:1992}, most chemical evolution models in the literature \citep[see][and references therein]{Molla:2014} explain the existence of radial abundance gradients in the discs of spiral galaxies by the combined effect of  radial variations of the star formation rate and the gas infall. As \citet{Henry:2010} discuss, the density threshold of star formation \citep[e.g.][]{Marcon-Uchida:2010} and the timescale for disc formation \citep[e.g.][]{Fu:2009} can regulate the shape of the abundance gradient, in the sense that radial changes of any of both quantities can vary the slope of the radial O/H gradients. Therefore, a better determination of the shape of the gradients and the absolute value of the chemical abundances at a given Galactocentric radius will help to constrain essential basic parameters of chemical evolution models.

The studies of radial abundance gradients in the Milky Way from \ion{H}{ii} regions data have focused mainly on O and N. Since the canonical work by \cite{Shaver:1983}, the Galactic radial O abundance gradient has been extensively studied in the last two decades \citep[e.~g.][]{Deharveng:2000, Esteban:2005, Quireza:2006, Rudolph:2006, Balser:2011, Esteban:2013, Esteban:2017, Fernandez-Martin:2017}.  Most of these studies have remarkably consistent slopes of the O gradient (between $-$0.040 dex kpc$^{-1}$ and $-$0.060 dex kpc$^{-1}$), even taking into account the different methodologies and wavelength coverage of the data. The most recent value reported in the literature for the radial O abundance gradient in our Galaxy is the one reported by \citet{Esteban:2018}, who computed a slope of $-0.042$ dex kpc$^{-1}$. \citet{Esteban:2018} also studied the radial gradient of N/H using a sample of very low-ionization \ion{H}{ii} regions, allowing them to avoid the use of an ICF for most of their sample. These authors compared also different ICF schemes for N and presented values of the slopes for N/H and N/O ratios in comparison with objects without an ICF scheme. They did not find large differences or trends between the objects using an ICF scheme and those ones without using it, and reported a radial N gradient with a slope of $-0.047$ dex kpc$^{-1}$ and an almost flat, within the uncertainties, N/O gradient, in agreement with what found by \citet{Shaver:1983}. However, other authors have found an enhancement of the N/O ratio in the inner part of the Galactic disc although without strong evidence for an overall linear radial gradient of this quantity \citep{Simpson:1995, Rudolph:2006}. Given the uncertainty of the ICF, the diverse methodologies used in the literature to compute the N abundances, and the different wavelength and Galactocentric distance ranges covered, it is difficult to arise to firm conclusions on the behaviour of the N/O ratio in our Galaxy. 

Regarding C in the Milky Way, \citet{Esteban:2005} and \citet{Esteban:2013} computed the radial C/H and C/O gradients from recombination lines and obtained negative slopes for both ratios, which were consistent with C yields that increase with metallicity owing to stellar winds in MS and decrease with metallicity owing to stellar winds in LIMS \citep{Carigi:2005}.

Radial gradients of $\alpha$-elements (Ne, S and Ar; we also include Cl in this group as argued before) from \ion{H}{ii} region data are relatively scarce in the literature although some studies have reported radial gradients for one or several $\alpha$-elements using either optical or IR spectra \citep[e. g.][]{Shaver:1983, Martin-Hernandez+2002, Esteban:2015, Fernandez-Martin:2017}. In general, the results obtained by all these authors for Ne, S and Ar are consistent with flat radial $\alpha$/O gradients; however the dispersion at a given Galactocentric distance is generally large, which can be attributed to uncertainties in the assumed ICF schemes or to the methodologies adopted for physical conditions and chemical abundance computations (i.~e. obtained from observations using different wavelength ranges that do not cover the same volume of the nebula). However, from deep, high-resolution spectra, these dispersion can be significantly reduced. \citet{Esteban:2015}, using published data for a sample of nine \ion{H}{ii} regions, reported for the first time the radial gradients for Cl/H and Cl/O ratios without using any ICF scheme for Cl. These authors found a slope of $-0.043$ dex kpc$^{-1}$ for the Cl/H radial gradient, in agreement with what found from planetary nebulae (PNe)  \citep{Henry:2004}. 
On the other hand, \cite{Esteban:2015} reported a flat slope value of $-0.001$ dex kpc$^{-1}$ for Cl/O, indicating  a lockstep evolution of both elements. 

The relatively large amount of deep, high-quality spectra of Galactic \ion{H}{ii} regions collected by our group in the last years by using modern high-sensitivity spectrographs attached to large (8-10m) ground-based telescopes and covering a wide range of Galactocentric distances \citep[e.~g.][]{Esteban:2004, Garcia-Rojas:2004, Garcia-Rojas:2005, Garcia-Rojas:2006, Garcia-Rojas:2007, Esteban:2013, Esteban:2017, Esteban:2018} have allowed us to develop detailed analysis of the physical conditions and chemical abundance patterns of different elements using both collisional excitation lines (CELs) and, in se\-ve\-ral cases,  recombination lines \citep[RLs, used to compute C/H and C/O ratios, see][] {Esteban:2005, Esteban:2013}. In these studies, we have significantly reduced the scatter previously found in abundance gradients of O, C and N at a given Galactocentric distance by minimizing the observational uncertainties \citep{Rodriguez:2017, Garcia-Rojas:2020b}. Special efforts have been devoted in the last years in the detection of the temperature-sensitive auroral lines such as [N~{\sc ii}] $\lambda$5755 and/or [O~{\sc iii}] $\lambda$4363 in objects in the inner and outer Galactic disc \citep{Esteban:2017, Fernandez-Martin:2017, Esteban:2018}; this is a challenge owing to a one or a combination of several factors: the intrinsic faintness of the object, the relatively low ionization degree of the gas, the high extinction by dust and, in the case of the inner objects, their relatively high metallicity. 

In this work, we have collected a large sample of high-quality medium-to-high resolution optical spectra of Galactic \ion{H}{ii} regions and present a re-analysis of the elemental abundances of C, N, O, Ne, S, Cl and Ar by using different ICF schemes proposed for \ion{H}{ii} regions, and studying effects of the temperature structure in the nebula with the aim of improving the computation of radial abundance gradients of these elements that can be used to better constrain chemical evolution models of the Milky Way. The structure of this paper is as follows. In \S~\ref{sec:sample} we describe the sample of spectra that were collected from the literature and present a reassessment of the distances of the objects using {\it Gaia} parallaxes of the second data release (DR2) \citep{gaiadr2}. In \S~\ref{sec:calculations}, we present the values of physical conditions, electron density and temperature and computations of ionic abundances. In \S~\ref{sec:ICFs-description} we discuss the results obtained by using different ICF schemes for our sample. In \S~\ref{sec:Abundance-gradients} we present the radial abundance gradients of the different elements and a comparison with results for other objects and galaxies. Finally, in \S~\ref{sec:conclusions} we summarize our main conclusions.


\section{Sample}
\label{sec:sample}

\input{Table1.tex} 

The sample comprises 33 Galactic \ion{H}{ii} regions with  high or  intermediate-resolution spectra from \citet{Esteban:2004, Esteban:2013, Esteban:2017, Esteban:2018, Garcia-Rojas:2004, Garcia-Rojas:2005, Garcia-Rojas:2006, Garcia-Rojas:2007, Garcia-Rojas:2014} and  \citet{Fernandez-Martin:2017}. These are, to our knowledge, the highest signal-to-noise ratio spectra for each individual H~{\sc ii} region of our sample. This is the meaning we give to the term "deep" spectrum in the present paper. In Table~\ref{tab:sample}, we list the objects of our sample and the references of their original spectroscopical data. 

\citet{Esteban:2017} and \citet{Esteban:2018} give the Galactocentric distance, $R_{\rm G}$, of most of the sample objects (all except IC~5146, Sh~2-132 and Sh~2-156). Those authors adopted the mean values of kinematic and stellar distances given in different published references and an uncertainty corresponding to their standard deviation. M\'endez-Delgado et al. (2020) have made a revision of the distances taking into account new  determinations based on {\it Gaia} DR2 parallaxes \citep{gaiadr2}. After a detailed discussion for each nebulae and considering the uncertainties, M\'endez-Delgado et al. (2020) adopt the $R_{\rm G}$ values obtained from {\it Gaia} data for objects with heliocentric distances smaller than 5 kpc and the values given by \citet{Esteban:2017} and \citet{Esteban:2018} for the rest except for few particular cases. Since our sample include more objects than the one studied by M\'endez-Delgado et al. (2020) and in the sake of homogeneity, we have applied the same methodology to revise the distances of the objects that are not included in M\'endez-Delgado et al. 

In the case of IC~5146, Sh\,2-61, Sh\,2-132, Sh\,2-156, Sh\,2-175, Sh\,2-219, Sh\,2-235, Sh\,2-237, Sh\,2-257, Sh\,2-271, Sh\,2-285 and Sh\,2-297, which are located at heliocentric distances smaller than 5 kpc, we have adopted the distance based on parallaxes from \textit{Gaia} DR2 data taken from the database of \citet{bailer-jonesetal18}. For 
Sh\,2-90 and Sh\,2-266, with heliocentric distances about or larger than 5 kpc, we assumed the distances quoted by \citet{Esteban:2018}, which have lower uncertainties and are based on several different but consistent kinematical and photometrical distances. We describe the details for each object below.

{\it IC~5146}. \citet{avedisovakondratenko84} found that this nebulae is ionized by the star BD$+$46~3474, the brightest star of the region. We use the $R_{\rm G}$ obtained from its {\it Gaia} DR2 parallax, which is consistent with the photometrical distances determined by \citet{herbig:2002} and \citet{harvey:2008}

{\it Sh\,2-61}. \citet{huntermassey90} report 4 stars as the ionizing sources of the nebula. We determine the mean distance corresponding to the Gaia DR2 parallaxes of those 4 sources, which is marginally consistent with to the one given by  \citet{Esteban:2018} -- $R_{\rm G}$ = 5.7$\pm$ 0.3 kpc -- within the errors. 

{\it Sh\,2-90}. \citet{samaletal14} identified 4 stars as the most probable ionizing sources of this \ion{H}{ii} region. Two of the stars give anomalously large parallaxes probably because they are very close to a field star. The other two stars give distances with very large uncertainties but in complete agreement with the one given by \citet{Esteban:2018}. We adopt the $R_{\rm G}$ given by these last authors for the object. 

{\it Sh\,2-132}. We have obtained the {\it Gaia} DR2 parallaxes for 11 stars that \citet{Harten:1978} consider associated with the \ion{H}{ii} region. The $R_{\rm G}$ assumed for this object is the weighted average of the individual distances of that group of stars. 

{\it Sh\,2-156}. We have determined and adopted the $R_{\rm G}$ of the nebula from the {\it Gaia} DR2 parallax for the O7 star that \citet{Lynds:1983} consider the ionizing source of the nebula. The rest of stars located inside a circle of 10 arcsec radius around the star show very similar parallaxes.  

{\it Sh\,2-175}. This nebula is ionized by a single star, LS~I~$+$64 26. We use the distance obtained from its {\it Gaia} DR2 parallax. 

{\it Sh\,2-219}. The ionization source of this \ion{H}{ii} region is the star LS~V~$+$47 24. Its {\it Gaia} DR2 parallax give a $R_{\rm G}$ in agreement with that reported by \citet{Esteban:2018}. 

 {\it Sh\,2-235}. \citet{camargoetal11} found that the star BD$+$35~1201 is its main ionizing source. We have assumed the distance derived from the {\it Gaia} DR2 parallax for that star. 

 {\it Sh\,2-237}. We analyzed the {\it Gaia} DR2 data of 10 of the stars identified by \citet{pandeyetal13} as members of the cluster associated to the nebula. Eight of them give consistent distances, which are very similar and more precise than the distance reported by \citet{Esteban:2018}. We adopted the mean distance obtained from the parallaxes of the stars associated to the nebula. 
 
 {\it Sh\,2-257}. \citet{ojhaetal11} proposed that HD 253327 is the ionizing star of the nebula, but its associated {\it Gaia} DR2 source indicates that its heliocentric distance is a factor of ten lower than in several previous studies. However, \citet{ojhaetal11}  and other authors agree on that Sh~2-257 belongs to the same molecular complex as Sh~2-254, 
 Sh~2-255, Sh~2-256 and Sh~2-258 and all of them should be located at the same distance. We have adopted the distance determined from the Gaia parallax of LS\,19, the ionizing star of Sh~2-255, the one with the best data, as representative of Sh\,2-257. This is in complete agreement with the mean value of photometrical and kinematical $R_{\rm G}$ determinations given in \citet{Esteban:2018} for this object. 
 
{\it Sh\,2-266}. This nebula is ionized by the Of star MWC\,137. Due to its large heliocentric distance, the parallax is very uncertain, but the distance of $R_{\rm G}$ is 13.6$^{+1.8}_{-1.3}$ kpc derived from that parallax is consistent with that given by \citet{Esteban:2018}, based on the kinematical data of \citet{Russeil:2007} and photometrical measurements by \citet{Mehner:2016}. We assume the distance given by \citet{Esteban:2018} as representative of Sh~2-266. 

 {\it Sh\,2-270}. The {\it Gaia} DR2 parallax of the ionizing star of this nebula identified by \citet{neckelstaude84} is very uncertain, giving a $R_{\rm G}$ =  11.9$^{+2.5}_{-1.6}$ kpc. The spectrophotometrical determination obtained by \citet{foster:2015} and the kinematical one by \citet{Russeil:2007} give considerably larger distances. We adopted the mean of these last values, which is the one assumed by \citet{Esteban:2018} for this object. The larger value of the distance makes the O/H ratio of Sh~2-270 more consistent with the expected according to the radial abundance gradient.

 {\it Sh\,2-271}. \citet{persietal87} identified the ionizing star, which lies in the centre of the nebula. We adopt the distance based on {\it Gaia} DR2 data for that star (source 3331937432404833792) as representative of the object. 

 {\it Sh\,2-285}. \citet{moffatetal79} and \citet{rollestonetal94} identified the two ionizing stars of this \ion{H}{ii} region. The distances of both stars determined from their {\it Gaia} DR2 
 parallaxes 
 are similar but with very different degree of accuracy. We adopt the distance of the star with the lowest uncertainty. 

{\it Sh\,2-297}. \citet{mallicketal12} identified the ionizing star of the \ion{H}{ii} region. We have adopted the distance derived from the {\it Gaia} DR2 parallax of that star for the nebula. 

\begin{figure} 
    \begin{center}
    \includegraphics[width=0.4\textwidth, trim=30 0 30 0,  clip=yes]{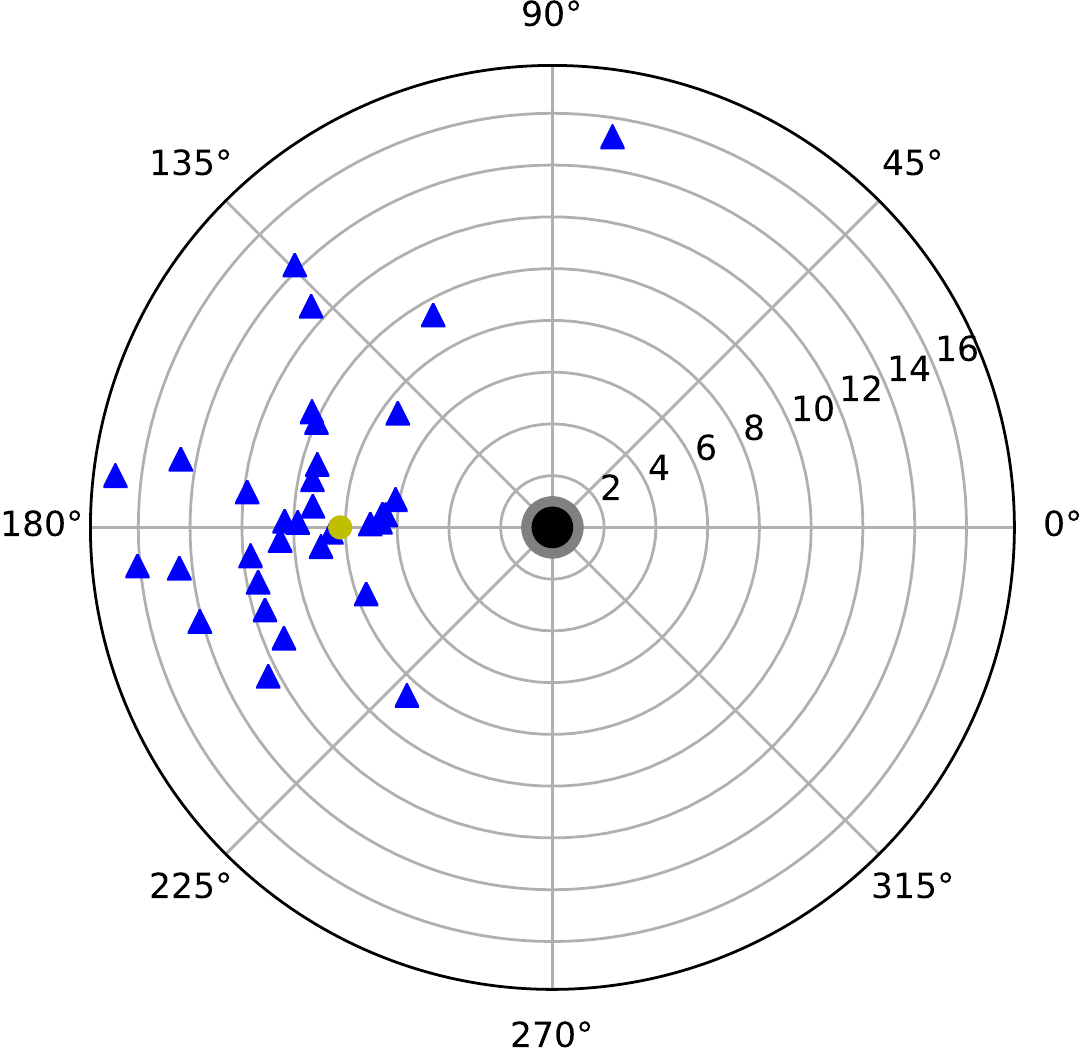}
    \caption{Spatial distribution of our sample of 33 \ion{H}{ii} regions (triangles) onto the Galactic plane with respect to the centre of the Milky Way. The Sun position is displayed by a small circle. The concentric circles indicate increasing Galactocentric distances (in kpc).}  
    \label{fig:dist_map}
   \end{center}
    \end{figure}

In Table~\ref{tab:sample}, we include the final adopted  $R_{\rm G}$ for our sample of Galactic \ion{H}{ii} region. The objects cover a rather wide range of $R_{\rm G}$ from 6.15 to 17.0 kpc. Our revision of distances considering $Gaia$ DR2 parallaxes provides a slightly narrower range of $R_{\rm G}$ in comparison with that obtained by  \citet{Esteban:2018}, that was from 5.1 to 17 kpc. It is important to note that the revision of the distances affect especially to those objects located nearer the Galactic Centre. In fact, all the nebulae with $R_{\rm G}$ smaller than 8 kpc show now larger values of $R_{\rm G}$. This is an interesting systematic effect that merits to be further investigated but, unfortunately, is out of the scope of the present paper. Fig.~\ref{fig:dist_map} shows the spatial distribution of the 33 \ion{H}{ii} regions onto the Galactic plane.

\section{Physical conditions and ionic abundances}
\label{sec:calculations}

\subsection{Electron temperature and density}

\input{Table2.tex} 

We have used the Python package \textsc{PyNeb} version 1.1.5  \citep{Luridiana:2015} to determine the physical conditions and chemical abundances in a homogeneous way using the line intensities compiled from the literature. To minimize the effects of atomic data on the computation of nebular abundances \citep{JuandeDios:2017} we have carefully selected the atomic data set used in this work, which is listed in Table~ \ref{tab:atomic_data}. The electron density, $n_{\rm{e}}$, was calculated using the  [\ion{O}{ii}] $\lambda3729/\lambda3726$ and/or [\ion{S}{ii}] $\lambda6717/\lambda6731$ line intensity ratios. For objects where it was possible to measure both densities we used their mean value in our calculations. In most of the nebulae we obtain low densities, $n_{\rm{e}} <100$ cm$^{-3}$, we assumed $n_{\rm{e}}=100$ cm$^{-3}$ for them \citep{Osterbrock:2006}. We have used a two-zone scheme characterized by $T_{\rm e}$([\ion{N}{ii}])  and $T_{\rm e}$([\ion{O}{iIi}]) derived from the [\ion{N}{ii}]~($\lambda6548+\lambda6584)/\lambda5755$ and [\ion{O}{iii}]~($\lambda4959+\lambda5007)/\lambda4363$ line intensity ratios, respectively. The intensity of [\ion{N}{ii}] $\lambda5755$ may have a possible contribution from the recombination of N$^{2+}$ (\citealt{Rubin:1986,Stasinska:2005, Liu:2000}). We have not considered such contribution owing that it has been proven to be  small in the H~{\sc ii} regions in our sample (see for example, \citealt{Garcia-Rojas:2006,Esteban:2017, Esteban:2018}), and because this correction is rather uncertain. 

We have used the temperature relation proposed by \citet[their equation 3]{Esteban:2009} to estimate $T_{\rm e}$([\ion{O}{iii}]) from $T_{\rm e}$([\ion{N}{ii}]) when the auroral $\lambda$4363 line is not detected. For IC~5146, we have done the calculations separately for the spectra of apertures 2, 3 and 4 observed by  \citet{Garcia-Rojas:2014}. These authors consider such apertures as representative of the main volume of the nebula. 
Our final values for the physical conditions and ionic abundances of IC~5146 are the weighted average ones obtained for those apertures. The final results for $n_{\rm e}$([\ion{O}{ii}]), $n_{\rm e}$([\ion{S}{ii}]), $T_{\rm e}$([\ion{N}{ii}]) and $T_{\rm e}$([\ion{O}{iii}]) for each nebulae are included in Table~\ref{tab:sample}.

\subsection{Ionic abundances}
\label{sec:ionic}

\input{Table3.tex} 

For our calculations, we have adopted $T_{\rm e}$([\ion{N}{ii}]) as representative of the N$^{+}$, O$^{+}$, S$^{+}$ and Cl$^{+}$ emitting regions and $T_{\rm e}$([\ion{O}{iii}]) for the C$^{2+}$, O$^{2+}$, Ne$^{2+}$, Cl$^{3+}$ and Ar$^{3+}$ ones. Following the recommendation of \citet{Dominguez-Guzman:2019}, we also adopted $T_{\rm e}$([\ion{N}{ii}]) to calculate the Cl$^{2+}$ abundance and the mean value of $T_{\rm e}$([\ion{N}{ii}]) and $T_{\rm e}$([\ion{O}{iii}]) to calculate S$^{2+}$ and Ar$^{2+}$ abundances. 

We calculated C$^{2+}$/H$^+$ abundances using the intensity of the pure recombination \ion{C}{ii} $\lambda4267$ line for a sample of 11 \ion{H}{ii} regions. We used the effective recombination coefficients of \citet{Davey:2000} for \ion{C}{ii} and \citet{Storey:1995} for 
\ion{H}{i} and $T_{\rm e}$([\ion{O}{iii}]) as representative of the C$^{2+}$ zone.
The ionic abundance of N$^{+}$ were calculated using $\lambda\lambda6548, 6584$ for the total sample.  
For oxygen, we calculate the ionic abundances of O$^{+}$, O$^{2+}$ using the line intensities of [\ion{O}{ii}] $\lambda\lambda3726, 3729$ and [\ion{O}{iii}] $\lambda\lambda4959,5007$, respectively, for all the objects of our sample. For Sh\,2-209, we used the line intensities of [\ion{O}{ii}] $\lambda\lambda$7320,7330 to calculate O$^{+}$ abundance since the $\lambda\lambda3726, 3729$ lines could not be measured in the spectrum due to the faintness and high extinction coefficient of this nebula. 
In Table~\ref{tab:O-N-ab} we include the values of C$^{2+}$, N$^{+}$, O$^{+}$ and O$^{2+}$ abundances and the degree of ionization measured as O$^{2+}$/(O${^+}$ + O$^{2+}$) = O$^{2+}$ for each object.

The Ne$^{2+}$ abundances were calculated for 15 regions with measurements of [\ion{Ne}{iii}] $\lambda\lambda3869, 3967$ lines. For  Sh\,2-83, Sh\,2-100, Sh\,2-128, Sh\,2-156 and Sh\,2-212, we only used the line intensity of [\ion{Ne}{iii}] $\lambda3869$ to calculate the Ne$^{2+}$/H$^+$ ratio, since the [\ion{Ne}{iii}] $\lambda3967$ line is blended with H7 in the spectra of these objects. 

The S$^{+}$ and S$^{2+}$ abundances were calculated for 33 and 23 Galactic \ion{H}{ii} regions, respectively. We used the [\ion{S}{ii}] $\lambda\lambda6717, 6731$ lines to calculate S$^{+}$/H$^+$. For S$^{2+}$/H$^+$, we used the [\ion{S}{iii}] $\lambda6312$ line as well as [\ion{S}{iii}] $\lambda\lambda9069, 9532$ lines in the spectra where the wavelength coverage was wide enough to include them. However, the [\ion{S}{iii}] $\lambda$9069, 9532 lines can be affected by strong telluric emission and absorption features \citep{Stevenson:1994, Noll:2012}. We discarded to use these lines in the targets where they are reported as affected by telluric emission/absorption in the original reference. For the remaining objects, we inspected whether the observed intensity ratio $I$([\ion{S}{iii}] $\lambda$9532)/$I$([\ion{S}{iii}] $\lambda$9069) deviated $\pm$20\% from the theoretical value of 2.47 (as computed using the atomic data adopted in this work); we discarded the [\ion{S}{iii}] $\lambda$9069 line when the ratio was larger than the theoretical one and we discarded [\ion{S}{iii}] $\lambda$9531 line when the ratio was lower. Following these criteria, we have used the [\ion{S}{iii}] $\lambda$9069 line to estimate the S$^{2+}$/H$^+$ ratio in IC\,5146, M8, Sh\,2-132, Sh\,2-156 and NGC\,3576, and [\ion{S}{iii}] $\lambda$9532 in the cases of NGC\,3603 and Sh\,2-298. We used both [\ion{S}{iii}] red lines for the rest of the regions for which we measure those lines. The [\ion{S}{iii}] $\lambda$6312 line, which is very sensitive to the electron temperature and much fainter than the [\ion{S}{iii}] $\lambda\lambda9069, 9532$ lines, was only considered when it was the only [\ion{S}{iii}] line available in the spectra.

We have calculated the Cl$^{2+}$ abundances for 25 Galactic \ion{H}{ii} regions by using the intensities of [\ion{Cl}{iii}] $\lambda\lambda5518, 5538$ lines. For nine of those objects, it was also possible to derive the Cl$^{3+}$ and/or Cl$^{+}$ abundances using the faint  [\ion{Cl}{iv}] $\lambda8046$ and/or [\ion{Cl}{ii}] $\lambda9139$ lines, respectively (see \citealt{Esteban:2015}).

We calculated the Ar$^{2+}$ abundances using the [\ion{Ar}{iii}] $\lambda\lambda7135,7751$ lines for a sample of 26 \ion{H}{ii} regions. For 10 of them we could also calculate the Ar$^{3+}$ abundance using the [\ion{Ar}{iv}] $\lambda4711$ and/or [\ion{Ar}{iv}] $\lambda4740$ lines. In Table~\ref{tab:ionic-abundances}, we show our results for the abundances of the observed ionic species of Ne, S, Cl and Ar.

\input{Table4.tex} 





The uncertainties associated with each value of $n_{\rm e}$, $T_{\rm e}$ and the ionic abundances were calculated using Monte Carlo simulations, generating 500 random values for each line intensity ratio assuming a Gaussian distribution with a sigma equal to the associated uncertainty.

\section{A Comprehensive analysis of ICF schemes and total abundances}
\label{sec:ICFs-description}


We determined the total O abundance by adding the contribution of O$^+$ and O$^{2+}$, O/H = O$^+$/H$^+$ + O$^{2+}$/H$^+$. IC 5146, Sh\,2-175, Sh\,2-270 and Sh\,2-288 do not show measurable [\ion{O}{iii}] lines due to their very low degree of ionization. For those objects we assume that O/H $\approx$ O$^+$/H$^+$ following \citet{Esteban:2018}. For the rest of the chemical elements, we have to use ICFs to derive their total abundances. The use of reliable ICFs is essential to properly estimate the contribution of unobserved ions in \ion{H}{ii} regions and other ionized nebulae. The total abundance of a  particular element, X, can be derived as X/H = X$^{i+}$/H$^{+} \times\,$ICF, where X$^{i+}$/H$^+$ is the ionic abundance calculated from the observed emission lines. The uncertainty quoted for total abundances include -- apart from the error of the ionic abundances -- the contribution due to the uncertainties associated with the ICFs, which is not a common procedure in most works. We use error propagation to estimate the uncertainty associated with the abundance determinations made with the different ICF schemes where the value of such uncertainty is not provided, and the analytical expressions of ADIS20 to derive the errors implied for their ICFs.


For the total N abundance, we have used two ICF schemes, the most common one based on the similarity of ionization potentials of N$^+$ and O$^+$,  N/O~$\simeq\mbox{N}^+/\mbox{O}^+$~\citep{Peimbert:1969} and the new ICF by \citet[][hereinafter ADIS20]{Amayo:2020} (see \S~\ref{sec:N-ICFs}). For objects of very low degree of ionization where we do not detect [\ion{O}{iii}] lines, we assume that N/H $\approx$ N$^+$/H$^+$ \citep{Esteban:2018}.


To calculate the total abundances of C, Ne, S, Cl and Ar, we have analysed different published ICF schemes based on the ionization potentials of given ions, photoionization models or observational data (e.g.\@ \citealt{Peimbert:1969, Izotov:2006, Esteban:2015}). We have also used the new set of ICFs based on photoionization models of giant extragalactic \ion{H}{ii} regions by ADIS20  and the ICF for Cl of Dom\'inguez-Guzm\'an et al. (in preparation) derived from an empirical relation obtained making use of very deep spectra of Galactic and Magellanic Clouds \ion{H}{ii} regions.

%


\subsection{Ionization correction factors for \ion{H}{ii} regions in the literature}
\label{sec:ICFs}

\begin{figure} 
    \begin{center}
    \includegraphics[width=0.35\textwidth, trim=30 0 30 0,  clip=yes]{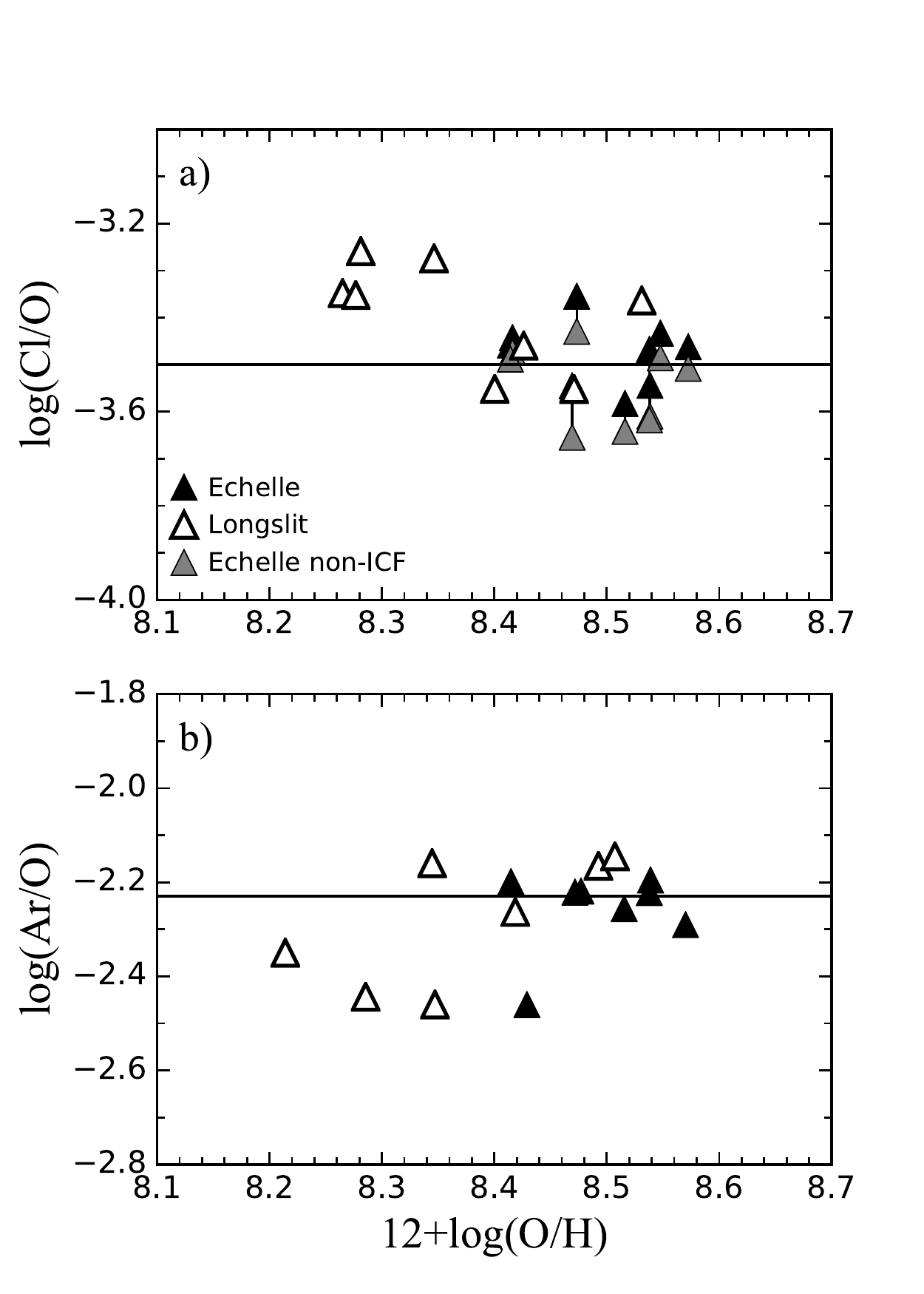}
    \caption{Cl/O and Ar/O abundance ratios as a function of O/H for a sample of Galactic \ion{H}{ii} regions. The empty symbols show regions observed with longslit spectra and filled symbols regions observed with echelle spectra. The gray symbols in panel (a) shows regions where Cl abundance has been computed without using an ICF; they are connected with a solid line with points corresponding to Cl abundance computed using an ICF for the same region. The Cl/O and Ar/O total abundances are calculated using the ICF scheme by \citet{Izotov:2006}.}  
    \label{fig:comp_ICF}
   \end{center}
    \end{figure}

In what follows we are going to briefly describe different ICF schemes available in the literature, whose results applied to our sample will be discussed in \S~\ref{sec:ICF-HII}. As we pointed out in \S~\ref{sec:intro}, in this paper we test three types of ICF schemes: based on ionization potential similarities, empirical ICFs based on observations of both Galactic and extragalactic \ion{H}{ii} regions, and ICFs based on photoionization models of integrated \ion{H}{ii} regions; these last sets of ICFs are, in principle, better suited for correcting elemental abundances in extragalactic objects, where the majority of the volume of the ionized nebula is observed. In Galactic \ion{H}{ii} regions, which have large angular extensions, the spectra usually cover a small volume of the nebula. Therefore, some of the available ICF schemes could introduce a bias in the calculations of the total abundances if the observed volume is not representative of the whole ionization structure of the nebula. In our sample, we have two different sets of observations: i) longslit spectra covering a relatively large volume on each nebula, for which it would seem appropriate to consider that the whole ionization structure is being observed, and ii) echelle spectra, that only cover a small volume of the nebula. In Fig.~\ref{fig:comp_ICF} we plot the Cl and Ar abundances obtained by applying the ICFs by \citet{Izotov:2006} in results based on longslit spectra (open symbols) or echelle spectra (filled symbols) of our sample of Galactic \ion{H}{ii} regions. Those ICFs are based on photoionization models of extragalactic \ion{H}{ii} regions. In the case of Cl, in Fig.~\ref{fig:comp_ICF} we also plot nebulae for which Cl/H has been computed by simply adding the observed ionic abundances (gray symbols). In that plot, we can see that apart from a possible metallicity dependence that can affect any kind of observation (more apparent in the case of Cl/O), the behaviour of the abundance ratios obtained from echelle or longslit spectra is basically the same.
A similar behaviour is also found for the S/O ratio.

For C, \citet{Garnett:1999} proposed an ICF(C$^{2+}$) scheme based on a grid of photoionization models of extragalactic \ion{H}{ii} regions built by \citet{Garnett:1995} to correct for the presence of C$^+$. Recently, \citet{Berg:2019} created analytic functions for different metallicities of the ICF for C based on a set of photoionization models customized to match the properties of their sample. These functions depend on the ionization parameter, that is also fitted as a function of the log([\ion{O}{iii}] $\lambda$5007/[\ion{O}{ii}] $\lambda$3727) ratio. Unfortunately, the expressions were optimized for low-metallicity objects ($Z$ from 0.05 to 0.4 $Z_{\odot}$, that corresponds to O/H ratios between 7.5 and 8.0 in the usual logarithmic scale). In this paper we have used their expression for $Z$ = 0.4 Z$_{\odot}$, which is only slightly lower than the average metallicity of our sample.  

\citet{Peimbert:1969} proposed ICFs for N, S and Ne based on the similarity of ionization potentials of different ions. These authors proposed the following ICFs for these elements: N/O = N$^{+}$/O$^{+}$, Ne/O = Ne$^{2+}$/O$^{2+}$ and S/O = (S$^+$ + S$^{2+})$/O${^+}$. 

\citet{Dors:2013} proposed an ICF(Ne$^{2+}$) based on photoionization models of extragalactic \ion{H}{ii} regions. They provided an analytical expression (their Eq. 14) to be applied for optical data.  
\citet{Perez-Montero:2007} calculated ICF schemes for Ne and Ar using a grid of photoionization models that cover different physical properties of extragalactic \ion{H}{ii} regions. Both ICF schemes depend on the ionization degree of the nebula, given as O$^{2+}$/O.
%

\citet{Stasinska:1978} used photoinization models to compute ICF(S$^+$+S$^{2+}$). This ICF depends on the ionization degree, as given by O$^{+}$/O, and of an  $\alpha$ parameter which they proposed to have a value of 3. \citet{Dors:2016} also used a grid of photoionization models of \ion{H}{ii} regions and propose the same expression for the ICF(S$^+$+S$^{2+}$) than \citet{Stasinska:1978}, but with $\alpha=3.27\pm0.01$. They recommend the use of this ICF for \ion{H}{ii} regions with O$^{+}$/O $> 0.2$, being its application less suitable for higher ionization degrees.

\cite{Izotov:2006} provided a set of ICFs based on the grid of photoionization models by \citet{Stasinska:2003} for Ne, S, Cl and Ar among other elements. Those ICFs depend on the metallicity and the ionization degree as given by $\omega$ = O$^{2+}$/O or $\upsilon$ = O$^{+}$/O. They proposed 3 different ICF fits for each element depending on the metallicity: low (12 + log O/H $\leq$ 7.2), high (12 + log O/H $\geq$ 8.2) and intermediate metallicity (7.2 $<$ 12 + log O/H $<$ 8.2) that in practice should be considered for linear interpolation with low-metallicity fits when 7.2 $<$ 12 + log O/H $<$ 7.6, and with high-metallicity fits when 7.6 $<$ 12 + log O/H $<$ 8.2. Regarding the ICF schemes we are interested in, the ICF(Ne$^{2+}$) is valid for the whole $\omega$ and $\upsilon$ ranges, while ICF(Cl$^{2+}$), ICF(S$^{+}$ + S$^{2+}$), ICF(Ar$^{2+}$) and ICF(Ar$^{2+}$ + Ar$^{3+}$) are restricted for values of $\upsilon \leq 0.8$.

For Cl, \citet{Esteban:2015} proposed an empirical ICF based on observational data of Galactic \ion{H}{ii} regions. This ICF is valid only for the range covered by that sample, $1 <$ O/O$^{2+}$ $<7$. The uncertainty associate to the ICF(Cl$^{2+}$/H$^{+}$) is 0.03 dex. 

%
%

 ADIS20 have used a grid of photoionization models available in the Mexican Million Models database \citep[3MdB][]{Morisset:2009} under the ``BOND'' reference \citep{Vale-Asari:2016} to propose a new set of ICFs for C, N, Ne, S, Cl and Ar. Their final grid was selected by applying several filters to resemble the properties of a large observational sample of extragalactic \ion{H}{ii} regions. Such ICF schemes depend on the ionization degree given by $\omega$ = O$^{2+}$/O. They also provide parametric formulas to calculate the uncertainty associated to the ICF of each element. These ICFs are, in principle, valid for all ionization degrees but, as it is commented below, results obtained from this ICF scheme for objects with $\omega \lesssim$ 0.1 should be taken with caution.
%

Finally, we analyze the empirical ICF of Dom\'inguez-Guzm\'an et al. (in preparation) for Cl based on deep observations of \ion{H}{ii} regions of the Milky Way and Magellanic Clouds. This ICF(Cl$^{2+}$/O$^+$) also depends of the ionization degree given by log([\ion{O}{iii}] $\lambda$5007/[\ion{O}{ii}] $\lambda$3727). This ICF is valid for the whole range of ionization degree of our sample. 

ICFs derived from photoionization model grids are aimed to cover a wide range of the parameter space of properties of extragalactic \ion{H}{ii} regions; unfortunately, very soft ionizing continua typical of extremely low-ionization objects are not normally taken into account in these grids, as the stellar cluster ionizing these regions can never reach the extremely soft ionizing radiation field provided by the early-type B stars or late-type O stars ionizing some of the objects in our sample. In fact, a large fraction of our sample ($\sim$27\% objects) show ionization degrees O$^{2+}$/(O$^{+}$+O$^{2+}$) $\leq$ 0.1. Therefore, we expect some limitations of the ICFs based on photoionization models when trying to reproduce the properties of these very low ionization nebulae.
In the following section, we present a detailed analysis of the suitability of the different ICF schemes described above in the determination of the total abundances of our sample of Galactic \ion{H}{ii} regions. We have studied the behaviour of the abundance ratios of C/O, N/O, Ne/O, S/O, Cl/O and Ar/O as a function of O/H using the available ICF schemes to decide which is the most appropriate for each element.  As we have taken into account the validity ranges provided for each ICF, the number of objects studied can differ within the different combinations of element/ICF. 
 One of the criteria used to analyze the suitability of the different ICFs is the comparison of the average abundance ratios of the elements with respect to O obtained for the nebulae with the solar value. However, it must be reminded that to make such comparison appropriately, the nebular ratios should be corrected for the fraction of atoms of the given element embedded
in dust grains. 

\cite{Mesa-Delgado:2009} estimated that the fraction of O 
embedded in dust grains in the Orion Nebula should be about 0.12 dex. \citet{Peimbert:2010} proposed that, in \ion{H}{ii} regions, the depletion of O increases with increasing O/H, from about 0.08 dex in the metal poorest objects, to about 0.12 dex in metal-rich ones. We consider that 0.10 dex is a good representative value for our sample. C is also expected to be depleted in dust, especially in polycyclic aromatic hydrocarbons and graphite. The C abundance in the gas and dust phases of the diffuse interstellar medium is something very poorly understood \citep[e.g.][]{Sofia:2011, Jenkins:2014}. However, \citet{Esteban:1998} estimated a dust depletion factor for C of $\sim$0.10 dex in the Orion Nebula, similar to that of O. In their study of C abundances in low metallicity star-forming regions, \citet{Esteban:2014} conclude that nebular C/O ratios can be directly compared with solar or stellar abundances without correction. 

\citet{Gail:1986} and \citet{Jenkins:2014} indicate that N is not expected to be a major constituent of dust grains, due to its inclusion in the highly stable gas form of N$_2$. Therefore, nebular N/O ratios should only be corrected for dust depletion in O. Following \citet{Esteban:2014}, we recommend to subtract $\sim$0.1 dex to the log(N/O) values obtained for \ion{H}{ii} regions when comparing with the solar ratio. The cases of Ne and Ar pose no problem at all, because they are noble gases and they are no part of molecules in the interstellar gas. As in the case of the N/O ratio, we should subtract 0.1 dex to the measured Ne/O and Ar/O ratios in  \ion{H}{ii} regions to compare with the solar ratios. Finally, in the cases of S and Cl the situation is more complicated. Customarily, S has been considered not to be depleted in dust grains \citep[e.g.][]{Sofia:1994} but some authors have questioned this assumption \citep[e.g.][]{Jenkins:2009}. The Cl abundance is difficult to calculate in diffuse and dense interstellar clouds 
and its observed trend is rather irregular \citep{Jenkins:2009}. Therefore, the fraction of S and Cl depleted in dust grains in the ISM and in \ion{H}{ii} regions in particular is actually unknown. Taking into account this, and being conservative, we will consider an additional uncertainty of 0.1 dex when comparing the nebular and solar S/O and Cl/O ratios.


\subsection{Selecting a set of ICF schemes for Galactic \ion{H}{ii} regions}
\label{sec:ICF-HII}

 We have carried out a detailed analysis of different ICFs for C, N, Ne, S, Cl, and Ar to select the most appropriate one to calculate the total abundances for our sample of Galactic \ion{H}{ii} regions. In principle, as O, Ne, S and Ar are $\alpha$-elements, the behaviour of Ne/O, S/O and Ar/O ratios {\it versus} O/H is expected to be flat. A similar behaviour is expected for Cl because it is produced by single-particle capture by a isotope of S or Ar \citep[see][]{Esteban:2015}. We have computed the average abundance ratio and its dispersion implied by the different ICFs for our sample of Galactic \ion{H}{ii} regions analyzing the possible dependencies with the ionization degree and the temperature structure \citep{Dominguez-Guzman:2019}. For C and N, we also use the C/O, N/O {\it versus} O/H relation to analyze the dispersion implied for the most used ICFs of these elements and the new ICFs by \citet{Berg:2019} and ADIS20. Finally, we compare our results with the recommended solar values of \citet{Lodders:2019} based on photospheric computations (C, N and O), meteoric results (S and Cl) or a combination of solar wind determinations and other considerations (Ne and Ar). We adopt the solar 12+log(O/H) of 8.73$\pm$0.04 as recommended by \citet{Lodders:2019}.

In Table.~\ref{Tab:ICFs}, we include a summary of the different references for the ICFs analyzed here as well as the average value and standard deviation computed for C/O, N/O, Ne/O, S/O, Cl/O and Ar/O ratios in each case. Bold numbers indicate the final adopted value of each abundance ratio. In the last row of Table.~\ref{Tab:ICFs} we include the solar values for the different abundance ratios compiled by \citet{Lodders:2019} for comparison.
 
 

\subsubsection{Carbon}

\input{Table5.tex}

\label{sec:ICF-C}

The total C abundances of the objects have been calculated from the C$^{2+}$/H$^+$ ratio determined using the \ion{C}{ii} $\lambda4267$ recombination line (hereafter RL) and applying an ICF. Ionic abundances derived from RLs have the advantage of being almost independent of $T_{\rm e}$ \citep{Osterbrock:2006,Peimbert:2017}, however they are always larger than the ionic abundances derived from CELs of the same ion. This is the well known abundance discrepancy problem and its origin is still under debate \citep[e.g.][]{Garcia-Rojas:2007, Esteban:2018b}. Only two of our sample objects have determinations of their 
C$^{2+}$ abundances obtained from UV CELs, the Orion Nebula \citep{Walter:1992} and M8 \citep{Peimbert:1993}. The small difference between the $R_{\rm G}$ values of these objects and the large uncertainty related to the C/H ratio of M8 (due to its rather low ionization degree), make it unnecessary to perform any analysis based on C abundances determined from CELs. 

For all of the nebulae having C$^{2+}$/H$^+$ ratios -- except Sh~2-152 -- we have also determined the O$^{2+}$ abundance obtained from \ion{O}{ii} RLs. We have compiled the O$^{2+}$/H$^+$ and O/H ratios calculated using RLs from \citet{Esteban:2005}, \citet{Esteban:2013} and \citet{Esteban:2017} and the physical conditions calculated in \S~\ref{sec:ionic}. In the case of Sh\,2-152, we have estimated its corresponding O/H$_{(\rm RLs)}$ assuming the average value of the difference O/H$_{(\rm RLs)}$ $-$ O/H$_{(\rm CELs)}$ = 0.19 $\pm$ 0.05 dex determined for the rest of the objects. 
Once we have the O/H ratio of the nebulae determined from RLs, we can calculate their C/O ratio free of the effect of the abundance discrepancy. Following the procedure outlined before, we calculate C abundances for 11 \ion{H}{ii} regions of our sample. We analyze the dispersion implied by the ICFs of \citet{Garnett:1999}, \citet{Berg:2019} and ADIS20, using the C/O {\it versus} O/H relation calculated using RLs for deriving C and O abundances. We have used the O$^{2+}$ and O$^+$ abundances determined from CELs to calculate the ionization degree owing to the dependence of the ICFs on this parameter. Table~\ref{tab:O-N-ab} also includes the O$^{2+}$ ionic abundances determined from RLs that have been used to derive C/O.


    \begin{figure} 
    \begin{center}
    \includegraphics[width=0.4\textwidth, trim=35 0 35 0, clip=yes]{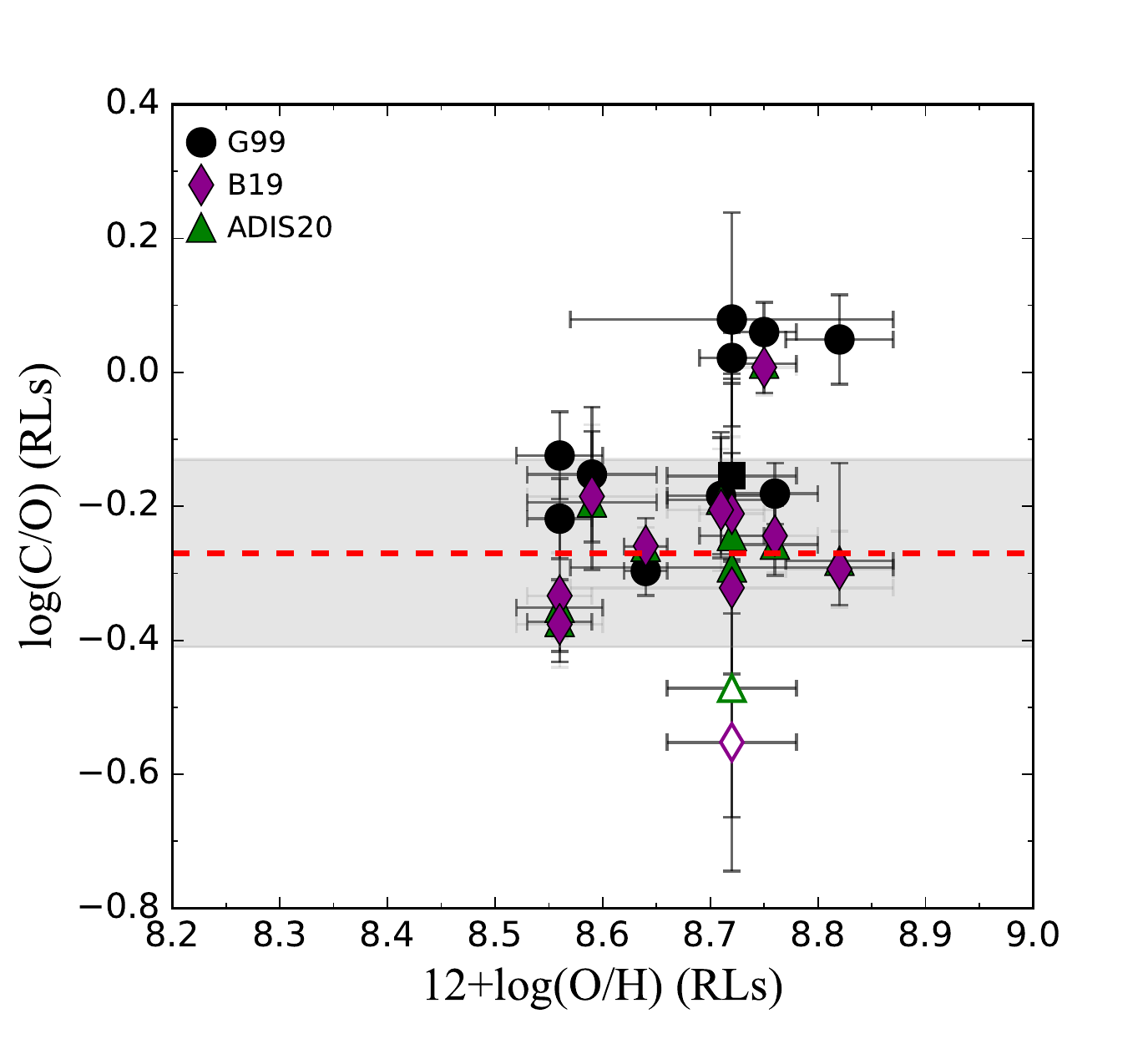}
    \caption{The C/O abundance ratio as a function of O/H for a sample of 11 Galactic \ion{H}{ii} regions. The different symbols show our calculations using the ICFs of \citealt{Garnett:1999} (G99, black dots), \citet{Berg:2019} (B19, purple diamonds) and ADIS20 (green triangles). The dashed line and the gray band shows the average value and the dispersion of log(C/O) implied by the ICF of \citet{Berg:2019}. The empty symbols identifies Sh\,2-152, for which O$^{2+}$ could not be measured from RLs, (see Sec.~\ref{sec:ionic}).}  
    \label{fig:C-icfs}
    \end{center}
    \end{figure}

Fig.~\ref{fig:C-icfs} shows the C/O abundance ratio as a function of O/H for the sample of 11 Galactic \ion{H}{ii} regions with C and O RL measurements for the different ICF schemes analyzed here; results using the scheme by \cite{Garnett:1999} are shown as black dots, those by \citet{Berg:2019} are plotted as purple diamonds and those computed using the scheme by ADIS20 as green triangles. Sh\,2-152 is identified in Fig.~\ref{fig:C-icfs} with empty symbols because, as commented before, the value of O/H (RLs) has been estimated indirectly because no \ion{O}{ii} RLs are detected in this object. In general the different ICF schemes provide similar dispersion of 0.12-0.14 dex in the C/O {\it versus} O/H relation. ICF schemes of \citet{Berg:2019} and ADIS20 provide very similar  average values (see Table~\ref{Tab:ICFs}). However, the ICF of \citet{Garnett:1999} although shows a dispersion similar to the other schemes, gives a substantially larger average log(C/O) of  $-1.00\pm0.13$. We should also note that M8, M16, and M20 show much larger values of C/O using the ICF by \citet{Garnett:1999} than using the other schemes. Additionally, all the ICF schemes provide a high (C/O $\gtrsim$ 1) ratio for M17 (see Table~\ref{tab:O-N-ab}).

In previous works, \citet{Esteban:2005} and \citet{Esteban:2013} reported C abundances for eight \ion{H}{ii} regions from this sample using the ICF scheme of \citet{Garnett:1999}. Our estimates of the C/O ratio for M16 and M20 are higher by 0.11 dex, respectively  than those reported by \citet{Esteban:2005} using the same ICF, while the rest of the sample show differences lower than 0.06 dex \citep[see also][]{Esteban:2013}. On the other hand, the ICFs of \citet{Berg:2019} and ADIS20 show results that are very consistent between them, providing small abundance differences for C/H and C/O ratios for this sample. A similar result was recently reported by \citet{Esteban:2020} using \ion{H}{ii} regions of the nearby spiral galaxies M101 and M31. 

We have adopted the C/H and C/O ratios obtained using the ICF scheme of \citet{Berg:2019} for our sample objects, and these are the values included in Tables~\ref{tab:total-abundances} and~\ref{tab:total-abundances-ratios}, respectively.
This scheme gives values very similar to the one proposed by ADIS20, although slightly higher dispersions. We decided not to adopt the values obtained with the ICF scheme by ADIS20 because it is still not published at the moment of the writing of this paper and an interested reader could not reproduce our results. We report an average value of log(C/O) = $-0.27\pm0.14$ (dashed line in panel (a) of  Fig.~\ref{fig:C-icfs}) for this sample. Such value is in excellent agreement with the solar value reported by \citet{Lodders:2019} of $-0.26\pm0.09$. As it has been discussed in \S~\ref{sec:ICFs}, no correction for dust depletion is needed to compare the nebular and solar C/O ratios. 

\subsubsection{Nitrogen}
\label{sec:N-ICFs}

  \begin{figure*} 
    \begin{center}
    \includegraphics[width=0.85\textwidth, trim=35 0 35 0,  clip=yes]{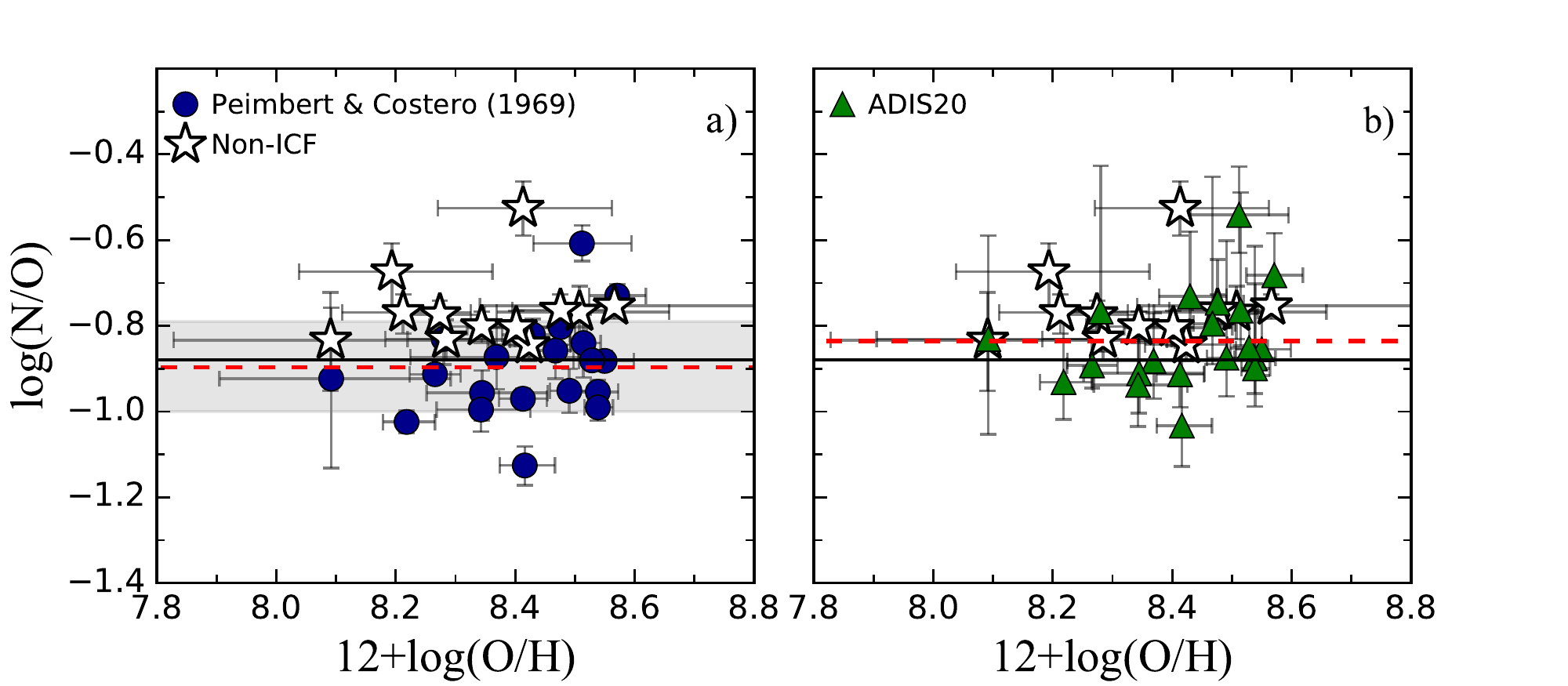}
    \caption{The N/O abundance ratio as a function of O/H for a sample of Galactic \ion{H}{ii} regions. Panel (a) N/O ratio using the classical ICF(N$^{+}$/O$^{+}$) of \citet{Peimbert:1969}.  Panel (b) N/O ratio using the  ICF(N$^{+}$/O$^{+}$) of ADIS20. Empty stars show the regions whose N/H ratio have been derived without using an ICF . The black solid line shows the solar value of N/O from \citet{Lodders:2019}. The red dashed line shows the average value of N/O of the objects. The gray band show the average and dispersion values of the sample without using the ICF of \citet{Peimbert:1969}.}  
    \label{fig:NO-icfs}
    \end{center}
    \end{figure*}

We compute the N/H and N/O ratios for a sample of 33 \ion{H}{ii} regions. For 12 of them, we calculate N/H and N/O without using an ICF, as the faintness or even absence of nebular [\ion{O}{iii}] lines in their spectra  makes the expected contribution of the N$^{2+}$/H$^+$ ratio smaller than the uncertainty of the N$^+$ abundance \citep{Esteban:2018}. In columns 8 and 11 of Table~\ref{tab:O-N-ab}, we show the N/H and N/O ratios computed without using an ICF. For the rest of the sample, we consider two ICF schemes, the first is the most common one that assumes the relation N/O = N$^+$/O$^+$ \citep{Peimbert:1969}. We also use the new ICF(N$^{+}$/O$^{+}$) proposed by ADIS20 for comparison. In columns  6, 7, 9 and 10 of Table~\ref{tab:O-N-ab} we show the results for N/H and N/O ratios using both ICF schemes.  A particular case is that of Sh\,2-212, which shows an extremely low value of log(N/O) $\approx$ $-1.57\pm0.14$ dex. From the fluxes reported in \cite{Esteban:2017}, Sh\,2-212 shows [\ion{N}{ii}] $\lambda$6583/$\lambda$6548 $\sim$ 4.8, which is much higher than the theoretical value, and indicates some problem in line measurements, probably in [\ion{N}{ii}] $\lambda$6548 line; however the extremely low N/O ratio remains whatever combination of [\ion{N}{ii}] lines used to compute $T_e$ and N$^{+}$/H$^+$ ratio. \citet{Fernandez-Martin:2017} report log(N/O) = $-$1.16 for this object, more consistent to the expected value. We do not have a clear explanation for this odd result from our spectrum of Sh\,2-212, in any case, we discarded the N/O value obtained for this region in the determination of our average values. 

In Fig.~\ref{fig:NO-icfs}, we show the N/O abundance ratio as a function of O/H for our sample of Galactic \ion{H}{ii} regions. Panels (a) and (b) show the results for the ICFs analyzed here. Empty stars are those regions of our sample whose N abundances were calculated without using an ICF. In panel (a) of Fig.~\ref{fig:NO-icfs}, we can see that N/O ratios obtained with the ICF scheme of \citet{Peimbert:1969} are systematically slightly lower than those determined without an ICF. \citet{Esteban:2018} found a similar result when using other ICFs schemes for N, as those of \citet{Izotov:2006} or \citet{Mathis:1991}. However, panel (b) of Fig.~\ref{fig:NO-icfs} shows that the log(N/O) values obtained using the ICF of ADIS20 are more consistent with those determined without ICF. Another important aspect we have to remark in Fig.~\ref{fig:NO-icfs} is the lack of any clear trend in the distribution of N/O ratios with respect to O/H  in the range of 8.00--8.60 in 12+log(O/H), as it is usually observed in other spiral galaxies (e.g., \citealt{Esteban:2020}), where we see an increase of the N/O ratio at high O/H ratios due to the secondary production of N \citep[see e.~g.][]{Matteucci:2014}. We will discuss this issue in \S~\ref{sec:comparison_other}.

The red dashed line in both panels of Fig.~ \ref{fig:NO-icfs} shows the average log(N/O) but only considering  the data points calculated using the ICF represented in each panel. We have found average values of log(N/O) of  $-0.90$ and  $-0.81$ using the ICF of \citet{Peimbert:1969} and  ADIS20, respectively, with a dispersion of 0.11 dex in both cases. These values are consistent with the solar log(N/O) of $-$0.88 $\pm$ 0.14 recommended by \citet{Lodders:2019} within the errors. On the other hand, the average log(N/O) for the objects whose N abundance has been obtained without an ICF is $-$0.76 $\pm$ 0.09. As stated in \S~\ref{sec:ICFs}, we should subtract 0.1 dex from the average nebular log(N/O) in order to compare it with the solar value. We can see that removing 0.1 dex to the average log(N/O) values obtained for the non-ICF sample or the ICF by ADIS20 we obtain a consistent comparison with the solar ratio. The opposite situation is  obtained with the  average log(N/O) determined with the ICF of \citet{Peimbert:1969}, since solar and nebular N/O ratios become inconsistent.


\subsubsection{Neon}
\label{sec:ICF-Ne}

    \begin{figure*} 
    \begin{center}
    \includegraphics[width=0.85\textwidth, trim=35 0 35 0, clip=yes]{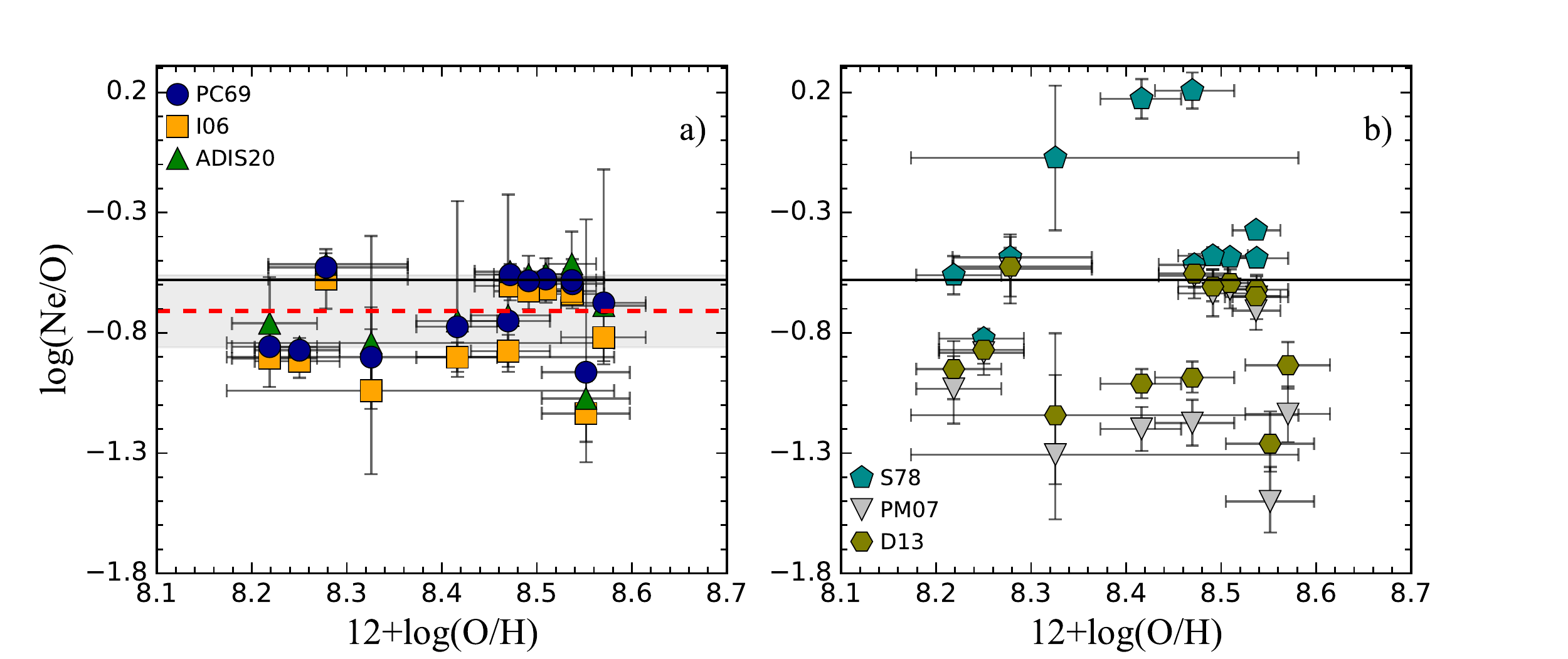}
    \caption{The Ne/O abundance ratio as a function of O/H for our sample of Galactic \ion{H}{ii} regions. The symbols show the different ICFs used to calculate the total Ne/O abundance ratio. Panel (a) shows our calculations using the ICFs of \citealt{Peimbert:1969} (PC69, blue circles), \citet{Izotov:2006} (I06, orange squares) and ADIS20 (green triangles). Panel (b) shows our calculations using the ICFs of \citet{Stasinska:1978} (S78, cyan pentagons)
     \citet{Perez-Montero:2007} (PM07, grey down-triangles) and \citet{Dors:2013} (D13, brown hexagons). The black solid line shows the solar Ne/O ratio from \citet{Lodders:2019}. The red dashed line and gray band in panel (a) show the mean and dispersion values of the data using the ICF of \citet{Peimbert:1969}.} 
    \label{fig:Ne-icfs}
    \end{center}
    \end{figure*}

\input{Table6.tex} 
\input{Table7.tex} 

 For this element, we have analyzed the ICFs proposed by \citet{Peimbert:1969}, \citet{Stasinska:1978}, \citet{Izotov:2006}, \citet{Perez-Montero:2007}, \citet{Dors:2013} and ADIS20. We have discarded M20 when we apply the ICF of \citet{Stasinska:1978} because it is out of the range of validity of this ICF. On the other hand, we have removed the results of Sh\,2-298 and Sh\,2-156 from the analysis since the values of the Ne/O ratios for these nebulae are considerably much higher and lower, respectively than for the rest of the objects. The high spectral resolution of Sh\,2-298 spectra reported by \citet{Esteban:2017} allows us to discard any blend between the [\ion{Ne}{iii}] $\lambda\lambda3869,3967$ lines and other ones, so the quoted line intensities should be confident. The explanation of the abnormally large Ne/H ratio determined for  Sh\,2-298 may come from the fact that the ionizing source of this ring nebula is a hot Wolf-Rayet star (HD56925) of the WN4 sequence \citep[although no overabundances have been previously reported in the nebula, see][]{Esteban:1989A}. The ionizing continua coming from a WN4 star is much harder than that considered as ionizing sources of models of normal \ion{H}{ii} regions. Therefore, the ionization balance for this nebula may not be well covered by the parameter space of the photoionization models used to elaborate the ICF schemes; this should be specially important for ions with high ionization potentials (IP), like Ne$^{2+}$ with an IP=40.96 eV and ICFs could be overestimating the contribution of lower ionization stages as Ne$^+$. In the case of Sh\,2-156 \citet[observed by][]{Fernandez-Martin:2017}, we have to take into account that the nebula shows one of the lowest ionization degrees of all the sample objects with detected [\ion{Ne}{iii}] lines and the lowest Ne$^{2+}$/H$^+$ ratio of the whole sample. Considering the large and uncertain extinction coefficient $c$(H$\beta$) = 1.57 $\pm$ 0.11 of the object, and the faintness of $\lambda$3869 line (the [\ion{Ne}{iii}] $\lambda$3967 line is not usable because of blending with H7 $\lambda$3970), it seems likely that the intensity ratio of $\lambda$3869 line quoted by \citet{Fernandez-Martin:2017} may be inaccurate.

Fig.~\ref{fig:Ne-icfs} shows the Ne/O abundance ratio as a function of O/H for our sample of objects. For a better comparison between the Ne/O ratios, we divided the plot in two panels, (a) and (b), that show our calculations using the different ICFs for Ne using the same scale in each plot. The solid line shows the solar Ne/O ratio \citep{Lodders:2019} and the dashed line in panel (a) shows the average value of Ne/O obtained with the ICF of \citet{Peimbert:1969}. 

In Table~\ref{Tab:ICFs} we show the average values and the standard deviation obtained with each ICF in the Ne/O {\it versus} O/H diagram. Our results show a similar scatter (between 0.15 and 0.19 dex) using the ICFs of \citet{Peimbert:1969}, \citet{Izotov:2006} and ADIS20, and these results are shown in panel (a) of Fig.~\ref{fig:Ne-icfs}. The average values of log(Ne/O) using these ICFs range from $-0.69$ to $-0.79$.  
%
%
In Fig.~\ref{fig:Ne-icfs}, 
we note an apparent slight dependence on metallicity in our sample implied by the use of the ICFs of \citet{Peimbert:1969}, \citet{Izotov:2006} and ADIS20. In principle, that dependence might be due to the ionization degree since in regions of low ionization the dispersion of Ne$^{2+}$/O$^{2+}$ is considerably enhanced \citep{Kennicutt:2003, Izotov:2006,Perez-Montero:2007, Croxall:2016a, Dominguez-Guzman:2019}. In Panel (b) of  Fig.~\ref{fig:Ne-icfs}, we present the results obtained by using the ICFs of \citet{Stasinska:1978}, \citet{Perez-Montero:2007} and \citet{Dors:2013}, which show dispersions between 0.27 and 0.39 dex, much larger than those obtained with the ICFs included in panel (a). The results derived using the ICF by 
\citet{Stasinska:1978} seem to show a strong dependence with metallicity.  

In this work, we have adopted the ICF proposed by \citet{Peimbert:1969} to obtain our final Ne/H and Ne/O ratios because it shows a good general behaviour in the Ne/O {\it versus} O/H relation and the smallest dispersion. Using that ICF, we have computed the Ne abundance in 13 \ion{H}{ii} regions of our sample (all the objects included in Table~\ref{tab:ionic-abundances} except Sh\,2-156 and Sh\,2-298). The average value of log(Ne/O) = $-0.71\pm0.15$, is marginally in agreement within the uncertainties with the solar value of log(Ne/O) = $-0.58\pm0.12$ given by \citet{Lodders:2019}, although this relative consistency breaks if we consider a decrease of 0.1 dex in log(Ne/O) to account for depletion of O onto dust grains. In any case, it should be consider that solar Ne abundance is very uncertain \citep{Lodders:2019}. In fact, there is still a debate on its value owing to the difficulties found in reconciling helioseismological measurements with the most recent photospheric abundances of light elements such as C, N, O, and Ne, which are needed to properly estimate the depth of the convection zone \citep[known as the ``solar modeling problem'', see][and references therein]{Drake:2005, Buldgen:2019}. Measurements of the Ne/O ratio in solar-like stars point to values much larger than the adopted one \citep{Drake:2005}. This would led to a much better agreement with helioseismology measurements but also to a higher disagreement with our adopted nebular Ne/O ratio. In Tables~\ref{tab:total-abundances} and~\ref{tab:total-abundances-ratios} we include the Ne/H and Ne/O ratios implied by the ICF of \citet{Peimbert:1969} for each object, respectively.

%
%

\subsubsection{Sulfur}
\label{sec:ICF-S}

  \begin{figure*} 
    \begin{center}
    \includegraphics[width=0.95\textwidth, trim=35 0 35 0, clip=yes]{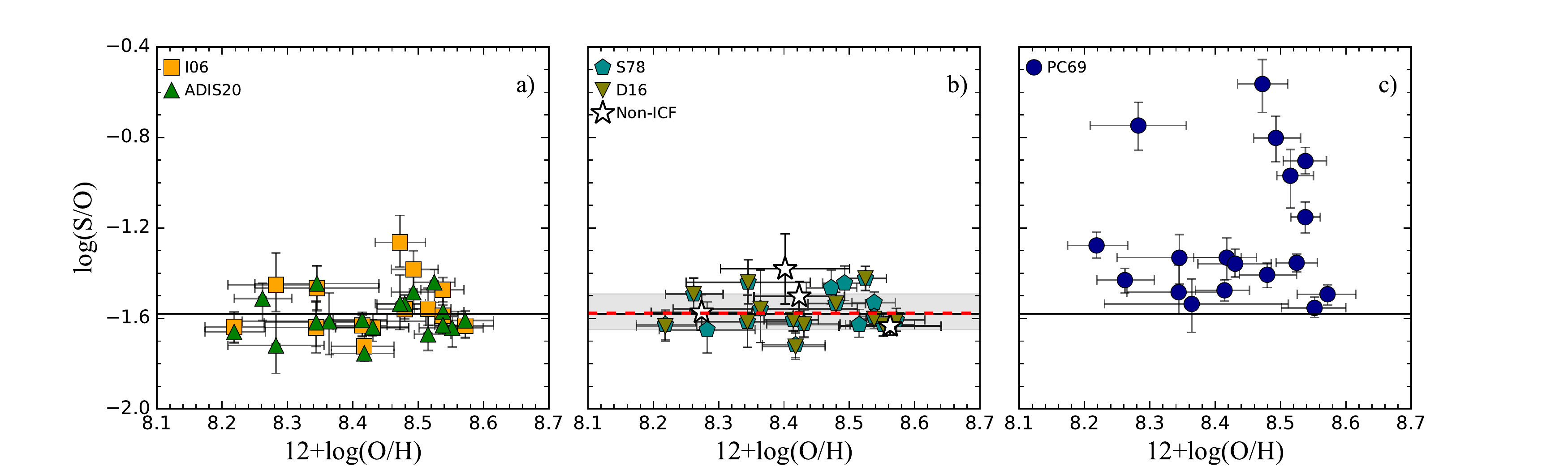}
    \caption{The S/O abundance ratio as a function of O/H for the sample of Galactic \ion{H}{ii} regions. The symbols show the different ICFs considered in the analysis. Panel (a) shows our calculations using the ICFs of \citealt{Izotov:2006} (I06, orange squares) and ADIS20 (green triangles), panel (b) shows our calculations using the ICFs of \citet{Stasinska:1978} (S78, cyan pentagons) and \citet{Dors:2016} (D16, brown down-triangles) and panel (c) shows our calculations using the ICFs of \citet{Peimbert:1969} (P69, blue circles). The empty stars show objects whose S/O ratio has been determined without using an ICF scheme. The red dashed line and gray shadow in panel (b) show the average and dispersion values of the S/O ratio obtained using the ICF by \citet{Dors:2016}. The black solid line shows the solar S/O ratio from \citet{Lodders:2019}}. 
    \label{fig:S-icfs}
    \end{center}
    \end{figure*}

We have considered the ICFs by \citet{Peimbert:1969}, \citet{Stasinska:1978}, \citet{Izotov:2006}, \citet{Dors:2016} and ADIS20 and the ionic abundances of S$^+$ and S$^{2+}$ given in Table~\ref{tab:ionic-abundances} to determine the S/H and S/O abundance ratios. As the ICFs by \citet{Izotov:2006}  and \citet{Dors:2016} are restricted to a given validity range, the number of objects for which we can apply with these ICFs is somewhat reduced from the group of objects with S$^+$ and S$^{2+}$ abundances available. We have obtained an extremely low value of the S/O ratio for Sh\,2-212, independently of the ICF used, ranging from $-1.94$ to $-2.11$. This behaviour is very similar to what found in the N/O ratio in this region, and it is clear that the main drivers of the low abundances are not the ICFs but the extremely low ionic abundances obtained for N$^+$ and S$^+$. We have discarded flux calibration problems in the wavelength region where [\ion{N}{ii}] and [\ion{S}{ii}] lines lie because the extinction derived with H$\alpha$ is in agreement with that derived from H$\delta$, and the total O/H ratio derived for this object is completely consistent with the assumed distance, which is very reliable (see \S~\ref{sec:sample}). Under this considerations, Sh\,2-212 is a clear outlier and hence, we have discarded the S/O ratio obtained for this region in the computation of the average values.


In Fig.~\ref{fig:S-icfs} we show the S/O ratio as a function of O/H for our sample of Galactic \ion{H}{ii} regions for the different ICF schemes considered for S. We have divided the figure in three panels for the sake of clarity. 
Panel (a) of Fig.~\ref{fig:S-icfs} shows the results for the ICF of \citet{Izotov:2006} (orange squares) and ADIS20 (green triangles) for a sample of 14 and 16 objects, respectively. The red dashed line shows the average value of log(S/O) using the ICF of \citet{Dors:2016}. Empty stars correspond to \ion{H}{ii} regions of very low ionization degree (O$^{2+}$/(O$^{+}$+O$^{2+}$) $\leq$ 0.03: IC~5146, Sh~2-235, Sh~2-257 and Sh~2-271) where we measure [\ion{S}{iii}] lines. In these objects, we expect that S$^{3+}$ should be absent and we can assume: S/H = (S$^+$ + S$^{2+}$)/H$^+$. Therefore, an ICF is not needed for determining their S abundance. These four objects are not considered for the calculation of the average and dispersion of the S/O values determined from each ICF scheme. Panel (b) shows the S/O abundance ratios implied by the ICFs of \citet{Stasinska:1978} (cyan pentagons) and \citet{Dors:2016} (brown down-triangles) and panel (c) shows our calculations using the ICFs of \citet{Peimbert:1969} (blue circles). 

The ICF scheme proposed by \citet{Peimbert:1969} provides the largest dispersion (0.30 dex) in the S/O {\it versus} O/H diagram, while the dispersions are much lower for the other ICFs, ranging from 0.08 to 0.12 dex. The ICFs of \citet{Stasinska:1978} and \citet{Dors:2016} show the lowest dispersion (0.08 dex). 
In Table~\ref{Tab:ICFs} we include the average values of log(S/O) using the different ICF schemes, which range from $-1.23$ to $-1.59$. Using the ICFs of \citet{Stasinska:1978} or \citet{Dors:2016} we obtain an average value of log(S/O) of $-1.57\pm0.08$, in excellent agreement with the solar abundance reported by \citet{Lodders:2019} of $-1.58\pm0.08$ dex (see Table~\ref{Tab:ICFs}). 

As we have stated in \S~\ref{sec:ionic} we have adopted the average value of $T_{\rm e}$([\ion{N}{ii}]) and $T_{\rm e}$([\ion{O}{iii}]) to calculate the S$^{2+}$ abundance rather than using the most widely accepted $T_{\rm e}$([\ion{O}{iii}]). We checked the effect of this scheme in the dispersion of S/O abundances and found that, for example, using the ICF of ADIS20 the dispersion slightly improves from 0.11 dex to 0.09 dex using this new temperature structure scheme. In the following we adopt the S/H and S/O ratios determined using the ICF of \citet{Dors:2016} as representative for the objects. This ICF gives the average S/O ratio closer to the solar one as well as the smallest dispersion. In Tables~\ref{tab:total-abundances} and~\ref{tab:total-abundances-ratios} we show the S/H and S/O ratios of 17 nebulae, respectively, which include the objects where an ICF was not used and for which the ICF of \citet{Dors:2016} was used, excluding Sh\,2-212.

\subsubsection{Chlorine}
\label{sec:Cl-ICF}
    \begin{figure*} 
    \begin{center}
    \includegraphics[width=0.8\textwidth, trim=35 0 35 0, clip=yes]{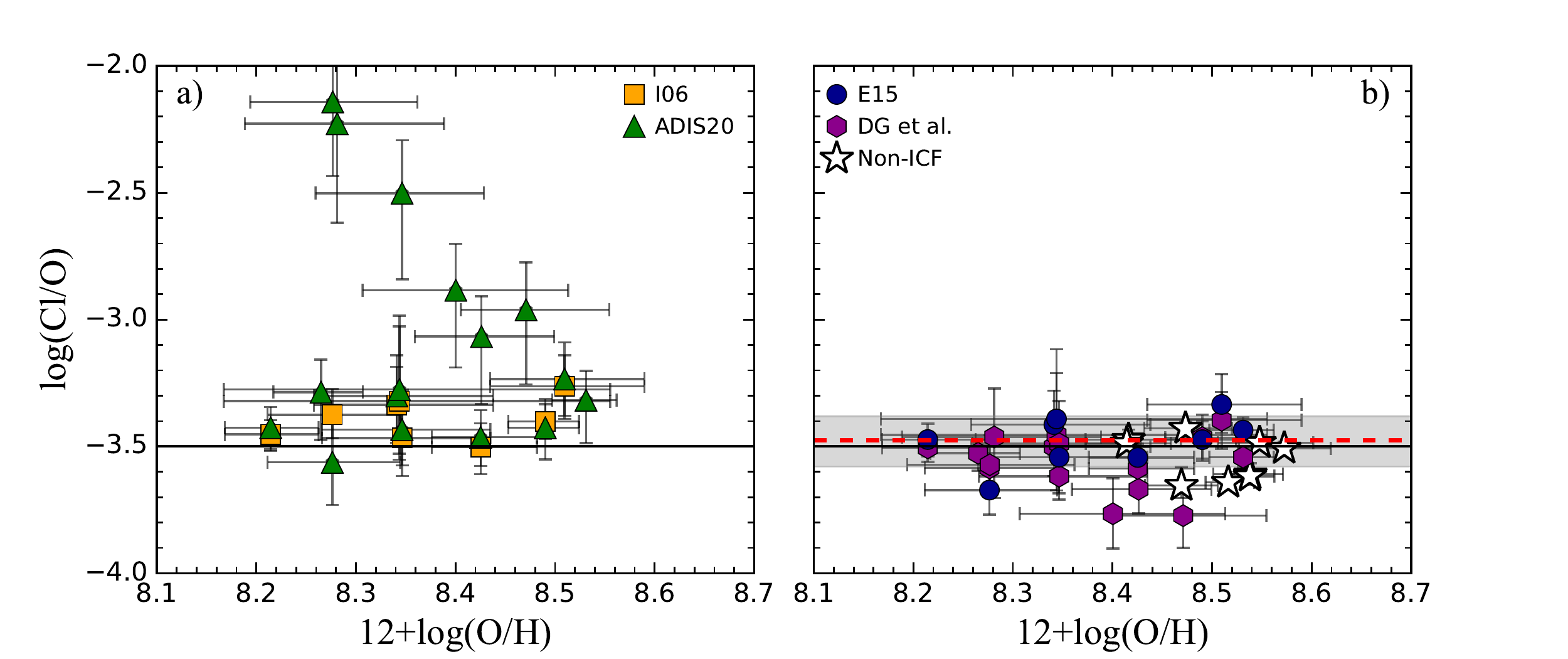}
     \caption{The Cl/O abundance ratio as a function of O/H for the sample of Galactic \ion{H}{ii} regions. The symbols show the different ICFs used to calculate the total Cl/O abundance ratio. Panel (a) shows our calculations using the ICFs of \citealt{Izotov:2006} (I06, yellow squares) and ADIS20 (green triangles). Panel (b) shows our calculations using the ICFs of \citet{Esteban:2015} (E15, blue circles) and Dom\'inguez-Guzm\'an et al. (in preparation) (DG, purple exagons). The black solid line in each panel shows the solar value of Cl/O recommended by \citet{Lodders:2019}. The red dashed line and grey band represents the average and standard deviation given by the ICF of \citet{Esteban:2015}. The empty stars show the Cl/O ratios of objects for which the Cl abundance has been determined without the use of an ICF.} 
    \label{fig:Cl-icfs}
    \end{center}
    \end{figure*}
We calculate the Cl$^{2+}$ abundances for a sample of 25 Galactic \ion{H}{ii} regions. We also obtain Cl$^{+}$ and Cl$^{3+}$ for five of the objects (M17, M42, NGC~2579, NGC~3576 and NGC~3603) permitting to determine the Cl/H ratios simply adding their Cl$^{+}$/H$^+$, Cl$^{2+}$/H$^+$ and Cl$^{3+}$/H$^+$ abundances. For M8, M16, M20 and Sh\,2-311, with O$^{2+}$/(O$^{+}$+O$^{2+}$) $\leq$ 0.26, we calculate the Cl abundance as Cl/H = Cl$^{+}$/H$^+$ $+$ Cl$^{2+}$/H$^+$ since the contribution of Cl$^{3+}$ is expected to be negligible \citep{Esteban:2015}. Therefore, for those two groups of nebulae, 9 in total, we calculate the Cl abundances without using an ICF, in the same way as \citet{Esteban:2015} did. In Table~ \ref{tab:ionic-abundances}, we present the results for Cl$^{+}$, Cl$^{2+}$ and Cl$^{3+}$ abundances (columns 5, 6 and 7). 
For the remaining 16 Galactic \ion{H}{ii} regions having only Cl$^{2+}$ abundance computations, we have used four different ICF schemes to calculate their Cl/H and Cl/O ratios: the one proposed by \citet{Izotov:2006}, the analytical ICF proposed by \citet{Esteban:2015} and the new ICFs proposed by ADIS20, and Dom\'inguez-Guzm\'an et al. (in preparation). Some of those ICFs are restricted to given validity ranges, which depend on the ionization degree of the nebulae (see \S~\ref{sec:ICFs-description}). Those \ion{H}{ii} regions outside the validity ranges were discarded from the analysis.  

Fig.~\ref{fig:Cl-icfs} shows the Cl/O abundance ratio as a function of O/H.  Panels (a) and (b) show the results for the different ICFs used to calculate Cl/O. The symbols identify the ICFs implied in this analysis. The black solid line shows the solar value of log(Cl/O) = $-3.50\pm0.09$ by \citet{Lodders:2019}. In panel (b) the empty stars correspond to the Galactic \ion{H}{ii} regions for which we have not used an ICF. We do not include this last group of nebulae for calculating the mean and dispersion of the Cl/O ratio. 

In Fig.~\ref{fig:Cl-icfs} we can see that the number of objects used with the different ICF schemes are not the same. In particular, the ICF of Dom\'inguez-Guzm\'an et al. (in preparation) and ADIS20 are used for the whole sample where Cl$^{2+}$ lines are detected, while the validity ranges for the rest of the ICFs limit the number of objects, ranging from 8 to 9 \ion{H}{ii} regions. Regarding the dispersion, the ICF of Dom\'inguez-Guzm\'an et al. gives a value of 0.11 dex; the ones obtained with \citet{Izotov:2006} and \citet{Esteban:2015} are  between 0.08 and 0.10 dex but the ICF by ADIS20 gives a significantly higher dispersion of 0.44 dex, which is mainly produced by the very low-ionization objects of our sample, for which ICFs based on photoionization models of giant \ion{H}{ii} regions may be inadequate (see \S~\ref{sec:ICFs}). We find average values of log(Cl/O) ranging from $-3.09$ to $-3.56$, all determinations except that obtained with ADIS20 are in good agreement with  the solar value of log(Cl/O) = $-3.50 \pm 0.09$ recommended by \citet{Lodders:2019}. We have selected the ICF proposed by \citet{Esteban:2015} as representative for the Cl abundances as it provides a reasonable dispersion and its average Cl/O ratio is consistent with the value obtained with the objects for which we do not apply an ICF. In panel (b) of  Fig.~\ref{fig:Cl-icfs}, the red dashed line and grey band represent the average value and the dispersion of the Cl/O ratio obtained using the ICF by \citet{Esteban:2015}, respectively. In Tables~\ref{tab:total-abundances} and~\ref{tab:total-abundances-ratios} we show the Cl/H and Cl/O ratios of 19 nebulae, respectively, which include the objects where an ICF was not used and for which the ICF of \citet{Esteban:2015} was used. 

\subsubsection{Argon}
\label{sec:Ar-ICF}

    \begin{figure*} 
    \begin{center}
    \includegraphics[width=0.75\textwidth, trim=35 0 35 0, clip=yes]{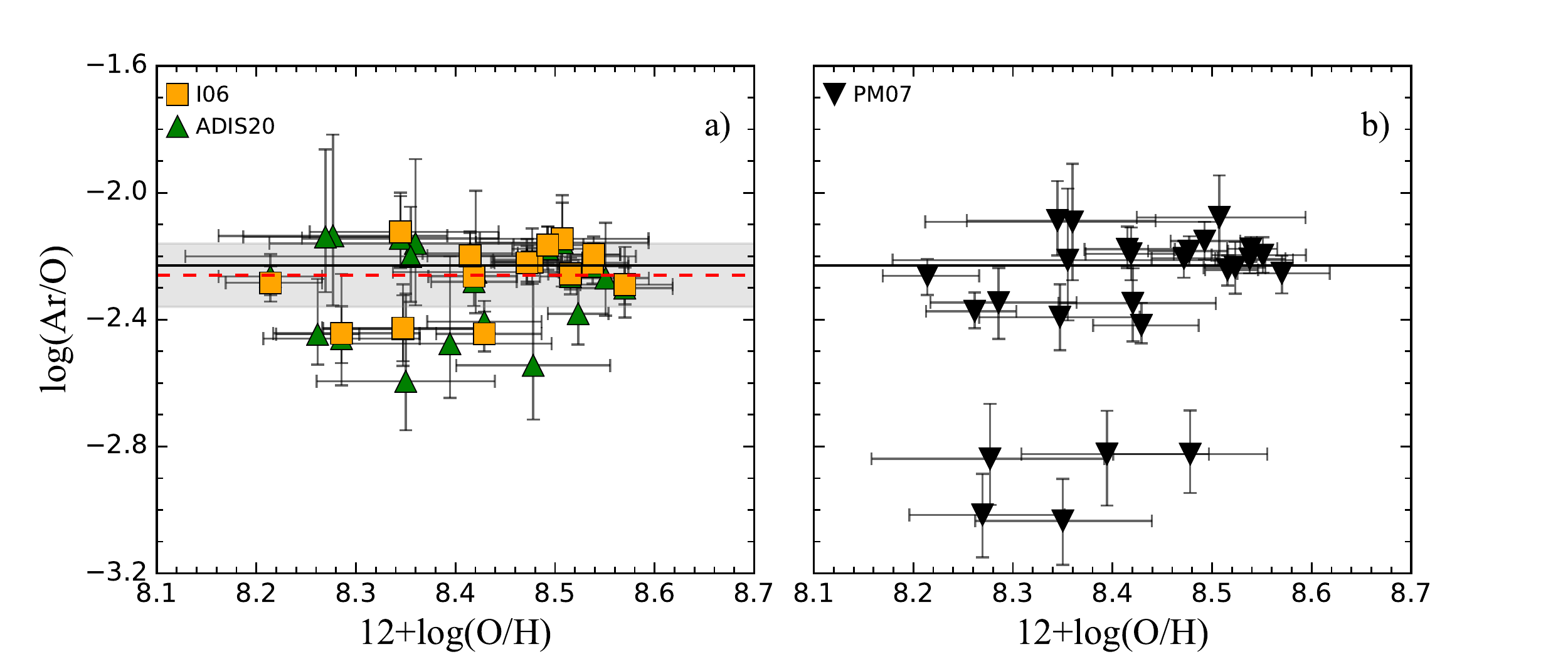}
    \caption{The Ar/O abundance ratio as a function of O/H for our sample of Galactic \ion{H}{ii} regions. The symbols show the different ICFs used in the analysis. Panel (a) shows our calculations using the ICFs of \citealt{Izotov:2006} (I06, orange squares) and ADIS20 (green triangles). Panel (b) shows our calculations using the ICFs of \citet{Perez-Montero:2007} (PM07, black down-triangles). The black solid line shows the solar value of \citet{Lodders:2019}. The dashed line and gray band in panel (a) show the average and dispersion values obtained using the ICF of \citet{Izotov:2006}.} 
    \label{fig:Ar-icfs}
    \end{center}
    \end{figure*}

We have computed the Ar$^{2+}$ abundances in 26 Galactic \ion{H}{ii} regions of our sample and Ar$^{3+}$ abundances in 10 of them. In Table~\ref{tab:ionic-abundances}, we present the Ar$^{2+}$ and Ar$^{3+}$ abundances for this sample (columns 8-9). For this element, we have compared the results obtained using three different ICF schemes: the ones by \cite{Izotov:2006}, \citet{Perez-Montero:2007} and ADIS20. As for other elements, the validity ranges of the ICFs have been also considered, reducing the number of objects in some cases (see below). It is worth mentioning that, similarly to other ions, we have used the mean value between $T_{\rm e}$([\ion{O}{iii}]) and $T_{\rm e}$([\ion{N}{ii}]) as representative for Ar$^{2+}$ as recommend by \citet{Dominguez-Guzman:2019} due to the application of such temperature reduces the dispersion in the Ar/O versus O/H diagram for all the ICF schemes. 

Fig.~\ref{fig:Ar-icfs} shows the Ar/O abundance ratio as a function of O/H for our sample of Galactic \ion{H}{ii} regions. Panels (a) and (b) show the values obtained using the ICF of \citet{Izotov:2006} (orange squares), ADIS20 (green triangles) and \citet{Perez-Montero:2007} (black down-triangles), respectively. The red dashed line represents the average value of our sample using the ICF of \citet{Izotov:2006} and the black solid line the solar value of log(Ar/O) = $-2.23\pm0.12$ by \citep{Lodders:2019}. We find a dispersion of 0.11 dex and 0.14 dex using the ICFs by \citet{Izotov:2006} and ADIS20 for 15 and 26 objects, respectively. The results using the ICFs of \citet{Perez-Montero:2007} provides the largest dispersion, of 0.29 dex, but this is because those authors do not provide a validity range for their ICF. The objects with log(Ar/O) $\leq$ $-$2.7 in Panel (b) of Fig.~\ref{fig:Ar-icfs} depart considerably from the rest of the sample. They are Sh\,2-219, Sh\,2-237, Sh\,2-257, Sh\,2-271 and Sh\,2-297, \ion{H}{ii} regions with very low  ionization degree. These objects were excluded in panel (a) because they are outside the validity ranges of the ICF by \citet{Izotov:2006}. We have recomputed the dispersion implied by the ICF of \citet{Perez-Montero:2007} discarding these low-ionization nebulae, obtaining a significantly lower dispersion (0.10 dex), of the same order of the values obtained with the other ICFs.

 
In Table~\ref{Tab:ICFs} we include the average and dispersion values of Ar/O obtained with each ICF scheme used, finding average log(Ar/O) values that range from  $-2.27$ to $-2.36$, all in good agreement with the solar value of log(Ar/O) = $-2.23\pm0.12$ reported by \citet{Lodders:2019}. However, Ar is a noble gas and we expect a log(O/H) about 0.10 dex less in the ionized gas due to depletion onto dust grains (see \S~\ref{sec:ICFs}), therefore, we should subtract that amount to the nebular Ar/O to compare with the solar ratio. Doing so, we can see that -- as also happens with Ne/O -- the agreement with the solar Ar/O would become worse, calling into question our assumptions for the fraction of O depleted onto dust. Our final adopted Ar abundances are those obtained using the ICF of \citet{Izotov:2006}.The results for Ar/H and Ar/O obtained for the 26 objects for which we can apply that ICF are listed in Tables~\ref{tab:total-abundances} and \ref{tab:total-abundances-ratios}, respectively.

\section{Radial abundance gradients}
\label{sec:Abundance-gradients}
\input{Table8.tex} 

We have performed least-squares linear fits to $R_{\rm G}$ -- or the fractional galactocentric distance with respect to the disc effective radius, $R_{\rm G}$/$R_{\rm e}$ -- of the abundances of the different elements for the sample objects collected in Tables~\ref{tab:total-abundances} and \ref{tab:total-abundances-ratios}. We have taken into account the uncertainties in the abundances and $R_{\rm G}$ using Monte Carlo simulations following the procedure of \citet{Esteban:2017} and \citet{Esteban:2018}. Table~\ref{tab:gradient-results} lists the parameters of the radial distributions of the abundance of each element as well as of their abundance ratio with respect to O. For each element we give the slope and intercept of the fits and the dispersion of the individual observational data around the fit. For some elements we have included two determinations of the gradients attending to the reasons outlined in each case. 
\subsection{C, N and O abundance gradients}
\label{sec:CNO-gradients}

\begin{figure*} 
    \begin{center}
     \includegraphics[width=0.9\textwidth, trim=35 0 35 0, clip=yes]{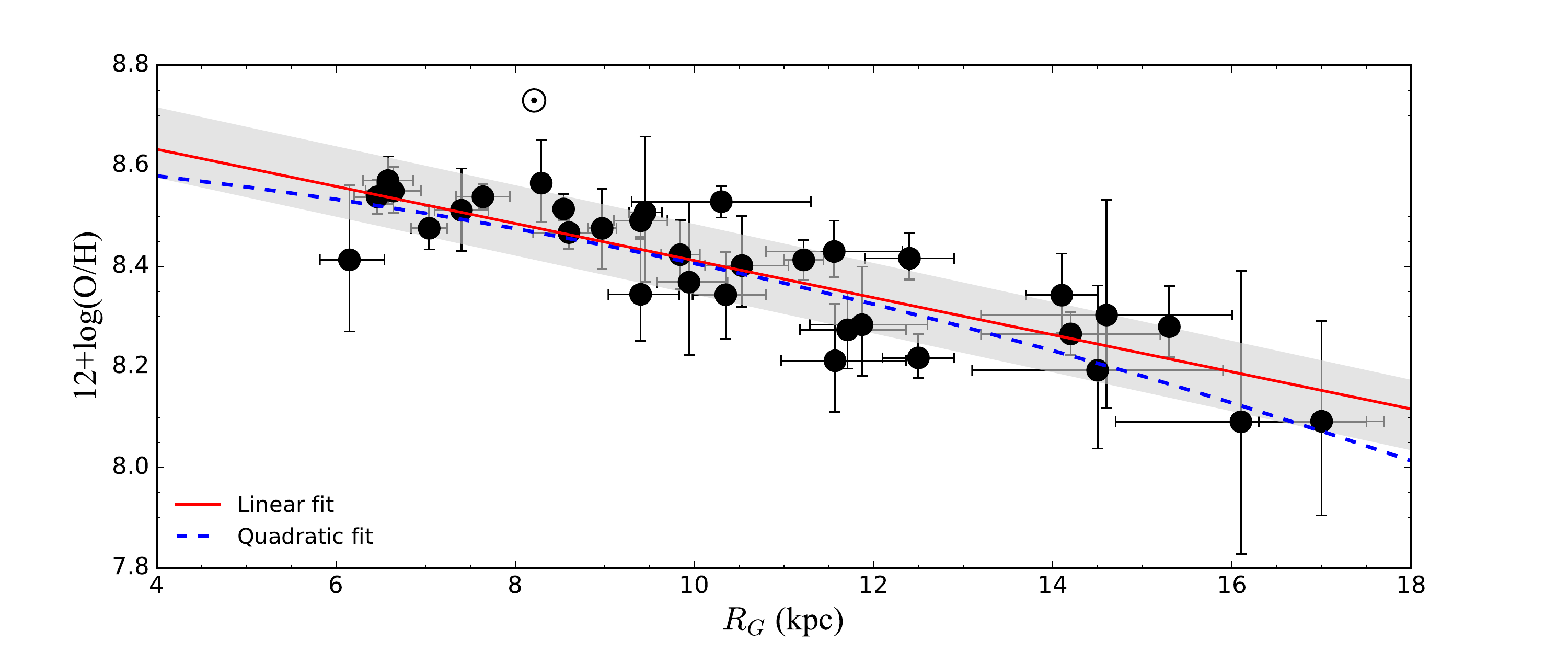}    
    \caption{The O abundances of 33 Galactic \ion{H}{ii} regions as a function of their $R_{\rm G}$. The solid and dashed lines indicate the least-squares linear fit and quadratic fit to the data. The solar symbol at 8.2 kpc indicate the photosphere abundance of O given by \citet{Lodders:2019}. The gray band represents the dispersion of the individual observational data around the fit, see Table~\ref{Tab:ICFs}.}  
    \label{fig:Oxygen-gradient}
    \end{center}
    \end{figure*}

The most recent determination of the radial O abundance gradient from Galactic \ion{H}{ii} regions is by \citet{Esteban:2018}, which is based on the same sample of objects as the present work. As it is commented in \S~\ref{sec:sample}, our revision of distances considering $Gaia$ DR2 parallaxes provides a range of $R_{\rm G}$  from 6.15 to 17 kpc for the sample, slightly different than the range considered by \citet{Esteban:2018}. In Table~\ref{tab:gradient-results} we can see that the slope of the radial O gradient with the new set of distances and a recalculation of physical conditions and abundances is of $-0.037\pm0.009$ dex kpc$^{-1}$ consistent with the value of $-0.041\pm0.006$ dex kpc$^{-1}$ obtained by \citet{Esteban:2018}. Fig.~\ref{fig:Oxygen-gradient}, shows the radial distribution of the O/H ratio of the sample objects and the corresponding radial gradient. 
In general, the values of the slope of the Galactic O abundance gradient determined in other recent studies based on optical or infrared spectra of \ion{H}{ii} regions are fairly similar. These values are between $-$0.040 and $-$0.060 dex kpc$^{-1}$ for objects covering a wide range of Galactocentric distances \citep{Deharveng:2000, Rudolph:2006, Quireza:2006, Balser:2011, Esteban:2015, Fernandez-Martin:2017, Esteban:2017}. The scatter around the fit for some of those determinations are about 0.16-0.20 dex or even higher \citep{Rudolph:2006, Quireza:2006}. It was claimed that such high dispersion could be real and not due to observational issues, implying that the gas is not well mixed. In this work, we have calculated a dispersion around the O abundance gradient of 0.07 dex, a value slightly larger than the one calculated by \citet{Esteban:2017} or \citet{Esteban:2018} of 0.05 dex, but consistent with the typical uncertainties of the O/H determinations in individual objects (see Table~\ref{tab:O-N-ab}). This result strongly suggest that O is rather well mixed in the Galactic ISM and discards the presence of significant azimuthal variations or gradients, at least in the quadrant of the Milky Way covered with the optical spectroscopical data used in this paper (see Fig.~\ref{fig:dist_map}).
On the other hand, \cite{Esteban:2018} reported an apparent break of the O/H gradient for the objects located at $R_{\rm G}$ $\leq$ 8 kpc implied by a flatter or even slightly positive slope of the gradient in the inner Galactic disc.  Chemical evolution models by \citet{Carigi:2019} reproduce that apparent inner flattening with an inside-out quenching of star formation rate in the 3$-$5 kpc range starting about 9 Gyr ago. However, the use of new determinations of Galactocentric distances at $R_{\rm G} < 8$ kpc discussed in \S\ref{sec:sample} (see also Mendez-Delgado et al. 2020) makes the possible presence of such break in the gradient less evident. In fact, a single data point, the comparatively lower O/H ratio of Sh\,2-61 -- the \ion{H}{ii} region of the sample with the smallest $R_{\rm G}$ -- is the only one that really contribute to the visual impression of a possible break.  In this sense, more deep spectroscopic observations of \ion{H}{ii} regions at $R_{\rm G} < 8$ kpc will provide a better constrain of the true behavior of the O abundance gradient in the inner part of the Galactic disc. This project is underway and the results will be presented in a future paper (Arellano-C\'ordova et al., in preparation).  

In Fig.~\ref{fig:Oxygen-gradient}, we have also included a quadratic fit to the 12+log(O/H) {\it versus} $R_{\rm G}$ distribution, which function is 12+log(O/H) = 8.64 $-$ 0.009 $\times$ $R_{\rm G}$ $-$ 0.0014 $\times$ $R_{\rm G}^2$. In the figure, we can see that the quadratic fit does not seem to improve the functional fitting of the data with respect to the simpler linear one. This is further supported by the fact that the dispersion of the data around the quadratic fit is about 0.07 dex, the same value we obtain in the case of the linear fit. In any case, the parabolic curve gives a slightly flatter gradient in the inner zones ($R_{\rm G}$ $<$ 8 kpc) and a steeper one in the external disc past 14 kpc.  \citet{Henry:2010} and \citet{Carigi:2019} also give quadratic fits to the O/H gradient of the Milky Way. The fitting parameters obtained by \citet{Henry:2010}, who include a large sample of data for \ion{H}{ii} regions and PNe, are fairly similar. The quadratic fit obtained by \citet{Carigi:2019} for their sample of \ion{H}{ii} region abundances determined from CELs using the direct method (the same used here, their equation 2), is similar for the external zones, but flatter in the inner Galaxy because they use the data by \citet{Esteban:2018} who, as it has been commented before, do not use revised distances using {\it Gaia} DR2 parallaxes.

In Fig.~\ref{fig:Oxygen-gradient}, we can see that the solar O/H ratio recommended by \citet{Lodders:2019} is 0.25 dex above the line representing the O/H gradient. This is an odd result, because one would expect the solar photospheric O/H ratio to be lower than the present-day one because the Sun was formed about 4.6 Gyr ago. Chemical evolution models of the Milky Way by \citet{Carigi:2005} estimates the O/H ratio in the solar vicinity has increased about 0.13 dex since the Sun was formed. However, this abundance difference due to age might be somewhat altered if Sun have migrated away from the $R_{\rm G}$ where it was formed \citep[e.g.][]{Wielen:1996, Portegies_Zwart:2009}. Recent dynamical evolution models of the Milky Way \citep{Martinez-Medina:2017} predict that the chemical effect may be negligible, because the displacement seems not significant, 7.7 $\pm$ 1.4 kpc at 1$\sigma$ for the $R_{\rm G}$ corresponding to the solar birthplace. To make things more tangled, the aforementioned increase in the local O/H ratio since the formation of the Sun is not consistent with the O abundances determined for young B-stars of the Orion OB 1 association and other zones of the solar vicinity, which are 8.76 $\pm$ 0.03 and 8.77 $\pm$ 0.05, respectively \citep{Nieva:2011, Nieva:2012},  consistent within the errors with the solar value of 8.73 recommended by \citet{Lodders:2019}.

In agreement with \citet{Carigi:2019}, we think that the offset of about 0.25 dex between the nebular and solar abundances may be explained by the combination of two effects: (a) the presence of some O depleted onto dust grains in the \ion{H}{ii} regions (in \S~\ref{sec:ICFs} we stated that this might account for 0.1 dex) and (b) the effect of abundance discrepancy, that would mean that the abundances determined from CELs are lower than true ones (see \S~\ref{sec:ICF-C}), better represented by the abundances determined from RLs. Considering the results of several previous papers on Galactic \ion{H}{ii} regions \citep{Esteban:2004, Esteban:2013, Garcia-Rojas:2004, Garcia-Rojas:2005,Garcia-Rojas:2007, Garcia-Rojas:2006}, we estimate that the average abundance discrepancy for O/H is about 0.20 dex. As we can see, the effects (a) and (b) point out in the same direction, and therefore they contribute to make the nebular O/H abundances about 0.3 dex lower than true ones. Considering the discussion above, the combination of (a), (b) and the abundance difference due to age (although rather uncertain)  can explain the observed 0.25 dex offset.

\begin{figure} 
    \begin{center}
     \includegraphics[width=0.45\textwidth, trim=35 0 35 0, clip=yes]{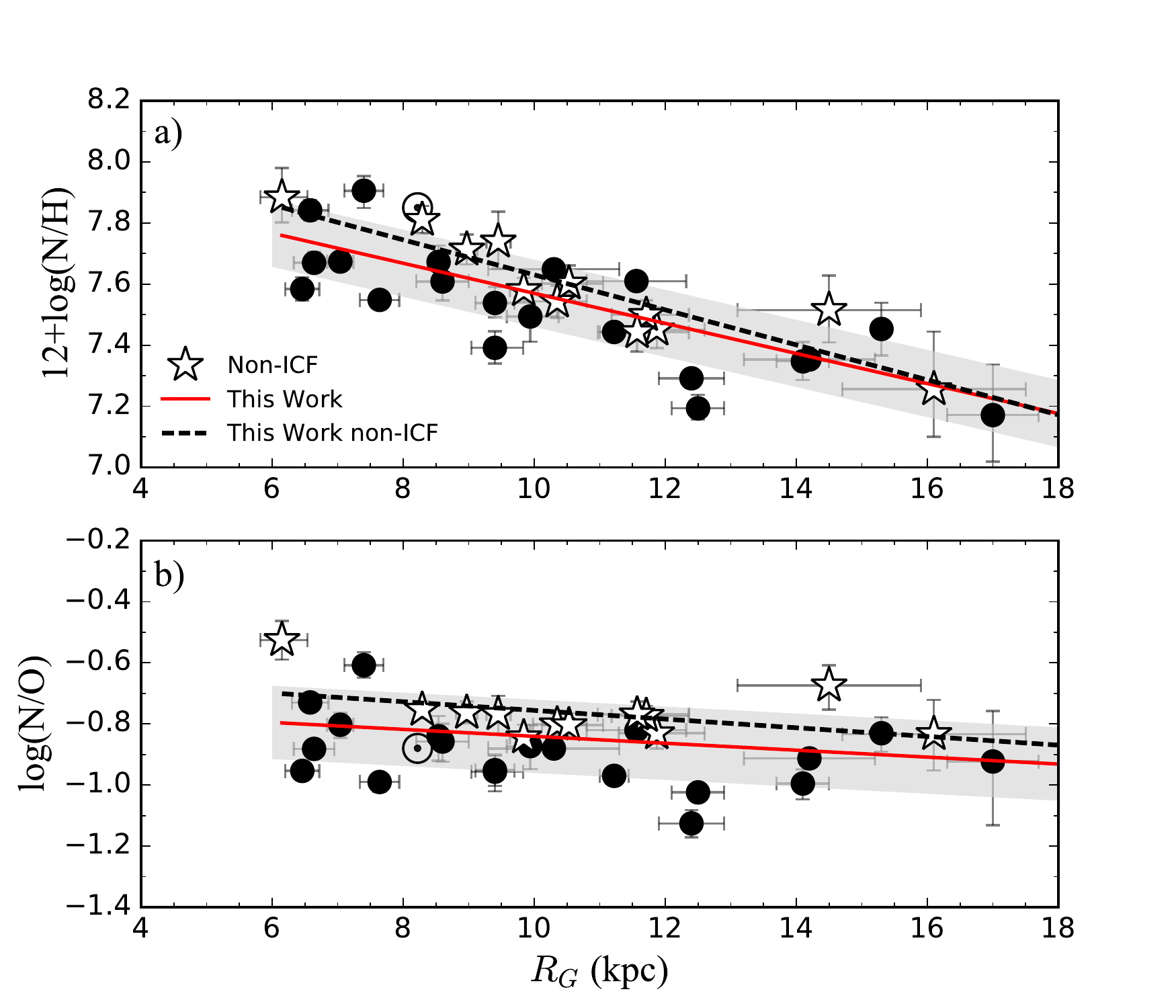}    
    \caption{The N abundance and N/O ratio as a function of $R_{\rm G}$ for \ion{H}{ii} regions in the Milky Way. Black circles represent our results using the ICF of \citet{Peimbert:1969} and empty stars represent the objects for which we do not use ICF for determining their N/H ratio. The solid red line indicates the least-squares linear fits to the data for all the sample objects. The black dashed line represents the same but only for the N/H determinations without ICF. The solar symbol at 8.2 kpc indicate the photosphere abundance of N/H and N/O given by \citet{Lodders:2019}. The gray band represents the dispersion of the individual observational data around the fit, see Table~\ref{Tab:ICFs}.}  
    \label{fig:nitrogen-gradient}
    \end{center}
    \end{figure}

Panels (a) and (b) of Fig.~\ref{fig:nitrogen-gradient} show the distributions of the N/H and N/O ratios of the \ion{H}{ii} regions included in Table~\ref{tab:O-N-ab} (with the exception of Sh\,2-212, see \S~\ref{sec:N-ICFs}) as a function of their $R_{\rm G}$, respectively. The N abundances have been calculated using the ICF proposed by \citet{Peimbert:1969}, except for those objects where -- following  \citet{Esteban:2018} and due to their very low ionization degree -- their N/H ratio was derived without using an ICF (empty  stars in Fig.~\ref{fig:nitrogen-gradient}). For the whole sample, We obtain a slope of  $-0.049\pm0.007$ dex kpc$^{-1}$ for the N abundance gradient and $-0.011\pm0.006$ dex kpc$^{-1}$ for the N/O one (see also Table~\ref{tab:gradient-results}). \citet{Esteban:2018} determine the N/H gradient using three different ICFs schemes \citep{Peimbert:1969, Mathis:1991, Izotov:2006} finding slopes between $-$0.047 and $-$0.050 and typical uncertainties of about $\pm$0.008, in agreement one each other and with our results. The same agreement is also applicable for the slope of the N/O gradient. Considering only the objects whose N/H ratio was derived without using an ICF, our slopes for N/H and N/O gradients are $-0.057\pm0.016$ dex kpc$^{-1}$ and $-0.014\pm0.010$ dex kpc$^{-1}$, respectively (the parameters of this fit are also included in Table~\ref{tab:gradient-results}). In Fig.~\ref{fig:nitrogen-gradient}, we can see that the solar N/H ratio recommended by \citet{Lodders:2019} is 0.19 dex above the line representing the N/H gradient. Since no significant dust depletion is expected for N (see \S~\ref{sec:ICFs}), this might be explained by  the effects of the abundance discrepancy. As we made in the case of O, considering previous published results for Galactic \ion{H}{ii} regions and assuming that the abundance discrepancy is produced by temperature fluctuations, the expected underestimation of the N/H ratio should be about 0.20 dex, in quite agreement with the observed offset. In the case of the N/O ratio, the offset between the recommended solar value and the gradient line is $-$0.06 dex and the depletion of O atoms onto dust grains can successfully account for it since temperature fluctuations are not expected to appreciably affect the N/O ratio.
%

 The slope of the N/O gradient is slightly negative and even almost flat within the uncertainties. This behaviour is consistent with some previous determinations based on optical spectra of \ion{H}{ii} regions \citep{Shaver:1983, Esteban:2018} but not with others based on FIR spectra. Contrary to what happens in the optical, [\ion{N}{iii}] lines are the only N lines observable in the FIR range and an ICF to correct for the presence of unseen N$^+$ is needed to derive the total N abundance in this case. \citet{Simpson:1995} and \citet{Rudolph:2006} obtained a clearly negative gradient or a step fit with two levels of N/O, indicating a substantial enhancement of the N/O ratio in the inner part of the Galactic disc. 
 As it was discussed by \citet{Esteban:2018}, an almost flat N/O gradient is an odd result; however, M31 shows a similar behaviour. The only determinations of the N/O gradient based on the measurement of $T_{\rm e}$ in \ion{H}{ii} regions of M31 also obtain rather flat slopes \citep{Zurita:2012,Esteban:2020}, with an average N/O ratio similar to the Galactic value. This result would indicate that the bulk of N should have a primary origin in these two galaxies. Different extensive studies using strong-line methods for estimating abundances indicate that the vast majority of spiral  galaxies show a negative radial N/O gradient \citep[e.g.][]{Pilyugin:2004,Perez-Montero:2016,Belfiore:2017}. However, \citet{Perez-Montero:2016} using spectra of \ion{H}{ii} regions of a sample of 350 spiral galaxies from the CALIFA survey, find that about 4-10\% of them display a flat or positive N/O gradient. Those authors  do not find differences in the average integrated properties of the subset of galaxies with flat or inverted N/O gradient and their whole sample, concluding the absence of special features that could explain this odd behavior. Although the origin of the observed trend between the N/O and O/H ratios is a controversial issue \citep[e.g.][]{McCall:1985, Thuan:1995, Henry:2000, Izotov:2006, Vincenzo:2016, Vincenzo:2018}, the presence of a flat N/O {\it versus} O/H trend in a galactic disc has a difficult -- and perhaps non-unique -- explanation. One possibility has been proposed by \citet{Vincenzo:2018}, who performed  hydrodynamical cosmological simulations of galaxies using the stellar yield set by \citet{Kobayashi:2011}, finding almost flat trends in N/O {\it versus} O/H diagrams when ignoring the contribution of failed supernovae. On the other hand, \citet{Vincenzo:2016} highlight the  importance of the star-formation efficiency (SFE) in regulating the value of the N/O ratio for a given O abundance. A lower SFE results in a lower production of oxygen per unit time by massive stars, decreasing the O/H ratio of the corresponding regions. As the N enrichment should be delayed with respect to O, the N/O ratio will increase. Therefore, radial changes of SFE along the galactic discs may -- in principle -- produce the observed behavior.

 \begin{figure} 
    \begin{center}
        \includegraphics[width=0.45\textwidth, trim=35 0 35 0, clip=yes]{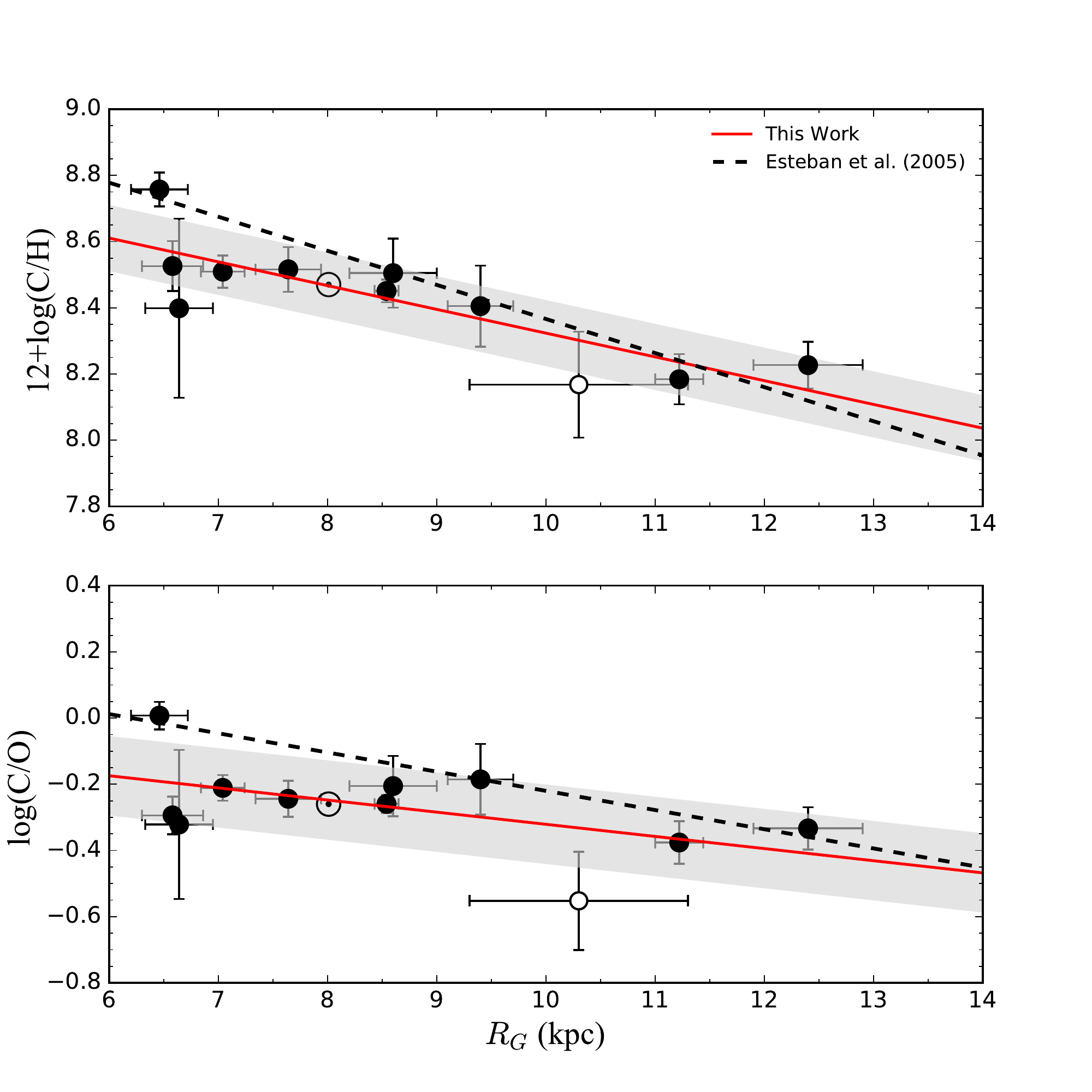} 
    \caption{The C abundance and C/O ratio as a function of $R_{\rm G}$ for \ion{H}{ii} regions in the Milky Way. The solar symbol at 8.2 kpc indicates the photosphere abundances for C given by \citet{Lodders:2019}. The red solid line indicate the least-squares linear fits to the data and the black dashed line the previous gradient determination by \citet{Esteban:2005}. The empty circle shows the Sh\,2-152, whose value of O/H determined from RLs was estimated by using an average value of the abundance discrepancy factor, see sec.~\ref{sec:ICF-C}. The gray band represents the dispersion of the individual observational data around the fit, see Table~\ref{Tab:ICFs}.}
    \label{fig:Carbon-gradient}
    \end{center}
    \end{figure}

Fig.~\ref{fig:Carbon-gradient} shows the C/H and C/O ratios for 11 \ion{H}{ii} regions as a function of their $R_{\rm G}$ using -- as it was stated in \S~\ref{sec:ICF-C} -- the ICF proposed by \citet{Berg:2019}. We have made a least-square linear fit to the data obtaining slopes of $-0.072\pm0.018$ dex kpc$^{-1}$ for C/H and $-0.037\pm0.016$ dex kpc$^{-1}$ for C/O. Using the ICF by \citet{Garnett:1999}, \citet{Esteban:2005} derived the gradients of C/H and C/O ratios using eight Galactic \ion{H}{ii} regions that form part of our own sample. These authors obtained a slope of $=-0.103\pm0.018$ dex kpc$^{-1}$ and $=-0.058\pm0.018$ dex kpc$^{-1}$ for the radial gradient of C/H and C/O, respectively, which are steeper than the ones we calculate. Fig.~\ref{fig:Carbon-gradient} also shows the comparison between the gradients for C/H and C/O ratios determined in this work and the ones obtained by \citet{Esteban:2005}. As we already discussed in \S~\ref{sec:ICF-C}, the main difference between the C/H ratios obtained by \citet{Esteban:2005} and our estimates is the ICF scheme used. 
Our new estimates of the C/H and C/O ratios show similar scatter around the computed gradients of of 0.10-0.11 dex, which is only marginally larger than the typical observational uncertainty of each magnitude (about $\pm$0.08 dex). It is easy to note that the value of the slope in both panels of Fig.~\ref{fig:Carbon-gradient} is rather dependent on the position of the \ion{H}{ii} region M17, which is located at the smallest $R_{\rm G}$ and shows the highest C/H and C/O ratios. In Fig.~\ref{fig:Carbon-gradient} it is evident that the solar C/H and C/O ratios recommended by \citet{Lodders:2019} coincide with the nebular values predicted by the gradients at the $R_{\mathrm G}$ of the Sun. This behaviour can be explained attending to two reasons. Firstly, as it was stated in \S~\ref{sec:ionic}, the C/H ratio has been determined from RLs (as well as the C/O ratio, because the O abundance used to derive that magnitude has also been determined from RLs), and the  abundance discrepancy would not affect the abundances and their ratios. Secondly, from what was discussed in \S~\ref{sec:ICFs}, we expect that dust depletion onto dust grains affects O/H and C/H ratios in a similar way.

\subsection{Ne, S, Cl and Ar abundance gradients}
\label{sec:alpha-gradients}
  \begin{figure*} 
    \begin{center}
    \includegraphics[width=0.99\textwidth, trim=35 0 35 0, clip=yes]{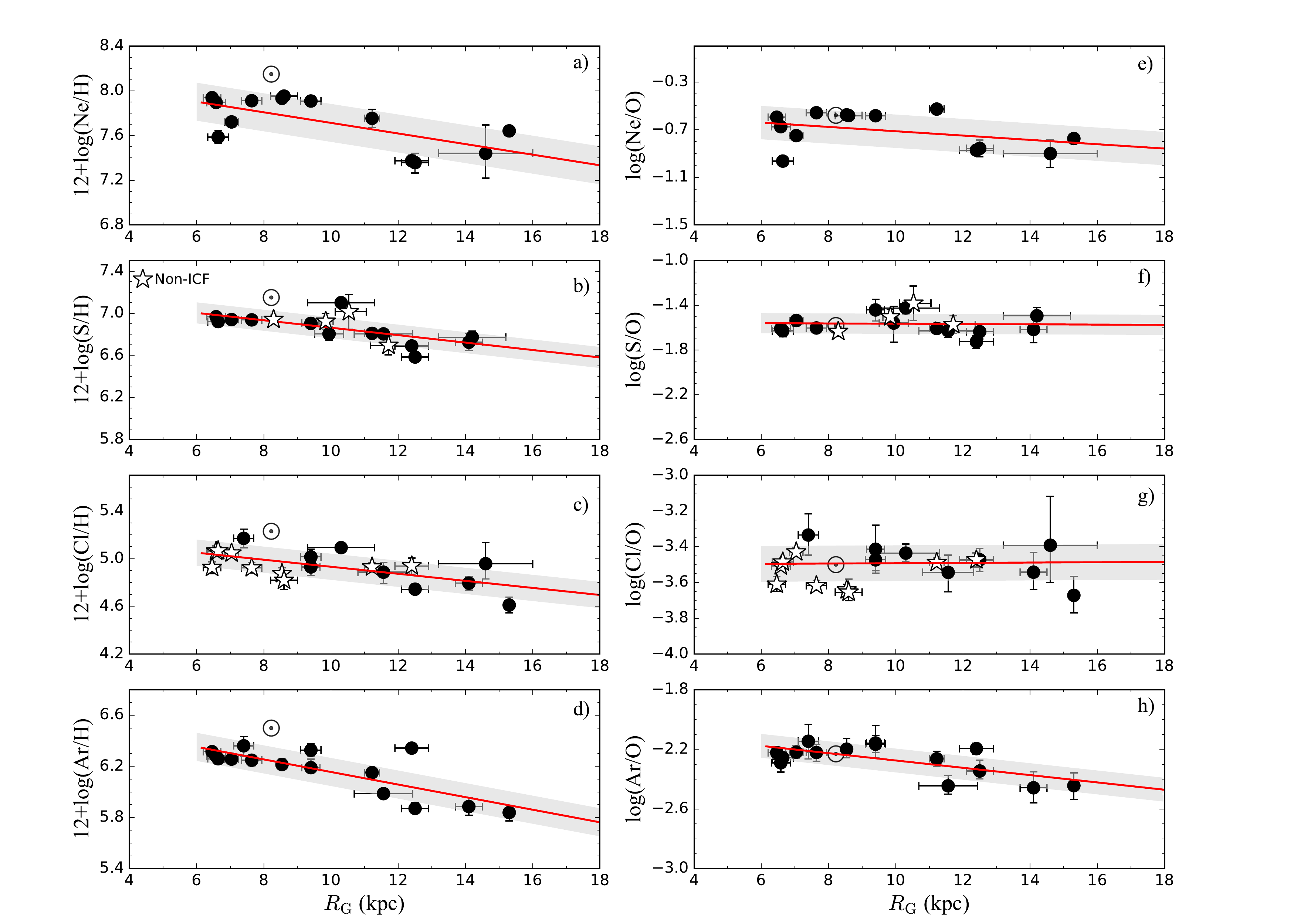}  
    \caption{Radial distribution of Ne, S, Cl and Ar abundances and Ne/O, S/O, Cl/O and Ar/O ratios as a function of $R_{\rm G}$ for \ion{H}{ii} regions in the Milky Way. The empty stars in some of the panels represent objects where the S or Cl abundances have been  determined without using an ICF scheme. The solar symbol at 8.2 kpc indicate the solar abundances recommended by \citet{Lodders:2019}. The red solid lines indicate the least-squares linear fits to the data. The gray band represents the dispersion of the individual observational data around the fit, see Table~\ref{Tab:ICFs}.} 
    \label{fig:gradients}
    \end{center}
    \end{figure*}
In Fig.~\ref{fig:gradients}, we present the radial distributions and gradients for Ne, S, Cl and Ar abundances as well as for the Ne/O, S/O, Cl/O and Ar/O abundance ratios for our sample of Galactic \ion{H}{ii} regions based on the total abundances shown in Table~\ref{tab:total-abundances}. We have identified with empty stars those nebulae in which the total abundance of a given element was calculated without using an ICF scheme. As in the elements discussed in \S~\ref{sec:CNO-gradients}, we performed a least-squares linear fit to compute the gradients taking into account the uncertainties in both axes using Monte Carlo simulations. Note that our total abundance determinations include the uncertainties of the ICF. The computed gradients are represented as red solid lines in the different panels of Fig.~\ref{fig:gradients}. The solar symbols indicate the solar abundance for each element -- or their abundance ratios with respect to O -- recommended by \citet{Lodders:2019}. Table~\ref{Tab:ICFs} includes the adopted solar Ne/O, S/O, Cl/O and Ar/O abundance ratios \citep{Lodders:2019}.

Panel (a) of Fig.~\ref{fig:gradients} shows the radial distribution of Ne/H ratios and the computed radial gradient. We obtain a slope of $-0.047\pm0.011$ dex kpc$^{-1}$ for this  gradient, with a dispersion around the fit of 0.17 dex, which is steeper but marginally consistent with our O/H gradient within the errors. However, the Ne/H gradient we obtain using the ICF scheme by ADIS20 gives a slope almost identical to that of O/H (see Table~\ref{tab:gradient-results}). \citet{Martin-Hernandez+2002} reported the Ne abundance gradient based on observations in the IR wavelengths for a sample of 25 \ion{H}{ii} regions covering $R_{\rm G}$ between 0 and 15 kpc. Those authors reported a slope of $-0.039\pm0.009$ dex kpc$^{-1}$, which is also consistent with our determinations using optical data. 
However, the intercept of our gradient is 0.3--0.4 dex lower than the value obtained by \citet{Martin-Hernandez+2002} using IR data. Offsets between IR and optical data have also been reported in other abundances and abundance ratios, such as the S/H and N/O ratios \citep{Rudolph:2006}. Other determinations of the Galactic Ne/H gradient from IR data are from \citet{Simpson:1990} and \citet{Simpson:1995}, who obtain a significantly steeper slope of $\approx -0.083$ dex kpc$^{-1}$.

The radial distribution of S/H ratios and the corresponding gradient are shown in panel (b) of Fig.~\ref{fig:gradients}. We obtain a slope of $-0.035\pm0.006$ dex kpc$^{-1}$ (very similar to the one we obtain with the ICF of ADIS20, as can be seen in Table~\ref{tab:gradient-results}), which is  consistent with the one of $-0.041\pm0.014$ dex kpc$^{-1}$ estimated by \citet{Rudolph:2006} using FIR lines and also very similar to the slope of our O/H gradient. We report a dispersion around the S/H gradient of 0.10 dex, somewhat larger than the individual observational uncertainties. Recently, \citet{Fernandez-Martin:2017} reported a slope of $-0.108\pm0.006$ dex kpc$^{-1}$ using optical spectra for \ion{H}{II} regions located at $R_{\rm G}$ between 5 and 17 kpc. That value of the slope is considerably much steeper than our determination and other previous estimates  from the literature \citep[e.g.][]{Shaver:1983,Simpson:1995,Afflerbach:1997,Rudolph:2006}. 
\citet{Esteban:2018} observed some \ion{H}{II} regions with a very low ionization degree -- O$^{2+}$/O < 0.03 -- but measurable [\ion{S}{III}] 6312 or 9069 \AA\ lines. For such nebulae is possible to assume that S/H $\approx$ S$^{+}$/H$^{+}$ $+$ S$^{2+}$/H$^{+}$, since the contribution of S$^{3+}$/H$^{+}$ in those objects is expected to be negligible. The objects with such properties are IC~5146, Sh\,2-235, Sh\,2-257 and Sh\,2-271, whose total abundances of S/H range from 6.70 to 7.01 but, unfortunately, cover a rather narrow range of $R_{\rm G}$ -- between 9.3 and 11.7 kpc -- and no confident gradient can be estimated with so small baseline. Inspecting tables~\ref{tab:O-N-ab} and \ref{tab:ionic-abundances} we can see that there is a sizable group of low-ionization degree \ion{H}{II} regions that lack of determination of their S$^{2+}$/H$^{+}$ ratios. This is because the rather faint auroral [\ion{S}{III}] 6312 \AA\ line could not be detected in those objects. We plan to obtain additional optical spectra covering the bright nebular [\ion{S}{III}] 9069, 9532 \AA\ lines of that group of nebulae for trying to increase the number of objects with S/H ratios determined without ICF and estimate a more precise gradient for this element.  

Panel (c) of Fig.~\ref{fig:gradients} shows the radial distribution of Cl/H for 18 \ion{H}{ii} regions of our sample. The empty stars indicate those regions whose  Cl/H ratio has been derived without using an ICF scheme and were previously studied by \citet{Esteban:2015} (see also \citealt*{Arellano-Cordova:2019}). We obtain a slope of $-0.030\pm0.007$ dex kpc$^{-1}$ and a scatter around the fit of 0.11 dex (see Table~\ref{tab:gradient-results}). This slope is flatter than 
the value of $-0.043\pm0.012$ dex kpc$^{-1}$ reported by \citet{Esteban:2015} considering only the 9 objects for which Cl abundance was determined without using an ICF scheme. This discrepancy is mainly due to our procedure to calculate the Cl$^{2+}$/H$^+$ ratio (see \S~\ref{sec:ICF-C}). Note that the radial gradient of Cl/H determined in this study includes a number of objects with direct determinations of abundances considerably larger than any other previous work.

%
Panel (d) of Fig.~\ref{fig:gradients} shows the radial abundance gradient of Ar/H as a function of $R_{\rm G}$ for 15 \ion{H}{ii} regions, those for which we can apply the ICF by \citet{Izotov:2006}. We obtained a slope of $-0.049\pm0.005$ and a dispersion around the fit of 0.11 dex. The slope we found applying the ICF scheme by ADIS20 is almost the same, $-0.046\pm0.013$. Previous determinations of the Ar/H gradient obtained steeper slopes, as those by \citet{Fernandez-Martin:2017} of $-0.074\pm0.006$ dex kpc$^{-1}$, or by \citet{Shaver:1983} of $-0.060\pm0.015$ dex kpc$^{-1}$. However, \citet{Martin-Hernandez+2002} using IR lines of \ion{H}{II} regions located at $R_{\rm G}$ in the range 0-13 kpc, reported a slope of $-0.045\pm0.011$ dex kpc$^{-1}$, in good agreement with our results. 

In all the left panels (a-d) of Fig.~\ref{fig:gradients}, we can see that the solar abundances recommended by \citet{Lodders:2019} are always above he nebular values predicted by our gradients at the $R_{\rm G}$ of the Sun. The value of the offset is 0.35, 0.22, 0.24 and 0.26 dex for Ne/H, S/H, Cl/H and Ar/H, respectively. As it has been discussed in the cases of the O and N gradients, those offsets can be explained by the effect of the abundance discrepancy (and temperature fluctuations if both phenomena are related) when determining abundances from CELs. Considering the results of several previous papers for Galactic \ion{H}{ii} regions \citep{Esteban:2004, Esteban:2013, Garcia-Rojas:2004, Garcia-Rojas:2005,Garcia-Rojas:2007, Garcia-Rojas:2006}, we estimate that the average underestimation factor of the Ne/H, S/H, Cl/H and Ar/H ratios when temperature fluctuations are considered is quite similar and about 0.20 dex, in reasonable agreement with the observed offsets. 

\subsubsection{Ne/O, S/O, Cl/O and Ar/O ratio gradients}

We have also determined the gradients of Ne/O, S/O, Cl/O and Ar/O abundance ratios as a function of $R_{\rm G}$. These gradients are shown through the right-hand panels (e-h) of Fig.~\ref{fig:gradients}. The parameters of our least-squares linear fits are listed in Table~\ref{tab:gradient-results} in columns 8 to 10. O, Ne, S and Ar are considered $\alpha$-elements, which are synthesized in massive stars by $\alpha$-particle capture. Therefore, their abundances should vary in lockstep and the abundance ratio of Ne, S and Ar with respect to O should be constant. In Table~\ref{tab:gradient-results}, we can note that our slope of the log(S/O) gradient is flat, but in the cases of log(Ne/O) and log(Ar/O) are slightly negative. However, the slope we obtain for log(Ne/O) with the ICF scheme by ADIS20 is almost perfectly flat, and that of log(Ar/O) becomes actually nearly flat. As it has been discussed in \S~\ref{sec:ICF-HII}, part of the trends of the abundance ratios are modulated by the ICF scheme and the $T_{\rm e}$ structure assumed. This suggest that the set of ICFs by ADIS20 seems to  better reproduce the behaviour of the abundances of the ionic species of Ne and Ar in our \ion{H}{II} regions.

Cl is an element of the group of halogens, which also includes F, Br and I. 
Its most common isotopes are $^{35}$Cl and $^{37}$Cl and both are expected to 
be produced during oxygen burning in stars \citep{Woosley:1995, Pagel:1997}. \citet{Clayton:2003} indicates that $^{35}$Cl can be produced by proton capture by a $^{34}$S nucleus, since $^{37}$Cl is produced 
by radioactive decay of $^{37}$Ar, which is formed by neutron capture by $^{36}$Ar. As we can see, Cl is produced by single proton or neutron captures by isotopes of $\alpha$-elements and, therefore, we also expect a lockstep evolution of O and Cl and a constant log(Cl/O) ratio across the Milky Way disc. Our result is entirely consistent with a flat gradient within the errors and also with the previous determination by \cite{Esteban:2015}. Those authors reported a slope of  $-0.001\pm0.014$ dex kpc$^{-1}$ for log(Cl/O) with their sample of 9 \ion{H}{ii} regions without using an ICF scheme. 

As it has been discussed in \S~\ref{sec:ICFs} and can be noted in Table~\ref{Tab:ICFs} and Fig.~\ref{fig:gradients}, the average values of the abundance ratios of Ne, S, Cl and Ar with respect to O obtained with our adopted ICFs are in good agreement with the solar ones compiled by \citet{Lodders:2019} within the uncertainties (the differences are smaller than 0.04 dex except in the case of Ne/O, whose difference is 0.1 dex). Our abundance ratios are also fairly consistent with the values obtained by \citet{Esteban:2020} from deep spectra of \ion{H}{ii} regions of the nearby massive spiral galaxies M31 and M101. Especially in the cases of S/O and Cl/O although \citet{Esteban:2020} use the ICF scheme by \citet{Izotov:2006}. However, as it is also discussed in \S~\ref{sec:ICFs}, the nebular Ne/O, Ar/O, and perhaps S/O and Cl/O ratios should be affected by differential depletion factors when comparing with solar abundance ratios. In the case of the nebular Ar/O, we have to subtract 0.1 dex to compare with the solar values, however, this is not supported by our results. For Ne/O the expected behaviour is similar to Ar/O, however, the Ne solar abundance is quite uncertain and still debatable (see \S~\ref{sec:ICF-Ne}). In the cases of S/O and Cl/O, due to the unknown depletion factors of S and Cl, the amount to be subtracted to the nebular ratio may be between 0 and 0.1 dex, however, panels (f) and (g) of Fig.~\ref{fig:gradients} suggest that the correction for depletion, if any, should be small.



\subsection{Comparison with PNe, young stellar objects and some spiral galaxies}
\label{sec:comparison_other}

PNe provide crucial information about the nucleosynthesis and chemical evolution of different elements. It is thought that PNe progenitors do not synthesize O and other $\alpha$-elements as Ne, S or Ar and, in principle, they can be considered as probes of the chemical composition of the ISM at the moment of the birth of the progenitors. Therefore, they can be used as tracers of the temporal evolution of the abundance gradients of the Galaxy. However, the recent results by \citet{Delgado-Inglada:2015} indicate the production of O in C-rich PNe at near-solar metallicties that can increase the O/H ratio up to 0.3  dex. This result calls into question the suitability of O in PNe as tracer of the ISM composition. Other critical problems in the determination of the abundance gradients from PNe data is the large uncertainties in the distances of these objects \citep{Frew:2016, Stanghellini+2018A, Pena:2019, Garcia-Rojas:2020}, the different range of ages of their progenitors and the radial migration along the lifetime of the progenitors. Most of the determinations of the Galactic O/H gradient based on data of young PNe -- with ages between 1 and 4 Gyr -- provide slopes which tend to be flatter than our value \citep{Liu:2004, Maciel:2015, Stanghellini+2018A, Pena:2019}, with typical values in the range between $-0.019$ and $-0.048$ dex kpc$^{-1}$ for $R_{\rm G}$ between 0 and 16 kpc. This behaviour has led some authors to propose the steeping of the O gradient with time \citep[e.g.][]{Stanghellini+2018A}. However, other authors as \citet{Henry:2004} or \citet{Maciel:2013} claim that the time variation of the O/H gradient may be masked by the aforementioned uncertainties related to the abundance and gradient determination with PNe. 
%
%
 As $\alpha$-elements, Ne, S and Ar should have abundance gradients similar to that of O/H, but as it has been discussed in \S~\ref{sec:ICF-HII}, their value may be affected by the ICF scheme used. For Ne and Ar, \citet{Pena:2019} reported a slope of $-0.026\pm0.014$ dex kpc$^{-1}$ and $-0.010\pm0.006$ dex kpc$^{-1}$, respectively, for a sample of young PNe, which are flatter than our slopes for these elements in \ion{H}{II} regions. \citet{Henry:2004} studied a large sample of PNe providing the radial gradients for the different elements studied here but for a sample of objects covering progenitors of a wide range of ages. Those authors reported values of the slope for the gradients of Ne, S, Cl and Ar abundances fairly similar to ours, from $-0.030$ to $-0.048$ dex kpc$^{-1}$. On the other hand, the results obtained by \citet{Stanghellini+2018A} for their compilation of PNe data from the literature show slopes of Ne and Ar which are somewhat flatter than ours, $-0.018$ to $-0.030$ dex kpc$^{-1}$, respectively. As we can see, the comparison with available data from PNe does not give conclusive evidence of a time variation of the abundance gradients in the Milky Way. 

Cepheids are also important tracers of the chemical composition of the ISM, further, their $R_{\rm G}$ is well-determined -- they are considered standard candles -- and the chemical abundances of a large number of elements can be estimated \citep{Andrievsky:2002b}. These stars are relatively young, younger than the PNe progenitors but older than \ion{H}{ii} regions. The different gradients of O/H obtained from Cepheids show a wide range of variation, from $-0.023$ to $-0.080$ dex kpc$^{-1}$ \citep[see the compilation by][and references therein]{Molla:2019}. A recent analysis of results of a large sample of Galactic Cepheids -- with O/H ratios with average uncertainties of about 0.12 dex -- obtains a slope for the O/H gradient of $-0.055\pm0.003$ dex kpc$^{-1}$, somewhat steeper than ours \citep{Maciel:2019}. Measurements of C/H ratios in Cepheids give slopes of the gradient between $-0.055$ dex kpc$^{-1}$ and $-0.080$ dex kpc$^{-1}$ \citep{Andrievsky:2002a, Luck:2011, Luck:2014}, which are quite in agreement with our result for \ion{H}{ii} regions. Also, we obtain consistent results with the slopes of the N/H gradient derived by \citet{Luck:2011} and \citet{Luck:2014} from Cepheids, which values are between $-0.046$ dex kpc$^{-1}$ and $-0.051$ dex kpc$^{-1}$. In the case of the S abundance gradient, our value of the slope is in agreement with the one of $-0.051$ dex kpc$^{-1}$ reported by \citet{Andrievsky:2002a} but not with that obtained with \citet{Luck:2014}, who report a slope of $-0.093$ dex kpc$^{-1}$, significantly steeper than our result. 

As \ion{H}{ii} regions, young OB stars can also be tracers of the present-day chemical composition of the ISM. The slopes of the gradients of N/H, O/H and S/H obtained by \citet{Daflon:2004} are consistent with our determinations within the uncertainties, but not in the case of C/H. For this element, \citet{Daflon:2004} obtain a significantly flatter slope of  $-0.037\pm0.010$ dex kpc$^{-1}$. In contrast, our result for C/H is consistent with the gradient derived by \citet{Rolleston:2000} for OB stars, who reported a slope of $-0.070$ dex kpc$^{-1}$.  

\input{Table9.tex} 

Finally, we compare the Galactic gradients of Table~\ref{tab:gradient-results} with those ones obtained for other nearby spiral galaxies. To make that comparison appropriately, we have recalculated the Galactic gradients using the fractional Galactocentric distance with respect to the disc effective radius -- $R_{\rm G}$/$R_{\rm e}$ -- of the \ion{H}{ii} regions. The convenience of this normalization compared to $R_{\rm G}$/$R_{\rm 25}$ is discussed in \citet{Esteban:2020}. The corresponding slopes of such re-normalization are shown in columns 4 and 9 of Table~\ref{tab:gradient-results}.  The effective radius, $R_{\rm e}$, contains the half of the total light emitted by the system \citep[e.g.][]{Sanchez:2012, Molla:2019, Esteban:2020}. We have considered the value of $R_{\rm e} = 4.55$ kpc adopted by \citet{Esteban:2020} for the Milky Way, which is obtained applying the relation $R_{\rm e} =$ $1.678\times R_{\rm d}$ \citep{Sanchez:2014} to the value of the scale length, $R_{\rm d}$, obtained by \cite{Licquia:2016} (see more details in \citealt{Esteban:2020}). In Table~\ref{tab:comp_grad} we include  the slopes of the O/H, C/H, S/H and N/O gradients with respect to $R_{\rm G}$/$R_{\rm e}$ for the Milky Way -- our results -- and recent determinations for the following spiral galaxies:  M33 and NGC~300 \citep{ToribioSanCipriano:2016}, M101 and M31 \citep{Esteban:2020} and four galaxies of the CHAOS project: NGC~628, NGC~3184, NGC~5194 and M101  \citep{Berg:2020}. All these determinations from the literature have been selected because they are based on deep spectroscopy of \ion{H}{ii} regions which abundances are based on the direct method ($T_{\rm e}$-sensitive method). In the table, we also give the morphological type, $R_{\rm e}$ and the absolute $B$-band magnitude ($M_{\rm B}$) of the galaxies. The values of $R_{\mathrm e}$ have been taken from \citet{Esteban:2020} and \citet{Berg:2020}. 

From Table~\ref{tab:gradient-results}, we can see that the slope of the O/H gradient of the different galaxies is rather similar, with a range between $-$0.07 and $-$0.24 dex $(R_{\rm G}/R_{\rm e})^{-1}$, with an average value of  $-0.15 \pm 0.05$ dex $(R_{\rm G}/R_{\rm e})^{-1}$. The galaxy with the flattest O/H gradient, NGC~5194, may be affected by interactions with its companion, NGC 5195, resulting in radial migration and mixing of the gas \citep[see discussion in][]{Croxall:2015}. Removing that galaxy -- the only interacting one of the sample -- the average slope is slightly modified to $-0.17 \pm 0.04$ dex ($R_{\rm G}/R_{\rm e})^{-1}$. We do not find any correlation of the value of the slope with $M_{\rm B}$ or the morphological type of the galaxy. The average slope of this small sample of galaxies is steeper than the characteristic slope obtained from large surveys using strong-line methods for estimating the O/H ratio. In fact, \citet{Sanchez-Menguiano:2016}, using the O3N2 calibrator, obtain an average slope of $-0.075 \pm 0.016$ dex $(R_{\rm G}/R_{\rm e})^{-1}$ for a sample of 122 face-on galaxies of the CALIFA survey, while \citet{Sanchez-Menguiano:2018} for a sample of 102 spiral galaxies observed with MUSE and using the same calibrator, obtain a characteristic slope of $-0.10 \pm 0.03$ dex $(R_{\rm G}/R_{\rm e})^{-1}$. 

The slope of the C/H gradient is very similar for the galaxies included in Table~\ref{tab:gradient-results} except in the case of M31. However, the value for this galaxy is rather uncertain because it is based on only two \ion{H}{ii} regions and other reasons outlined in \citet{Esteban:2020}. It is remarkable that the slope is the same, $-0.33$ dex $(R_{\rm G}/R_{\rm e})^{-1}$, in the Milky Way, M33 and M101, precisely those galaxies whose C/H gradient is determined with more objects and having better determinations of  individual C abundances. The slopes of the S/H gradients do not show large variations, the mean value of the sample of Table~\ref{tab:gradient-results} is  $-0.23 \pm 0.07$ dex $(R_{\rm G}/R_{\rm e})^{-1}$, somewhat steeper than the O/H one, although this is not the case for the Milky Way, where we find the same slopes for both $\alpha$-elements. The slope of the N/O gradient shows a comparatively larger variation than the other ones included in Table~\ref{tab:gradient-results}. The mean slope of N/O is $-0.21 \pm 0.10$ dex $(R_{\rm G}/R_{\rm e})^{-1}$. The object with the flattest slope is the Milky Way. Removing it, the mean slope of N/O becomes slightly steeper $-0.24 \pm 0.08$ dex $(R_{\rm G}/R_{\rm e})^{-1}$. In \S~\ref{sec:CNO-gradients} we discussed about the infrequency of flat N/O gradients in galaxies and its possible explanations. 

In their comparison between the abundances of M31 and M101 and other galaxies including the Milky way, \citet{Esteban:2020} find that the \ion{H}{ii} regions of the Milky Way show O, C and N abundances larger than other galaxies for the same $R_{\rm G}/R_{\rm e}$ ratio. This fact is confirmed with our new results and indicates that the Milky Way seems to have an average metallicity at $R_{\mathrm e}$ larger than other more massive and luminous galaxies as M31 or M101, in contradiction with what we would expect from the mass-metallicity relation. \citet{Esteban:2020} proposed that this odd behavior might be solved if the true $R_{\mathrm e}$ is about two times larger than the most recent determinations. This seems also a likely possibility attending to other reasons collected in \citet{Esteban:2020}. The adoption of a larger $R_{\mathrm e}$ for the Milky Way would not affect the values of the slopes we obtain for the gradients.

\section{Summary and conclusions}
\label{sec:conclusions}
We compile deep spectroscopical data for a sample of 33 Galactic \ion{H}{ii} regions from the literature to carry out a reanalysis of the radial abundance gradients of  C, N, O, Ne, S, Cl and Ar of the Milky Way. The physical conditions and chemical abundances of the nebulae have been calculated in a homogeneous way using an updated and carefully selected set of atomic data. All the nebulae have direct determinations of $T_{\rm e}$, permitting to derive their precise ionic abundances. We have made a revision of the $R_{\rm G}$ of the objects using {\it Gaia} parallaxes of the second data release, DR2 \citep{gaiadr2,bailer-jonesetal18}. The sample covers a range of $R_{\rm G}$ from 6 to 17 kpc. Apart from O, we can determine the total abundances of N, Cl and S without using an ICF for several nebulae, being the number of objects relatively large in the cases of N and Cl. We compare the suitability of different ICF schemes for the determination of the total abundances of the elements studied here taking into account: (a) the dispersion of the distribution of abundance ratios with respect to O (i.e. N/O, Ne/O, etc) {\it versus} the O/H ratio; (b) the comparison of the average values of those abundance ratios with the solar ones and (c) the comparison with abundances determined without ICF. We have selected the results of particular different ICF schemes as the final adopted total abundances of the different elements. Since the results of the recent ICF scheme by ADIS20 provides robust (or the most robust in some cases) results for most elements (C, N, Ne, S and Ar) we decide not to use them because the scheme is not published at the moment of the writing of this paper. In any case, the abundances obtained with the finally selected ICFs have similar level of confidence. For $\alpha$-elements (we include also Cl because it is a byproduct of $\alpha$-elements), for which we expect a constant abundance ratio with respect to O, we obtain the following average values: log(Ne/O) = $-0.71 \pm 0.15$, log(S/O) = $-1.58 \pm 0.08$, log(Cl/O) = $-3.48 \pm 0.10$ and log(Ar/O) = $-2.27 \pm 0.11$,  similar to the ratios obtained very recently for M31 or M101 and fairly consistent with the solar values. This last result does not support the presence of a sizable fraction of O depleted onto dust grains in the ionized gas of \ion{H}{ii} regions. The ICF by \citet{Berg:2019} gives an average log(C/O) = $-0.27 \pm 0.14$, entirely consistent with the solar value. In the case of log(N/O), the average ratio obtained for objects whose N abundances has been derived without an ICF ($-0.76 \pm 0.09$) is somewhat higher than the one obtained using an ICF ($-0.90 \pm 0.11$, more consistent with the solar value). Only the first case is consistent with the claimed 0.10 dex depletion of O in \ion{H}{ii} regions.


We have made a reassessment of the radial abundance gradients for the different elements in our Galaxy. Our new gradient of O (with a slope of  $-0.037 \pm 0.009$ dex kpc$^{-1}$) is in agreement with the most recent determinations, but our revised $R_{\rm G}$ based on {\it Gaia} DR2 parallaxes produce a systematic shift of the values of the innermost objects towards larger $R_{\rm G}$. This implies that the possible flattening of the gradient at $R_{\rm G} <$ 7-8 kpc suggested by \citet{Esteban:2018} might be a spurious effect due to incorrect previous determinations of $R_{\rm G}$.  Moreover, we have found that a quadratic fit to the radial O/H distribution does not improve the fitting, and that a linear gradient seems sufficient to represent the data. The slope of the N/H gradient is similar to that of O/H, implying a rather flat gradient of log(N/O) (with a slope of $-0.011 \pm 0.006$ dex kpc$^{-1}$), a result which is consistent with previous determinations for the Milky Way and M31 but rather uncommon in other spiral galaxies. The dispersion around the fit for the O and N gradients (when no-ICF correction is made for N) is of 0.07 dex, not substantially larger than the typical uncertainties of individual abundance determinations, implying that the gas of the ISM is well-mixed,  discarding the existence of significant azimuthal gradients, at least in the observed quadrant of the Milky Way. On the other hand, we find slopes ranging from $-0.030$ to $-0.049$ dex kpc$^{-1}$ for Ne, S, Cl and Ar abundance gradients, with actually flat slopes for log(S/O) and log(Cl/O) and relatively flat for log(Ne/O) and log(Ar/O), although this 
behaviour varies with the ICF scheme used. For example, using the ICF scheme by ADIS20, the log(Ne/O) gradient becomes flat, and the log(Ar/O) also flattens. Note that our new determination for the Cl abundance gradient includes a significant larger number of objects than previous determinations. We also highlight that the  dispersion around the fit of the S, Cl and Ar gradients are about 0.11 dex, again of the order of the observational errors considering the additional uncertainty associated with the ICFs. This result further reinforces the absence of significant inhomogeneities in the ISM gas across the Galactic disc. In the case of Ne, the dispersion is much higher, of 0.17-0.19 dex, mainly due to the intrinsic uncertainties of the ICF schemes for this element. We determine the C gradient with a sample of 11 \ion{H}{ii} regions with measurements of C$^{2+}$ and O$^{2+}$ RLs. The value of the slope is  $-0.072 \pm 0.018$ dex kpc$^{-1}$, much steeper than the gradients of the rest of elements. We find that for all the elements whose abundance has been derived from the intensity of CELs (all except C), the solar abundance ratios are always above the nebular ones predicted by our gradients at the $R_{\rm G}$ of the Sun by factors between 0.2 and 0.3 dex.   These offsets may be produced by the abundance discrepancy (perhaps related to temperature fluctuations), implying that nebular abundances determined from CELs may be lower than the true ones, perhaps better represented by the abundances determined from RLs.

In general, the slopes of our gradients are consistent with those obtained from data of other kinds of objects (PNe, young OB stars and Cepheids). However, the large dispersion of the slope values obtained for those objects by the different research groups does not give conclusive results about the possible temporal evolution of the gradients.  



Finally, we compare the Galactic radial gradients of the O/H, N/H, S/H and N/O ratios normalized to $R_{\rm e}$ with those of other nearby spiral galaxies with direct determination of abundances of their \ion{H}{ii} regions. We find that the slope of the O abundance gradient is very similar to that of other non-interacting spirals, with a mean slope of $-0.17 \pm 0.04$ dex ($R_{\rm G}/R_{\rm e})^{-1}$, independently of their $M_{\rm B}$ or morphological type. However, this mean value of the slope is steeper than the characteristic one obtained in large surveys that use strong-line methods for estimating the O/H ratio for large samples of spirals. Those surveys find slope values between $-$0.075 and $-$0.10 dex $(R_{\rm G}/R_{\rm e})^{-1}$. It is remarkable that the slope of the C abundance gradient we find for the Milky Way, $-0.33$ dex $(R_{\rm G}/R_{\rm e})^{-1}$, is exactly the same that \citet{Esteban:2020} obtain for M33 and M101, precisely the external galaxies with better determinations of nebular C/H ratios. The slope of the N/O gradient shows a relatively large variation, from $-$0.05 to $-$0.36 dex $(R_{\rm G}/R_{\rm e})^{-1}$, being the Milky Way precisely the galaxy with the flattest slope of the N/O gradient. Our results confirm the findings of \citet{Esteban:2020}, who reported that the \ion{H}{ii} regions of the Milky Way show larger O, C and N abundances than other galaxies for the same $R_{\rm G}/R_{\rm e}$ ratio. This odd behavior might be solved if the true $R_{\mathrm e}$ is about two times larger than commonly accepted.  

\section*{Acknowledgements}
 We are very grateful to the referee, Dick Henry, for his valuable comments that have contributed to improve the final version of the paper. We thank A. Amayo, G. Delgado-Inglada, G. Stasi\'nska, G. Dom\'inguez-Guzm\'an and M. Rodr\'iguez for providing us their ICF computations prior to publication. KZA-C acknowledges support from Mexican CONACYT posdoctoral grant 364239.  We acknowledge support from the State Research Agency (AEI) of the Spanish Ministry of Science, Innovation and Universities (MCIU) and the European Regional Development Fund (FEDER) under grant with reference AYA2015-65205-P. JG-R acknowledges support from an Advanced Fellowship from the Severo Ochoa excellence program (SEV-2015-0548).  JEM-D thanks the support of the Instituto de Astrof\'isica de Canarias under the Astrophysicist Resident Program and acknowledges support from the Mexican CONACyT (grant CVU 602402). The authors acknowledge support under grant P/308614 financed by funds transferred from the Spanish Ministry of Science, Innovation and Universities, charged to the General State Budgets and with funds transferred from the General Budgets of the Autonomous Community of the Canary Islands by the MCIU.  This work has made use of data from the European Space Agency (ESA) mission
{\it Gaia} (\url{https://www.cosmos.esa.int/gaia}), processed by the {\it Gaia}
Data Processing and Analysis Consortium (DPAC,
\url{https://www.cosmos.esa.int/web/gaia/dpac/consortium}). Funding for the DPAC
has been provided by national institutions, in particular the institutions
participating in the {\it Gaia} Multilateral Agreement.




\bibliographystyle{mnras}
\bibliography{refs.bib} 






\bsp	
\label{lastpage}
\end{document}

%% file: Table1.tex
 \begin{table*}
 \caption{Sample of Galactic \ion{H}{ii} regions compiled from the literature and our calculated values of electron densities and temperatures.}
 \begin{tabular}{lcccccc}
 \hline
 \multicolumn{1}{l}{Region} & \multicolumn{1}{c}{$R_{\rm G}$} &
 \multicolumn{1}{c}{$n_{\rm e}$([\ion{O}{ii}])} &  \multicolumn{1}{c}{$n_{\rm e}$([\ion{S}{ii}])} &
 \multicolumn{1}{c}{$T_{\rm e}$([\ion{N}{ii}])} & \multicolumn{1}{c}{$T_{\rm e}$([\ion{O}{iii}])}  & \multicolumn{1}{c}{Reference} \\
 
  \multicolumn{1}{l}{} & \multicolumn{1}{c}{(kpc)} &
 \multicolumn{1}{c}{(cm$^{-3})$} &  \multicolumn{1}{c}{(cm$^{-3})$} &  \multicolumn{1}{c}{(K)} & \multicolumn{1}{c}{(K)} & \multicolumn{1}{c}{}\\
 \hline

IC~5146         &    8.29$\pm$0.10$^{\rm b}$                &  $<100$                            &     $<100$                            &     $ 7100\pm100$                       &  $5700\pm200^{\rm d}$                                           & 1  \\
M8              &    7.04 $\pm$ 0.20$^{\rm a}$               &   $ 1600 \pm600$       & 	$ 1300 \pm 200 $                  & 	$ 8400 \pm 100 $                    & 	$ 8000 \pm 100 $                             & 2  \\
M16             &    6.58 $\pm$ 0.28$^{\rm a}$               &   $ 1200 \pm 200 $                 & 	$ 1100 \pm 200 $                  & 	$ 8300 \pm 100 $                    & 	$ 7600 \pm 200 $                             & 3  \\
M17             &    6.46 $\pm$ 0.26$^{\rm a}$               &   $ 500 \pm 100 $                  & 	$ 400 \pm 100 $                   & 	$ 8900 \pm 200 $                    & 	$ 8000 \pm 100 $                             & 2  \\
M20             &    6.64 $\pm$ 0.31$^{\rm a}$               &   $ 300 \pm 100 $                  & 	$ 300 \pm 100 $                   & 	$ 8300 \pm 200$         			& 	$ 7800 \pm 300$                   & 3  \\
M42             &    8.54$\pm$ 0.11$^{\rm a}$                &   $ 6900 ^{+ 2400 }_{- 1500 }$     & 	$ 4900 ^{+ 2900 }_{- 1800 }$      & 	$ 10200 \pm 200$         			& 	$ 8300 \pm 100 $                             & 4  \\
NGC\,2579       &    12.4$\pm$ 0.5$^{\rm a}$                 &   $ 1200 \pm300$       & 	$ 800 \pm200$                     & 	$ 8600 \pm 300 $$^{\rm d}$                    & 	$ 9300 \pm 200$                   &  5 \\
NGC\,3576       &    7.64 $\pm$ 0.3$^{\rm a}$                &   $ 1700 \pm 200 $                 & 	$ 1100 \pm 300 $                  & 	$ 8800 \pm 200 $                    & 	$ 8400 \pm100  $                 &  6 \\
NGC\,3603       &    8.6 $\pm$ 0.4                           &   $ 2700 ^{+ 800 }_{- 500 }$       & 	$ 3100 ^{+ 1300 }_{- 700 }$       & 	$ 11200 \pm 500 $                   & 	$ 9000 \pm 100 $                             &  3 \\
Sh\,2-61        &    6.15 $^{+ 0.39 }_{- 0.33 }$$^{\rm b}$   &   $-$                              &     $ 1100 \pm200$                    & 	$ 7500 \pm 400$         			& $ 6300 \pm 500$$^{\rm d}$           &  7 \\
Sh\,2-83        &    15.3$\pm$ 0.1$^{\rm a}$                 &   $-$                              &     $ 300 \pm 100 $                   & 	$ 11900 \pm 600 $                   & $ 10400 \pm 500$                    &  8 \\
Sh\,2-90        &    7.4 $\pm$ 0.3$^{\rm c}$                 &   $-$                              &     $ 200 \pm 100 $                   & 	$ 8200 \pm 200 $                    & $ 7200 \pm 300 ^{\rm d}$                &  7 \\
Sh\,2-100       &    9.4 $\pm$ 0.3$^{\rm a}$                 &   $ - $                            &     $ 400 \pm 200 $                   &     $ 8600 \pm 300 $                    & $ 8200 \pm 100 $                                &  8 \\
Sh\,2-127       &    14.2$\pm$ 1.0$^{\rm a}$                 &   $-$                              &     $ 600 \pm 100 $                   &	    $ 9800 \pm 200 $                    & $ 9500 \pm 200 ^{\rm d}$                      &  8 \\
Sh\,2-128       &    12.5$\pm$ 0.4$^{\rm a}$                 &   $ - $                            &     $ 500 \pm 100 $                   & 	$ 10600 \pm 200 $                   & $ 9900 \pm 300 $                               &  8 \\
Sh\,2-132       &    9.94$^{+ 0.43 }_{- 0.36 }$$^{\rm b}$    &   $-$                              &     $ 300 \pm 100 $                   &	    $ 9300 \pm 500 $                    & $ 8800 \pm 600 ^{\rm d}$           &  9 \\
Sh\,2-152       &    10.3$\pm$ 1.0$^{\rm a}$                 &   $-$                              &     $ 700 \pm 100 $                   & 	$ 8200 \pm 100 $                    & $ 7300 \pm 100 ^{\rm d}$                       &  7 \\
Sh\,2-156       &    9.40$^{+ 0.27 }_{- 0.24 }$$^{\rm b}$    &   $-$                              &     $ 900 \pm 100 $                   &     $ 9400 \pm 400 $                    & $ 9000 \pm 400 $                   &  9 \\
Sh\,2-175 	    &    9.45 $^{+ 0.19 }_{- 0.18 }$$^{\rm b}$   &   $-$                              &     $ < 100 $                         &     $ 7000 \pm 1300 $                   & $ 5600 \pm 400 $$^{\rm d}$                     &  7 \\
Sh\,2-209       &    17.0$\pm$0.7$^{\rm a}$                  &  $-$                               &     $300 \pm300$                      &     $ 10600 \pm 800$                    & $ 10700 ^{+ 1000 }_{- 1200 }$               &  8 \\
Sh\,2-212       &    14.6$\pm$ 1.4$^{\rm a}$                 &   $-$                              &     $ < 100 $                         & 	$ 8300 \pm 700 $                    & $ 11100 \pm 1000$                   &  8 \\
Sh\,2-219       &    11.87$^{+ 0.73 }_{- 0.58 }$$^{\rm b}$   &   $-$                              &     $ < 100  $                        & 	$ 8600 \pm 300 $                    & $ 7800 \pm 500$$^{\rm d}$           &  7 \\
Sh\,2-235       &    9.84$^{+ 0.22 }_{- 0.21 }$$^{\rm b}$    &   $-$                              &     $ < 100  $                        & 	$ 8100 \pm 200 $                    & $ 7100 \pm 200 ^{\rm d}$                       &  7 \\
Sh\,2-237       &    10.35$^{+ 0.45 }_{- 0.37 }$$^{\rm b}$   &   $-$                              &     $ 400 \pm 100 $                   & 	$ 8700 \pm 300$                     & $ 8000 \pm 400$$^{\rm d}$           &  7 \\
Sh\,2-257       &    10.53$^{+ 0.52 }_{- 0.41 }$$^{\rm b}$   &   $-$                              &     $ 100 \pm100$                     & 	$ 7900 \pm 200 $                    & $ 6900 \pm 300 ^{\rm d}$                       &  7 \\
Sh\,2-266       &    14.5$\pm$ 1.4$^{\rm c}$                 &   $-$                              &     $ 300 \pm 200 $                   & 	$ 8300 \pm 400 $                    & $ 7400 \pm 600 ^{\rm d}$           &  7 \\
Sh\,2-270 	    &    16.1$\pm$1.4$^{\rm c}$                  &   $-$                              &     $   400 \pm 100 $                 &     $ 9300 \pm 1000$                    &  $ 8800 \pm 1300$$^{\rm d}$    &  7 \\
Sh\,2-271       &    11.71$^{+ 0.66 }_{- 0.53 }$$^{\rm b}$   &   $-$                              &     $ < 100  $                        & 	$ 8700 \pm 200 $                    & $ 7900 \pm 300 ^{\rm d}$                       &  7 \\
Sh\,2-285 	    &    11.57$^{+ 0.79 }_{- 0.60 }$$^{\rm b}$   &   $-$                              &     $ < 100 $	                      &     $ 8500 \pm 300 $                    & $ 7700 \pm 40 $$^{\rm d}$                     &  7 \\
Sh\,2-288       &    14.1$\pm$ 0.4$^{\rm a}$                 &   $-$                              &     $ 400\pm300$                      &	    $ 9400 \pm 300 $                    & $ 9200 \pm 500 $                              &  8 \\
Sh\,2-297       &    8.97 $^{+ 0.16 }_{- 0.15 }$$^{\rm b}$   &   $-$	                          &     $ < 100  $                        & 	$ 7800 \pm 200 $                    & $ 6700 \pm 300 ^{\rm d}$                   &  7 \\
Sh\,2-298       &    11.56$^{+ 0.87 }_{- 0.65 }$$^{\rm a}$   &   $-$                              &     $ < 100 $                         &     $ 11700 \pm500$                     & $ 11700 \pm 200 $                               &  8 \\
Sh\,2-311       &    11.22$\pm$ 0.22$^{\rm a}$               &   $ 300 \pm 100 $                  & 	$ 300 \pm 100 $                   & 	$ 9300 \pm 200 $                    & $ 8900 \pm 100 $                         &  10\\
 \hline                     
\end{tabular}
\label{tab:sample}
\begin{description} 

\item  References for the line intensities: (1) \citet{Garcia-Rojas:2014}, (2) \citet{Garcia-Rojas:2007}, (3) \citet{Garcia-Rojas:2006}, (4) \citet{Esteban:2004}, (5) \citet{Esteban:2013}, (6)  \citet{Garcia-Rojas:2004}, (7) \citet{Esteban:2018},
  (8) \cite{Esteban:2017}, (9) \citet{Fernandez-Martin:2017}, (10) \citet{Garcia-Rojas:2005}. Galactocentric distances taken from: 
 $^{\rm a}$ M\'endez-Delgado et al. (2020); $^{\rm b}$ {\it Gaia} DR2 parallaxes (see text for details); $^{\rm c}$ \citet{Esteban:2018}. $^{\rm d}$  $T_{\rm e}$([\ion{O}{iii}]) or $T_{\rm e}$([\ion{N}{ii}]) estimated using the temperature relation of \citet{Esteban:2009}.
 \end{description} 
 \end{table*}
 

%% file: Table2.tex
 \begin{table*}\footnotesize
 \caption{Atomic data used in this work.}
 \begin{center}
 \begin{tabular}{lcc}
 \hline
 \multicolumn{1}{l}{Ion} & \multicolumn{1}{c}{Transition Probabilities} &
 \multicolumn{1}{c}{Collision Strengths} \\
 \hline

O$^{+}$   &  \citet{FFT:2004} & \citet{Kisielius:2009}\\
O$^{2+}$  &  \citet{Wiese:1996}, \citet*{Storey:2000} & \citet{Storey:2014}\\
N$^{+}$   &  \citet{FFT:2004} & \citet{Tayal:2011}\\
Ne$^{2+}$  &  \citet{McLaughlin:2011} & \citet{McLaughlin:2011}\\
S$^{+}$   &  \citet{Podobedova:2009} & \citet{Tayal:2010}\\
S$^{2+}$  &  \citet{Podobedova:2009} & \citet{Grieve:2014}\\
Cl$^{+}$ &  \citet{MendozaZeippen:1983} & \citet{Tayal:2004}\\
Cl$^{2+}$ &  \citet{Fritzsche:1999} & \citet{Butler:1989}\\
Cl$^{3+}$ &  \citet{Kaufman:1986}, \citet{Mendoza:1982a}, \citet{Ellis:1984} & \citet{Galavis:1995}\\
Ar$^{2+}$ &   \citet{Mendoza:1983}, \citet{Kaufman:1986}  & \citet*{Galavis:1995}\\
Ar$^{3+}$ &   \citet{Mendoza:1982b}  & \citet{Ramsbottom:1997}\\
 \hline
 \end{tabular}
 \end{center}
 \label{tab:atomic_data}
 \end{table*}

%% file: Table3.tex
\begin{table*}
 \caption{Ionic abundances of carbon, oxygen and nitrogen for our sample of 33 Galactic \ion{H}{ii} regions.}
 \label{tab:O-N-ab}
 \begin{tabular}{l c c c c c c }
 \hline
 \multicolumn{1}{l}{Region} & \multicolumn{1}{c}{C$^{2+}$} &
 \multicolumn{1}{c}{N$^{+}$}  & \multicolumn{1}{c}{O$^{+}$} & \multicolumn{1}{c}{O$^{2+}$} & \multicolumn{1}{c}{O$^{2+}$} & \multicolumn{1}{c}{O$^{2+}$/O}\\
 
 \multicolumn{1}{l}{} & \multicolumn{1}{c}{ (RLs)} &
 \multicolumn{1}{c}{}  & \multicolumn{1}{c}{} & \multicolumn{1}{c}{} & \multicolumn{1}{c}{(RLs)} & \multicolumn{1}{c}{(CELs)}\\


\hline
IC\,5146    &   $-$                  & $ 7.81 \pm 0.05 $   &   $ 8.57\pm0.09   $                 &    $ - $                   &   $-$                     &   $\sim$0.0    \\                                 
M8 	        &   8.31$\pm$0.02        & $ 7.54 \pm 0.03 $   &   $ 8.35 \pm 0.06 $                 &    $ 7.89 \pm 0.03 $       &   8.23$\pm$0.03          &   0.26  \\
M16         &   8.39$\pm$0.04        & $ 7.73 \pm 0.04 $   &   $ 8.46 \pm0.06  $                 &    $ 7.92 \pm0.07$         &   8.30$\pm$0.04          &   0.22  \\
M17 	    &   8.73$\pm$0.03        & $ 6.84 \pm 0.05 $   &   $ 7.79 \pm 0.07 $                 &    $ 8.45 \pm 0.04 $          &   8.68$\pm$0.02          &   0.82  \\   
M20         &   8.19$\pm$0.05        & $ 7.60 \pm 0.04 $   &   $ 8.48 \pm 0.05 $                 &    $ 7.74 \pm0.09$         &   8.00$\pm$0.22           &   0.15   \\ 
M42         &   8.34$\pm$0.02        & $ 6.89 \pm 0.06 $   &   $ 7.73 \pm 0.12 $                 &    $ 8.44 \pm 0.01 $       &   8.57$\pm$0.01          &   0.84  \\
NGC\,2579   &   8.18$\pm$0.05        & $ 6.87 \pm 0.06 $   &   $ 7.99 \pm 0.11 $                 &    $ 8.21 \pm 0.04 $       &   8.46$\pm$0.03          &   0.62  \\
NGC\,3576   &   8.44$\pm$0.02        & $ 7.06 \pm 0.05 $   &   $ 8.05 \pm0.06$                   &    $ 8.37 \pm 0.02 $       &   8.62$\pm$0.05           &   0.68  \\
NGC\,3603   &   8.49$\pm$0.07        & $ 6.48 \pm 0.08 $   &   $ 7.34 \pm0.14$                   &    $ 8.43 \pm 0.04 $       &   8.69$\pm$0.05          &   0.93  \\
Sh\,2-61 	&   $-$                  & $ 7.89 \pm 0.10 $   &   $ 8.41\pm0.15$                    &    $ 5.83 \pm 0.26 $         &   $-$                     & 	0.002    \\         
Sh\,2-83 	&   $-$                  & $ 6.32 \pm 0.08 $   &   $ 7.15 \pm 0.13$                  &    $ 8.24 \pm0.09$           &   $-$                     & 	0.92   \\
Sh\,2-90    &   $-$                  & $ 7.75 \pm 0.05 $   &   $ 8.35 \pm 0.08 $                 &    $ 7.99 \pm 0.12 $       &   $-$                     &	0.30   \\
Sh\,2-100   &   8.36$\pm$0.08        & $ 6.78 \pm 0.07 $   &   $ 7.73 \pm 0.10 $                 &    $ 8.41 \pm 0.04 $       &   8.52$\pm$0.06            &    0.83   \\
Sh\,2-127   &   $-$                  & $ 7.29 \pm 0.03 $   &   $ 8.20 \pm 0.05$                  &    $ 7.39 \pm 0.05 $       &   $-$                      &	0.13   \\
Sh\,2-128   &   $-$                  & $ 6.81 \pm 0.04 $   &   $ 7.83 \pm 0.06 $                 &    $ 7.98 \pm0.07$         &   $-$                      &	0.58   \\
Sh\,2-132   &   $-$                  & $ 7.47 \pm 0.09 $   &   $ 8.35 \pm 0.16 $                 &    $ 7.12 \pm0.20$         &   $-$                      &	0.06   \\
Sh\,2-152   &   7.97$\pm$0.14        & $ 7.57 \pm 0.02 $   &   $ 8.45 \pm 0.03 $                 &    $ 7.73  \pm0.05$        &   $-$                       & 	0.16   \\
Sh\,2-156   &   $-$                  & $ 7.28 \pm 0.07 $   &   $ 8.23 \pm 0.14$                  &    $ 7.68 \pm 0.10$        &   $-$                       & 	0.22   \\
Sh\,2-175   &   $-$                  & $ 7.74 \pm 0.10 $   &   $ 8.51 \pm 0.15$                  &    $ - $                   &   $-$                       &    0.01    \\         
Sh\,2-209   &   $-$                  & $ 6.74 \pm 0.14 $   &   $ 7.66^{+0.33}_{-0.26} $          &    $7.88^{+0.22}_{-0.18}$  &   $-$                       & 	0.62   \\         
Sh\,2-212   &   $-$                  & $ 6.62 \pm 0.17 $   &   $ 8.18 ^{+ 0.30 }_{- 0.21 }$      &    $ 7.82\pm0.18$          &   $-$                       &    0.32 \\      
Sh\,2-219   &   $-$                  & $ 7.45 \pm 0.06 $   &   $ 8.28 \pm 0.12$                  &    $ 6.04  \pm0.20$        &   $-$                       & 	0.01   \\
Sh\,2-235   &   $-$                  & $ 7.56 \pm 0.04 $   &   $ 8.41 \pm 0.07 $                 &    $ 6.91 \pm 0.10 $       &   $-$                       & 	0.03   \\
Sh\,2-237   &   $-$                  & $ 7.54 \pm 0.05 $   &   $ 8.34 \pm0.09$                   &    $ 6.39  \pm0.13$        &   $-$                       & 	0.01   \\
Sh\,2-257   &   $-$                  & $ 7.60 \pm 0.06 $   &   $ 8.40 \pm 0.10$                  &    $ 6.57 \pm 0.15 $       &   $-$                       & 	0.01  \\
Sh\,2-266   &   $-$                  & $ 7.52 \pm 0.11 $   &   $ 8.19 \pm0.17$                   &    $ 6.06\pm 0.27$         &   $-$                       &    $\sim$0.0   \\     
Sh\,2-270   &   $-$                  & $ 7.26 \pm 0.19 $   &   $ 8.09 ^{+ 0.30 }_{- 0.26 }$      &    $ - $                   &   $-$                       &    $\sim$0.0    \\    
Sh\,2-271   &   $-$                  & $ 7.50 \pm 0.05 $   &   $ 8.27 \pm0.08$                   &    $ 5.85 \pm 0.12 $       &   $-$                       & 	0.004    \\
Sh\,2-285   &   $-$                  & $ 7.44 \pm 0.07 $   &   $ 8.21 \pm 0.11 $                 &    $ - $                   &   $-$                       &    $\sim$0.0    \\    
Sh\,2-288   &   $-$                  & $ 7.21 \pm 0.07 $   &   $ 8.21 \pm 0.10 $                 &    $ 7.76 \pm 0.12 $       &   $-$                       &    0.26   \\
Sh\,2-297   &   $-$                  & $ 7.71 \pm 0.05 $   &   $ 8.47 \pm 0.08 $                 &    $ 6.72 \pm 0.13 $       &   $-$                       & 	0.02   \\
Sh\,2-298   &   $-$                  & $ 7.32 \pm 0.07 $   &   $ 8.15 \pm 0.11$                  &    $ 8.11 \pm 0.03 $       &   $-$                       &	0.47   \\
Sh\,2-311   &   8.01$\pm$0.05        & $ 7.31 \pm 0.04 $   &   $ 8.28  \pm0.06$                  &    $ 7.83 \pm 0.03 $       &   8.56$\pm$0.04            &    0.26  \\                    

\hline                     
\end{tabular}
\begin{description}
\item  Note: in units of 12+log($X^{n+}$/H$^{+}$)
\end{description} 
\end{table*}

%% file: Table4.tex
 \begin{table*}\footnotesize
 \caption{Abundances of the observed ionic species of Ne, S, Cl and Ar.}
 \begin{tabular}{lcccccccc}
 \hline
 \multicolumn{1}{l}{Region} & \multicolumn{1}{c}{Ne$^{2+}$} &
 \multicolumn{1}{c}{S$^{+}$} &  \multicolumn{1}{c}{S$^{2+}$} &
 \multicolumn{1}{c}{Cl$^{+}$} & \multicolumn{1}{c}{Cl$^{2+}$} & 
  \multicolumn{1}{c}{Cl$^{3+}$} & \multicolumn{1}{c}{Ar$^{2+}$} &
  \multicolumn{1}{c}{Ar$^{3+}$} \\
 \hline      

IC~5146      &   $-$                        & $ 6.78 \pm 0.05$       &  $ 6.22 \pm 0.05 $                            &     $ -$                             &        $ - $                           &   $ -$	                       &      $ -$                                  &      $ -$        \\
M8           &   $7.14\pm 0.04$             & $ 6.06\pm0.04$                  &  $ 6.86 \pm 0.03$                             &     $ 4.22 \pm 0.03 $                &        $ 4.97 \pm 0.03 $               &   $ -$	                       &      $ 6.21 \pm 0.02$                      &      $ 3.93 \pm0.11$                    \\
M16          &   $7.24^\pm0.09$             & $ 6.44 \pm 0.04 $               &  $ 6.77 \pm 0.04$                             &     $ 4.43 \pm 0.05 $                &        $ 4.95 \pm 0.04 $               &   $ -$						   &      $ 6.23 \pm 0.02 $                     &      $ 4.14 ^{+ 0.16 }_{- 0.22 }$        \\
M17          &   $7.85\pm 0.04$             & $ 5.53\pm0.05$                  &  $ 6.82 \pm0.04$                              &     $ 3.60 \pm0.10$                  &        $ 4.90 \pm 0.06$                &   $ 3.15 ^{+ 0.16 }_{- 0.21 }$ &      $ 6.31 \pm 0.02 $                     &      $ 4.26 ^{+ 0.11 }_{- 0.15 }$        \\
M20          &   $6.78\pm 0.10 $            & $ 6.31 \pm 0.03 $               &  $ 6.77 \pm 0.03 $                            &     $ 4.46 \pm 0.04 $                &        $ 4.94 \pm 0.04 $               &   $ - $                        &      $ 6.25 \pm 0.04$                      &      $ 4.27 ^{+ 0.16 }_{- 0.25 }$        \\     
M42          &   $7.86\pm 0.02$             & $ 5.47 ^{+ 0.17 }_{- 0.11 }$    &  $ 6.65 \pm0.06$                              &     $ 3.51 \pm 0.08$                 &        $ 4.81 \pm 0.05 $               &   $ 3.66 \pm 0.06$             &      $ 6.25 \pm 0.04$                      &      $ 4.66 \pm 0.05$                    \\
NGC\,2579    &   $7.33\pm 0.05$             & $ 5.54 \pm0.07$                 &  $ 6.66 \pm 0.05 $                            &     $ 3.56 \pm 0.06$                 &        $ 4.92 \pm 0.07 $               &   $ 2.79 \pm 0.06 $            &      $ 6.19 \pm 0.03 $                     &      $ 3.96 \pm 0.12$                    \\
NGC\,3576    &   $7.79\pm 0.02$             & $ 5.79 \pm 0.05 $               &  $ 6.84 \pm0.05$                              &     $ 3.80 \pm 0.05 $                &        $ 4.88 \pm 0.04 $               &   $ 3.22 \pm 0.05 $            &      $ 6.33 \pm 0.03$                      &      $ 4.42 \pm 0.05$                    \\
NGC\,3603    &   $7.88\pm0.05$              & $ 5.07 \pm 0.09 $               &  $ 6.69 \pm0.08$                              &     $ 3.18 \pm 0.07 $                &        $ 4.76 \pm 0.08 $               &   $ 3.87 \pm 0.04 $            &      $ 6.22 \pm 0.05$                      &      $ 4.81 \pm 0.09$                     \\
Sh\,261      &   $-$                        & $6.77 \pm 0.09 $                &  $-$                                          &     $-$                              &        $-$                             &   $-$                          &      $-$                                   &      $-$                                  \\            
Sh\,2-83     &   $7.71\pm0.12$              & $ 5.20 \pm0.07$                 &  $ 6.25 \pm 0.12 $                            &     $-$                              &        $ 4.59 \pm0.09$                 &   $-$                          &      $ 5.70 \pm 0.06$                      &      $ -$                                \\
Sh\,2-90     &   $-$                        & $6.33 \pm 0.05 $                &  $-$                                          &     $-$                              &        $ 5.11 \pm 0.09$                &   $-$                          &      $ 6.27\pm 0.09$                       &      $-$                                 \\
Sh\,2-100    &   $7.83\pm 0.05$             & $ 5.53 \pm0.07$                 &  $ 6.86\pm0.11$                               &     $-$                              &        $ 4.99 \pm 0.07 $               &   $-$                          &      $ 6.29 \pm 0.05 $                     &      $4.79\pm0.10$                     \\
Sh\,2-127    &   $-$                        & $ 5.91 \pm 0.03 $               &  $ 6.69 \pm 0.06$                             &     $-$                              &        $ 4.73 \pm 0.05 $               &   $-$                          &      $ 5.67 \pm 0.05$                      &      $ -$                     \\
Sh\,2-128    &   $7.13\pm 0.11$             & $ 5.62 \pm 0.03 $               &  $ 6.44 \pm0.06$                              &     $-$                              &        $ 4.71 \pm 0.05$                &   $-$                          &      $ 5.84 \pm0.03$                       &      $ -$                     \\
Sh\,2-132    &   $-$                        & $ 6.38 \pm0.09$                 &  $ 6.58 \pm 0.05 $                            &     $-$                              &        $-$                             &   $-$                          &      $ 6.07\pm0.10$                        &      $-$                                 \\
Sh\,2-152    &   $-$                        &  $ 6.15 \pm 0.02 $              &  $ 6.93 \pm 0.04 $                            &     $-$                              &        $ 4.99 \pm 0.03$                &   $-$                          &      $ 6.11 \pm 0.03 $                     &      $-$                                 \\
Sh\,2-156    &   $6.47\pm0.13$              & $ 5.88 \pm0.08   $              &  $ 6.84 \pm0.05$                              &     $-$                              &        $ 4.85 \pm 0.09$                &   $-$                          &      $ 6.11 \pm 0.03 $                     &      $ - $                   \\
Sh\,2-175    &   $-$                        & $6.70\pm0.08$                    &  $-$                                          &     $-$                              &        $-$                             &   $-$                          &      $-$                                   &      $-$                                  \\            
Sh\,2-209    &   $-$                        & $5.51 \pm 0.11 $                &  $-$                                          &     $-$                              &        $-$                             &   $-$                          &      $-$                                   &      $-$                                  \\            
Sh\,2-212    &   $6.89\pm0.22$              & $ 5.16 \pm0.16$                 &  $ 6.56 ^{+ 0.30 }_{- 0.26 }$                 &     $-$                              &        $ 4.89\pm0.20$                  &   $-$                          &      $ 6.03 ^{+ 0.17 }_{- 0.07 }$          &      $ -$        \\
Sh\,2-219    &   $-$                        & $6.31 \pm 0.06 $                &  $-$                                          &     $-$                              &        $ 4.80 ^{+ 0.18 }_{- 0.30 }$    &   $-$                          &      $ 5.22 \pm 0.08 $                     &      $-$                                 \\
Sh\,2-235    &   $-$                        &  $ 6.92 \pm 0.08$      &  $ 6.65 \pm 0.08 $                            &     $-$                              &        $ 4.75 \pm 0.07 $               &   $-$                          &      $ 5.87 \pm 0.05 $                     &      $-$                                 \\
Sh\,2-237    &   $-$                        & $6.43\pm0.06$                   &  $-$                                          &     $-$                              &        $ 4.84 ^{+ 0.16 }_{- 0.22 }$    &   $-$                          &      $ 5.10 ^{+ 0.07 }_{- 0.11 }$           &      $-$                                 \\
Sh\,2-257    &   $-$                        & $ 7.01 \pm 0.17 $       &  $ 6.69 \pm0.19$                              &     $-$                              &        $ 4.64 \pm0.11$                 &   $-$                          &      $ 5.37 ^{+ 0.14 }_{- 0.08 }$          &      $-$                                 \\
Sh\,2-266    &   $-$                        & $6.61 \pm 0.10 $                &  $-$                                          &     $-$                              &        $-$                             &   $-$                          &      $-$                                   &      $-$                                  \\            
Sh\,2-270    &   $-$                        & $6.29 ^{+ 0.18 }_{- 0.14 }$     &  $-$                                          &     $-$                              &        $-$                             &   $-$                          &      $-$                                   &      $-$                                  \\            
Sh\,2-271    &   $-$                        & $ 6.70\pm0.09 $          &  $ 6.44 \pm 0.09 $                            &     $-$                              &        $ 4.71 \pm0.10$                 &   $-$                          &      $ 5.04 \pm 0.10$                      &      $-$                                 \\
Sh\,2-285    &   $-$                        & $5.48 \pm 0.07$                 &  $-$                                          &     $-$                              &        $-$                             &   $-$                          &      $-$                                   &      $-$                                  \\            
Sh\,2-288    &   $-$                        & $ 5.93 \pm 0.07 $               &  $ 6.61\pm0.11$                               &     $-$                              &        $ 4.73 \pm 0.07 $               &   $-$                          &      $ 5.80 \pm 0.06 $                     &      $ -$        \\
Sh\,2-297    &   $-$                        & $6.47 \pm 0.05$                 &  $-$                                          &     $-$                              &        $ 4.71 \pm0.11$                 &   $-$                          &      $ 5.44 \pm0.09$                       &      $-$                                 \\
Sh\,2-298    &   $7.85\pm 0.04$             & $ 6.48 \pm 0.06 $               &  $ 6.51 \pm0.06$                              &     $-$                              &        $ 4.84 \pm 0.10 $               &   $-$                          &      $ 5.95 \pm 0.03$                      &      $ 4.33\pm0.09$                      \\
Sh\,2-311    &   $7.06\pm0.04$                & $ 6.21 \pm 0.04 $               &  $ 6.66 \pm 0.03 $                            &     $ 4.30 \pm 0.04 $                &        $ 4.81 \pm 0.04 $               &   $-$                          &      $ 6.07 \pm 0.03$                      &      $ -$                                \\

\hline

 \end{tabular}
 \label{tab:ionic-abundances}
 \begin{description} 
\item Note: in units of 12+log($X^{n+}$/H$^{+}$).
\end{description} 
 \end{table*}

%% file: Table5.tex
\begin{table*}\footnotesize
\caption{Average values and standard deviation for Ne/O, S/O, Cl/O, Ar/O and C/O in our sample of Galactic \ion{H}{ii} regions using different ICF schemes. Bold numbers indicate the final adopted value of each abundance ratio. Solar abundance values are included for comparison.}
 \begin{center}
 \begin{tabular}{l c c c c c c  }
 \hline
 \multicolumn{1}{c}{ICF}  &  \multicolumn{1}{c}{C/O} & \multicolumn{1}{c}{N/O} & \multicolumn{1}{c}{Ne/O}  &  \multicolumn{1}{c}{S/O}  &  \multicolumn{1}{c}{Cl/O} & \multicolumn{1}{c}{Ar/O}  \\
 \hline
No ICF                                   & $-$                      & $\mathbf{-0.76\pm0.09}$     & $-$                              & $-1.52\pm0.11$            &   $-3.55\pm0.08$           &   $-$               \\

\citet{Peimbert:1969}                    & $-$                      & $\mathbf{-0.90\pm0.11}$     & $\mathbf{-0.71\pm0.15}$           & $-1.23\pm0.30$            &   $-$                     &   $-$               \\
\citet{Stasinska:1978}                   & $-$                      & $-$                         & $-0.29\pm0.39$                    & $-1.57\pm0.08$            & $-$                       &   $-$               \\
\citet{Garnett:1999}                     & $-1.00\pm0.13$           & $-$                         & $-$                               & $-$                       & $-$                       &   $-$              \\
\citet{Izotov:2006}                      & $-$                      & $-$                         & $-0.79\pm0.19$                    & $-1.55\pm0.12$            & $-3.40\pm0.08$            &   $\mathbf{-2.27\pm0.11}$    \\
\citet{Perez-Montero:2007}               & $-$                      & $-$                         & $-0.92\pm0.32$                    & $-$                       & $-$                       &   $-2.36\pm0.29$    \\
\citet{Dors:2013}                        & $-$                      & $-$                         & $-0.82\pm0.27$                    & $-$                       & $-$                       &   $-$               \\
\citet{Esteban:2015}                     & $-$                      & $-$                         & $-$                               & $-$                       & $\mathbf{-3.48\pm0.10}$   &   $-$               \\
\citet{Dors:2016}                        & $-$                      & $-$                         & $-$                               & $\mathbf{-1.58\pm0.08}$   & $-$                       &   $-$               \\
\citet{Berg:2019} 						 & $\mathbf{-0.27\pm0.14}$  & $-$                         & $-$                               & $-$                       & $-$                       &   $-$              \\
Amayo et al. 2020                        & $-0.26\pm0.12$           & $-0.81\pm0.11$              & $-0.69\pm0.17$                    &  $-1.59\pm0.09$           & $-3.09\pm0.44$            &   $-2.30\pm0.14$   \\
Dom\'inguez-Guzm\'an et al. (in prep.)   & $-$                      & $-$                         & $-$                               & $-$                       &  $-3.56\pm0.11$           &   $-$               \\

\hline
\multicolumn{1}{c}{Solar abundances}  &  \multicolumn{1}{c}{C/O} &  \multicolumn{1}{c}{N/O} &\multicolumn{1}{c}{Ne/O}  &  \multicolumn{1}{c}{S/O}  &  \multicolumn{1}{c}{Cl/O} &  \multicolumn{1}{c}{Ar/O} \\
 \hline

                                          
                                          
\citet{Lodders:2019}                      &   $-0.26\pm0.09$  & $-0.88\pm0.14$    & $-0.58\pm0.12$ & $-1.58\pm0.08$  & $-3.50\pm0.09$  &   $-2.23\pm0.12$     \\

 \hline
 \end{tabular}
 \end{center}
 \label{Tab:ICFs}
 \end{table*}

%% file: Table6.tex
\begin{table*}\scriptsize
 \caption{Total abundances of C, N, O, Ne, S, Cl and Ar  for Galactic \ion{H}{ii} regions calculated using the ICF schemes adopted for each element (see text for details).} \label{tab:total-abundances}
 \begin{tabular}{lccccccccc c c}
 \hline
 \multicolumn{1}{l}{Region} &
 \multicolumn{1}{c}{C/H}  &
  \multicolumn{3}{c}{N/H}  &
  \multicolumn{1}{c}{O/H}  &
 \multicolumn{1}{c}{O/H} &
 \multicolumn{1}{c}{Ne/H} &
 \multicolumn{1}{c}{S/H} &
 \multicolumn{1}{c}{Cl/H} &
 \multicolumn{1}{c}{Ar/H}\\

  \multicolumn{1}{l}{} &
 \multicolumn{1}{c}{}  &
  \multicolumn{1}{c}{(1)}  &
  \multicolumn{1}{c}{(2)}  &
  \multicolumn{1}{c}{(3)} &
  \multicolumn{1}{c}{(CELs)}  &
 \multicolumn{1}{c}{(RLs)} &
 \multicolumn{1}{c}{} &
 \multicolumn{1}{c}{} &
 \multicolumn{1}{c}{} &
 \multicolumn{1}{c}{}\\

 \hline      

IC\,5146       &   $-$                       &  $ - $                              &    $ -$                             &  $ 7.81 \pm 0.05 $            &   $ 8.57\pm0.09$                  &    $-$                        &     $-$                 &     $6.94\pm0.04$$^{\rm b}$          &  $-$                             &    $-$                               \\       
M8 	           &   $ 8.51 \pm 0.05 $         &  $ 7.67 \pm 0.02 $                  &   $ 7.73 \pm0.11$                   &  $-$                            &   $ 8.48 \pm 0.04 $             &    8.23$\pm$0.03              &     $ 7.72 \pm 0.05 $   &     $ 6.94 \pm 0.02 $                &   $ 5.05 \pm 0.03^{\rm c} $      &    $ 6.26 \pm 0.02 $                 \\
M16            &   $ 8.53 \pm 0.08 $         &  $ 7.84 \pm 0.03 $                  &   $ 7.89  \pm0.11$                  &  $-$                            &   $ 8.57 \pm 0.05 $             &    8.30$\pm$0.04              &     $ 7.9 \pm 0.05 $    &     $ 6.97 \pm 0.03 $                &   $ 5.06 \pm 0.04^{\rm c} $      &    $ 6.28 \pm 0.03 $                 \\
M17 	          &   $ 8.76 \pm 0.05 $       &  $ 7.58 \pm 0.04 $                  &   $ 7.66 ^{+ 0.26 }_{- 0.09 }$      &  $-$                           &   $ 8.54 \pm 0.03 $                &    8.68$\pm$0.02              &     $ 7.94 \pm 0.04 $   &     $ - $                            &   $ 4.93 \pm 0.05^{\rm c} $      &    $ 6.32 \pm 0.02 $                 \\
M20            &   $ 8.40 \pm 0.27 $         &  $ 7.67 \pm 0.04 $                  &   $ 7.70  \pm0.10$                   &  $-$                           &   $ 8.55\pm0.05$                &    8.00$\pm$0.22              &     $ 7.59 \pm 0.06 $   &     $ 6.92 \pm 0.03 $                &   $ 5.07 \pm 0.04^{\rm c} $      &    $ -$                              \\
M42            &   $ 8.45 \pm 0.03 $         &  $ 7.67 \pm 0.05 $                  &   $ 7.75 ^{+ 0.28 }_{- 0.08 }$      &  $-$                            &   $ 8.51\pm0.03$                &    8.57$\pm$0.01              &     $ 7.93 \pm 0.03 $   &     $ - $                            &   $ 4.87 \pm 0.05^{\rm c} $      &    $ 6.26 \pm 0.03 $                 \\
NGC\,2579      &   $ 8.23 \pm 0.07 $         &  $ 7.29 \pm 0.03 $                  &   $ 7.38 ^{+ 0.19 }_{- 0.10 }$      &  $-$                            &   $ 8.42\pm 0.05$               &    8.46$\pm$0.03              &     $ 7.37 \pm 0.05 $   &     $ 6.69 \pm 0.03 $                &   $ 4.94 \pm 0.07^{\rm c} $      &    $ 6.21 \pm 0.05 $                 \\
NGC\,3576      &   $ 8.52 \pm 0.07 $         &  $ 7.55 \pm 0.02 $                  &   $ 7.64 ^{+ 0.20 }_{- 0.09 }$      &  $-$                            &   $ 8.54\pm0.03$                &    8.62$\pm$0.05              &     $ 7.95 \pm 0.03 $   &     $ 6.94 \pm 0.04 $                &   $ 4.92 \pm 0.04^{\rm c} $      &    $ 6.34 \pm 0.03 $                 \\
NGC\,3603      &   $ 8.50 \pm 0.10 $         &  $ 7.61 \pm 0.06 $                  &   $ 7.67 ^{+ 0.34 }_{- 0.10 }$      &  $-$                            &   $ 8.47\pm0.04$                &    8.69$\pm$0.05              &     $ 7.91 \pm 0.04 $   &     $ - $                            &   $ 4.82 \pm 0.08^{\rm c} $      &    $ 6.25 \pm 0.04 $                 \\
Sh\,2-61 	     &   $-$                     &  $ - $                              &   $ -$                              &  $ 7.89 \pm0.10$                &   $ 8.41\pm0.15$                  &    $-$                        &     $-$                 &     $-$                              &   $-$                            &                                      \\
Sh\,2-83 	     &   $-$                     &  $ 7.45 \pm 0.09 $                  &   $ 7.51 ^{+ 0.34 }_{- 0.11 }$      &  $-$                            &   $ 8.28 \pm0.08$                 &    $-$                        &     $ 7.75 \pm 0.08 $   &     $ - $                            &   $ 4.61 \pm 0.07 $              &    $ 5.84 ^{+ 0.03 }_{- 0.07 }$      \\
Sh\,2-90       &   $-$                       &  $ 7.91 \pm 0.06 $                  &   $ 7.97 ^{+ 0.14 }_{- 0.11 }$      &  $-$                            &   $ 8.51 \pm 0.08 $             &    $-$                        &     $-$                 &     $-$                              &   $ 5.17 \pm 0.08 $              &    $ 6.36 \pm 0.07 $                 \\
Sh\,2-100      &   $ 8.40 \pm 0.12 $         &  $ 7.54 \pm 0.05 $                  &   $ 7.61 ^{+ 0.27 }_{- 0.09 }$      &  $-$                            &   $ 8.49\pm0.04$                &    8.59$\pm$0.06              &     $ 7.91 \pm 0.05 $   &     $ - $                            &   $ 5.02 \pm 0.06 $              &    $ 6.33 \pm 0.05 $                 \\
Sh\,2-127      &   $-$                       &  $ 7.35 \pm 0.03 $                  &   $ 7.37 \pm 0.09 $                 &  $-$                            &   $ 8.27 \pm 0.04 $             &    $-$                        &     $-$                 &     $ 6.77 \pm 0.06 $                &   $ 4.74 \pm 0.04 $              &    $ - $                             \\
Sh\,2-128      &   $-$                       &  $ 7.19 \pm 0.04 $                  &   $ 7.29 ^{+ 0.17 }_{- 0.09 }$      &  $-$                            &   $ 8.22 \pm 0.05$              &    $-$                        &     $ 7.36 \pm 0.08 $   &     $ 6.58 \pm 0.05 $                &   $-$                            &    $ 5.87 \pm 0.05 $                 \\
Sh\,2-132      &   $-$                       &  $ 7.49 \pm 0.09 $                  &   $ 7.48 \pm 0.17 $                 &  $-$                            &   $ 8.37 \pm0.16$               &    $-$                        &     $-$                 &     $ 6.80 \pm 0.06 $                &   $-$                            &    $ - $                             \\
Sh\,2-152      &  $ 8.17 \pm 0.16 $          &  $ 7.65 \pm 0.02 $                  &   $ 7.68 \pm 0.09 $                 &  $-$                            &   $ 8.53 \pm 0.03 $             &    8.72$\pm$0.06$^{\rm a}$    &     $-$                 &     $ 7.10 \pm 0.04 $                &   $ 5.09 \pm 0.04 $              &    $ - $                             \\
Sh\,2-156      &  $-$                        &  $ 7.39 \pm 0.06 $                  &   $ 7.43 \pm 0.15 $                 &  $-$                            &   $ 8.34 \pm0.11$               &    $-$                        &     $-$                 &     $ 6.90 \pm 0.03 $                &   $ 4.93 \pm 0.08 $              &    $ 6.19 \pm0.07$                   \\
Sh\,2-175      &  $-$                        &  $ - $                              &   $ -$                              &  $ 7.74 \pm 0.10 $              &   $ 8.51 \pm 0.15$              &    $-$                        &     $-$                 &     $-$                              &   $-$                            &   $-$                                \\
Sh\,2-209      &  $-$                        &  $- $                               &   $ - $                             &  $-$                            &   $8.08^{+0.23}_{-0.17}$        &    $-$                        &     $-$                 &     $-$                              &   $-$                            &   $-$                                \\
Sh\,2-212      &  $-$                      &  $ 6.78 \pm 0.11 $                  &   $ 6.84 \pm 0.25 $                 &  $-$                              &   $ 8.30 ^{+ 0.23 }_{- 0.18 }$  &    $-$                        &     $7.44 \pm 0.25 $    &     $-$                              &   $ 4.96 ^{+ 0.18 }_{- 0.13 }$   &    $ - $                             \\
Sh\,2-219      &  $-$                        &  $ - $                              &   $ - $                             &  $ 7.45 \pm 0.07 $              &   $ 8.28 \pm0.12$               &    $-$                        &     $-$                 &     $-$                              &   $-$                            &    $ - $                             \\
Sh\,2-235      &  $-$                        &  $ - $                              &   $ - $                             &  $ 7.56 \pm 0.04 $              &   $ 8.42 \pm 0.07 $             &    $-$                        &     $-$                 &     $6.92\pm0.08$$^{\rm a}$          &   $-$                            &    $ - $                             \\
Sh\,2-237      &  $-$                        &  $ - $                              &   $ - $                             &  $ 7.54 \pm 0.05 $              &   $ 8.34 \pm 0.09$              &    $-$                        &     $-$                 &     $-$                              &   $-$                            &    $ - $                             \\
Sh\,2-257      &  $-$                        &  $ - $                              &   $ - $                             &  $ 7.60 \pm 0.06 $              &   $ 8.40\pm0.10$                &    $-$                        &     $-$                 &     $7.01\pm0.17$$^{\rm a}$          &   $-$                            &    $ - $                             \\
Sh\,2-266      &  $-$                        &  $ - $                              &   $- $                              &  $ 7.52 \pm 0.11 $              &   $ 8.19\pm0.17 $               &    $-$                        &     $-$                 &     $-$                              &   $-$                            &   $-$                                \\
Sh\,2-270      &  $-$                        &  $ - $                              &   $ - $                             &  $ 7.26 \pm 0.19$               &   $ 8.09 ^{+ 0.30 }_{- 0.26 }$  &    $-$                        &     $-$                 &     $-$                              &   $-$                            &   $-$                                \\
Sh\,2-271      &  $-$                        &  $ - $                              &   $ - $                             &  $ 7.50 \pm 0.05 $              &   $ 8.27 \pm0.08$               &    $-$                        &     $-$                 &     $6.70\pm0.09$$^{\rm a}$          &   $-$                            &    $ - $                             \\
Sh\,2-285      &  $-$                        &  $ - $                              &   $ -$                              &  $ 7.44 \pm 0.07 $              &   $ 8.21 \pm0.11$               &    $-$                        &     $-$                 &     $-$                              &   $-$                            &                                      \\
Sh\,2-288      &  $-$                        &  $ 7.35 \pm 0.06 $                  &   $ 7.4 \pm 0.14 $                  &  $-$                            &   $ 8.34 \pm0.08$               &    $-$                        &     $-$                 &     $ 6.72 \pm 0.09 $                &   $ 4.79 \pm 0.06 $              &    $ 5.89 \pm 0.07 $                 \\
Sh\,2-297      &  $-$                        &  $ - $                              &   $ -$                              & $ 7.71 \pm 0.05 $               &   $ 8.48 \pm 0.08 $             &    $-$                        &     $-$                 &     $-$                              &   $-$                            &    $ -$                              \\
Sh\,2-298      &  $-$                        &  $ 7.61 \pm 0.03 $                  &   $ 7.70 ^{+ 0.15 }_{- 0.10 }$      &  $-$                            &   $ 8.43\pm0.06$                &    $-$                        &     $-$                 &     $ 6.80 \pm 0.04 $                &   $ 4.89 \pm 0.08 $              &    $ 5.99 \pm 0.04 $                 \\
Sh\,2-311 	    &  $ 8.18 \pm 0.08 $         &  $ 7.44 \pm 0.03 $                  &   $ 7.50  \pm0.11$                  &  $-$                            &   $ 8.41 \pm 0.04 $              &    8.56$\pm$0.04              &     $ 7.64 \pm 0.04 $   &     $ 6.81 \pm 0.02 $                &   $ 4.93 \pm 0.04^{\rm b} $      &    $ 6.15 \pm 0.02 $                 \\

\hline
\end{tabular}
\begin{description} 
\item[] Note: in units of 12+log(X/H).
\item[(1)] N abundances determined using the ICF scheme of \citet{Peimbert:1969}.
\item[(2)] N abundances determined using the ICF scheme of \citet{Amayo:2020}.   
\item[(3)] N abundances determined without ICF.                         

 \item $^{\rm a}$ Assumed value (see text for details).
\item $^{\rm b}$ S abundances determined without ICF. 
\item $^{\rm c}$ Cl abundances determined without ICF.
\end{description} 
 \end{table*}

%% file: Table7.tex
\begin{table*}
 \caption{Total abundance ratios of C/O, N/O, Ne/O, S/O, Cl/O, Ar/O and C/O for Galactic \ion{H}{ii} regions calculated using the ICF schemes adopted for each element (see text for details).} \label{tab:total-abundances-ratios}
 \begin{tabular}{lccccccccc  }
 \hline
 \multicolumn{1}{l}{Region} &
 \multicolumn{1}{c}{C/O}  &
  \multicolumn{3}{c}{N/O}  &
 \multicolumn{1}{c}{Ne/O} &
 \multicolumn{1}{c}{S/O} &
 \multicolumn{1}{c}{Cl/O} &
 \multicolumn{1}{c}{Ar/O}\\

  \multicolumn{1}{l}{} &
 \multicolumn{1}{c}{}  &
  \multicolumn{1}{c}{(1)}  &
  \multicolumn{1}{c}{(2)}  &
  \multicolumn{1}{c}{(3)} &
 \multicolumn{1}{c}{} &
 \multicolumn{1}{c}{} &
 \multicolumn{1}{c}{} &
 \multicolumn{1}{c}{}\\

 \hline      

IC\,5146           &   $-$                                 &    $ -$                             &     $ - $                                &  $ -0.75 \pm 0.03 $      &       $-$                   &   $-1.63\pm0.04$        &     $-$                           &   $-$                 \\
M8 	               &   $ -0.21 \pm 0.04 $                  &    $ -0.80 \pm 0.04 $               &     $ -0.75 \pm0.10$                     &  $-$                     &       $ -0.75 \pm 0.03 $    &   $ -1.53 \pm 0.05 $    &     $ -3.43 \pm 0.03 $            &   $ -2.22 \pm 0.05 $  \\
M16                &   $ -0.29 \pm 0.06 $                  &    $ -0.73 \pm 0.03 $               &     $ -0.68 \pm0.10$                     &  $-$                     &       $ -0.68 \pm 0.02 $    &   $ -1.61 \pm 0.05 $    &     $ -3.51 \pm 0.03 $            &   $ -2.29 \pm 0.05 $  \\
M17 	              &   $  0.01 \pm 0.04 $                  &    $ -0.95 \pm 0.03 $               &     $ -0.88 ^{+ 0.26 }_{- 0.08 }$        &  $-$                     &       $ -0.60 \pm 0.02 $    &   $ -$                  &     $ -3.61 \pm 0.04 $            &   $ -2.22 \pm 0.04 $  \\
M20                &   $ -0.32 \pm 0.23 $                  &    $ -0.88 \pm 0.02 $               &     $ -0.85 \pm0.08$                     &  $-$                     &       $ -0.96 \pm 0.03 $    &   $ -1.63 \pm 0.05 $    &     $ -3.49 \pm 0.02 $            &   $-$                 \\
M42                &   $ -0.26 \pm 0.03 $                  &    $ -0.84 \pm 0.07 $               &     $ -0.76 ^{+ 0.28 }_{- 0.08 }$        &  $-$                     &       $ -0.58 \pm 0.01 $    &   $ - $                 &     $ -3.64 \pm 0.03 $            &   $ -2.26 \pm 0.04 $  \\
NGC\,2579          &   $ -0.33 \pm 0.06 $                  &    $ -1.13 \pm 0.04 $               &     $ -1.03 ^{+ 0.18 }_{- 0.09 }$        &  $-$                     &       $ -0.87 \pm 0.02 $    &   $ -1.73 \pm 0.06 $    &     $ -3.47 \pm 0.04 $            &   $ -2.2 \pm 0.07 $   \\
NGC\,3576          &   $ -0.24 \pm 0.05 $                  &    $ -0.99 \pm 0.03 $               &     $ -0.90 ^{+ 0.20 }_{- 0.09 }$         &  $-$                    &       $ -0.58 \pm 0.01 $    &   $ -1.60 \pm 0.05 $    &     $ -3.62 \pm 0.03 $            &   $ -2.2 \pm 0.04 $   \\
NGC\,3603          &   $ -0.21 \pm 0.09 $                  &    $ -0.86 \pm 0.06 $               &     $ -0.79 ^{+ 0.34 }_{- 0.09 }$        &  $-$                     &       $ -0.56 \pm 0.02 $    &   $ - $                 &     $ -3.65 \pm 0.07 $            &   $ -2.22 \pm 0.05 $  \\
Sh\,2-61 	         &   $-$                                 &    $- $                             &     $ - $                                &  $ -0.53 \pm 0.06 $      &       $-$                   &   $-$                   &     $-$                           &   $-$                 \\
Sh\,2-83 	         &   $-$                                 &    $ -0.83 \pm 0.05 $               &     $ -0.77 ^{+ 0.34 }_{- 0.08 }$        &  $-$                     &       $ -0.53 \pm 0.04 $    &   $ - $                 &     $ -3.67 \pm 0.10 $            &   $ -2.44 \pm 0.09 $  \\
Sh\,2-90           &   $-$                                 &    $ -0.61 \pm 0.04 $               &     $ -0.54 \pm 0.11 $                   &  $-$                     &       $-$                   &   $-$                   &     $ -3.33 \pm 0.12 $            &   $ -2.15 \pm 0.11 $  \\
Sh\,2-100          &   $ -0.19 \pm 0.11 $                  &    $ -0.96 \pm 0.05 $               &     $ -0.88 ^{+ 0.27 }_{- 0.08 }$        &  $-$                     &       $ -0.58 \pm 0.03 $    &   $ - $                 &     $ -3.47 \pm 0.07 $            &   $ -2.17 \pm 0.06 $  \\
Sh\,2-127          &   $-$                                 &    $ -0.91 \pm 0.02 $               &     $ -0.89 \pm 0.07 $                   &  $-$                     &       $-$                   &   $ -1.49 \pm 0.07 $    &     $-$                           &   $-$                 \\
Sh\,2-128          &   $-$                                 &    $ -1.02 \pm 0.03 $               &     $ -0.93 ^{+ 0.16 }_{- 0.08 }$        &  $-$                     &       $ -0.86 \pm 0.07 $    &   $ -1.63 \pm 0.07 $    &     $ -3.47 \pm 0.06 $            &   $ -2.35 \pm 0.07 $  \\
Sh\,2-132          &   $-$                                 &    $ -0.88 \pm 0.07 $               &     $ -0.88 \pm 0.09 $                   &  $-$                     &       $-$                   &   $ -1.56 \pm 0.17 $    &     $-$                           &   $-$                 \\
Sh\,2-152          &   $ -0.55 \pm 0.15 $                  &    $ -0.88 \pm 0.02 $               &     $ -0.85 \pm 0.08 $                   &  $-$                     &       $-$                   &   $ -1.42 \pm 0.05 $    &     $ -3.44 \pm 0.05 $            &   $-$                 \\
Sh\,2-156          &   $-$                                 &    $ -0.96 \pm 0.06  $              &     $ -0.92 \pm 0.10 $                   &  $-$                     &       $-$                   &   $ -1.44 \pm 0.10 $    &     $ -3.41 \pm 0.13 $            &   $ -2.16 \pm 0.11 $  \\
Sh\,2-175          &   $-$                                 &    $ -0.76 \pm 0.06 $               &     $ -0.83 \pm 0.24 $                   &  $ -0.77 \pm 0.06 $      &       $-$                   &   $-$                   &     $-$                           &   $-$                 \\
Sh\,2-209          &   $-$                                 &    $ -$                             & 	   $ -$                                 &  $-$                     &       $-$                   &   $-$                   &     $-$                           &   $-$                 \\
Sh\,2-212          &   $-$                                 &    $ -1.57 \pm 0.13 $               &     $ -1.52 \pm 0.14 $                   &  $-$                     &       $ -0.90 \pm 0.12 $    &   $-$                   &     $ -3.39 ^{+ 0.27 }_{- 0.21 }$ &   $-$                 \\
Sh\,2-219          &   $-$                                 &    $ - $                            &     $ - $                                &  $ -0.83 \pm 0.04 $      &       $-$                   &   $-$                   &     $-$                           &   $-$                 \\
Sh\,2-235          &   $-$                                 &    $ - $                            &     $ - $                                &  $ -0.84 \pm 0.04 $      &       $-$                   &   $-1.50\pm0.07$        &     $-$                           &   $-$                 \\
Sh\,2-237          &   $-$                                 &    $ - $                            &     $ - $                                &  $ -0.80 \pm 0.03  $     &       $-$                   &   $-$                   &     $-$                           &   $-$                 \\
Sh\,2-257          &   $-$                                 &    $ - $                            &     $ - $                                &  $ -0.80 \pm 0.04  $     &       $-$                   &   $-1.38\pm0.16$        &     $-$                           &   $-$                 \\
Sh\,2-266          &   $-$                                 &    $ - $                            &     $-  $                                &  $ -0.67 \pm 0.07 $      &       $-$                   &   $-$                   &     $-$                           &   $-$                 \\
Sh\,2-270          &   $-$                                 &    $ - $                            &     $ -$                                 &  $ -0.83 \pm 0.11 $      &       $-$                   &   $-$                   &     $-$                           &   $-$                 \\
Sh\,2-271          &   $-$                                 &    $ - $                            &     $ -$                                 &  $ -0.77 \pm 0.03 $      &       $-$                   &   $-1.57\pm0.08$        &     $-$                           &   $-$                 \\
Sh\,2-285          &   $-$                                 &    $ - $                            &     $ - $                                &  $ -0.77 \pm 0.04 $      &       $-$                   &   $-$                   &     $-$                           &   $-$                 \\
Sh\,2-288          &   $-$                                 &    $ -1.00 \pm 0.05 $               &     $ -0.94 \pm 0.11 $                   &  $-$                     &       $-$                   &   $-1.62 \pm 0.12$      &     $ -3.54 \pm 0.11 $            &   $ -2.46 \pm 0.11 $  \\
Sh\,2-297          &   $-$                                 &    $ - $                            &     $ - $                                &  $ -0.76 \pm 0.04 $      &       $-$                   &   $-$                   &     $-$                           &   $-$                 \\
Sh\,2-298          &   $-$                                 &    $ -0.82 \pm 0.04 $               &     $ -0.73 ^{+ 0.14 }_{- 0.09 }$        &  $-$                     &       $-$                   &   $ -1.63 \pm 0.06 $    &     $ -3.54 \pm 0.10 $             &   $ -2.44 \pm 0.07 $ \\
Sh\,2-311 	        &   $ -0.38 \pm 0.06 $                  &    $ -0.97 \pm 0.02 $               &     $ -0.91 \pm0.10$                     &  $-$                     &       $ -0.77 \pm 0.01 $    &   $ -1.61 \pm 0.05 $    &     $ -3.49 \pm 0.02 $            &   $ -2.26 \pm 0.05 $  \\

\hline
\end{tabular}
\begin{description} 
\item[] Note: in units of log(X/O).
\end{description} 
 \end{table*}

%% file: Table8.tex
\begin{table*}\footnotesize
\caption{Radial gradients of O, N, C, S, Cl and Ar abundances and their abundance ratios with respect to O.}
 \begin{center}
 \begin{tabular}{c c c c c c c c c c c}
 \hline
\multicolumn{1}{c}{12+log($X$/H)}  & \multicolumn{1}{c}{$N$} &  \multicolumn{1}{c}{Slope}    &  \multicolumn{1}{c}{Slope}        &  \multicolumn{1}{c}{intercept} & \multicolumn{1}{c}{$\sigma$} &
\multicolumn{1}{c}{log($X$)} &  \multicolumn{1}{c}{Slope}   & \multicolumn{1}{c}{Slope}       &  \multicolumn{1}{c}{intercept} & \multicolumn{1}{c}{$\sigma$} \\

                             & \multicolumn{1}{c}{}    &  \multicolumn{1}{c}{(dex kpc$^{-1}$)} & \multicolumn{1}{c}{(dex ($R$/$R_{\rm e})^{-1}$)} & \multicolumn{1}{c}{}          & \multicolumn{1}{c}{}   & \multicolumn{1}{c}{}    &  \multicolumn{1}{c}{(dex kpc$^{-1}$)} &   \multicolumn{1}{c}{(dex ($R$/$R_{\rm e})^{-1}$)} 
							 & \multicolumn{1}{c}{}  \\
\hline
O              &  33       &  $-0.037\pm0.009$  &  $-0.17\pm0.04$    &  $8.78\pm0.08$ & 0.07 &  $-$            &  $-$                         &  $-$                  &         $-$        & $-$   \\
N                       &  32       &  $-0.049\pm0.007$  &  $-0.22\pm0.03$    &  $8.06\pm0.06$ & 0.11 &   N/O            &  $-0.011\pm0.006$        &  $-0.05\pm0.03$     &         $-0.73\pm0.06$ & 0.12  \\
N$^{\rm a}$            &  12       &  $-0.057\pm0.016$  &  $-0.26\pm0.07$    &  $8.20\pm0.16$ & 0.07 &   N/O$^{\rm a}$            &  $-0.014\pm0.010$      &  $-0.06\pm0.05$     &         $-0.61\pm0.10$ & 0.08  \\
C                       &  11       &  $-0.072\pm0.018$  &  $-0.33\pm0.08$     &  $9.04\pm0.17$ & 0.10 &  C/O            &  $-0.037\pm0.016$       &  $-0.17\pm0.07$     &         $0.05\pm0.15$ & 0.11   \\
\\                                                                                                                                                                    
                                                                                                                                                                    
Ne              &  13             &  $-0.047\pm0.011$  &  $-0.21\pm0.05$    &  $8.19\pm0.09$ & 0.17  & Ne/O          &  $-0.018\pm0.005$      &  $-0.08\pm0.02$    &      $-0.53\pm0.04$ & 0.14  \\

S              &  17             &  $-0.035\pm0.006$  &  $-0.16\pm0.03$    &  $7.22\pm0.06$ & 0.10  & S/O           &  $-0.001\pm0.008$      &  $0.00\pm0.04$     &      $-1.55\pm0.08$ & 0.09  \\

Cl              &  18             &  $-0.030\pm0.007$  &  $-0.14\pm0.03$    &  $5.23\pm0.07$ & 0.11  & Cl/O          &  $0.001\pm0.010$       &  $0.00\pm0.05$     &      $-3.50\pm0.09$ & 0.10  \\

Ar              &  15             &  $-0.049\pm0.005$  &  $-0.22\pm0.02$    &  $6.65\pm0.05$ & 0.11  & Ar/O          &  $-0.025\pm0.007$      &  $-0.11\pm0.03$    &       $-2.03\pm0.07$ & 0.08  \\

\\
\multicolumn{11}{c}{Using the ICFs of Amayo et al. (2020)}\\
\\
Ne             &  13             &  $-0.038\pm0.027$  &  $-0.17\pm0.12$    &  $8.12\pm0.29$ & 0.19  & Ne/O          &  $-0.003\pm0.025$      &  $-0.01\pm0.11$    &                 $-0.66\pm0.28$ & 0.17  \\
S              &  22    &  $-0.040\pm0.009$  &  $-0.18\pm0.04$    &  $7.25\pm0.09$ & 0.10  & S/O  &  $-0.006\pm0.007$      &  $-0.03\pm0.03$    &        $-1.52\pm0.07$ & 0.09  \\
Ar             &  26    &  $-0.046\pm0.013$  &  $-0.21\pm0.06$    &  $6.60\pm0.15$ & 0.13  & Ar/O          &  $-0.013\pm0.011$      &  $-0.06\pm0.05$    &        $-2.15\pm0.11$ & 0.13  \\


\hline
\end{tabular}
  \begin{description}
 \item $^{\rm a}$ N abundances calculated without ICF.
 \end{description}
 \end{center}
 \label{tab:gradient-results}
 \end{table*}

%% file: Table9.tex
\begin{table*} 
\caption{Comparison of the slopes of O, C, S and N/O gradients for the Milky Way and other spiral galaxies.} 
\centering
\label{tab:comp_grad} 
\begin{tabular}{lcccccccc} 
\hline 
 & Morphological &  $R_{\mathrm e}$ &  & \multicolumn{4}{c}{slope (dex ($R$/$R_{\rm e})^{-1}$)} & \\
Galaxy &  type & (kpc) & $M_{\mathrm B}$ & O/H & C/H & S/H & N/O & Reference \\
\hline                                                                                                                   
NGC~300          & Sc    & 2.5    & $-$18.1   & $-$0.14   & $-$0.20   & $-$       & $-$0.15   & 1 \\
M33              & SAcd  & 3.7    & $-$18.9   & $-$0.19   & $-$0.33   & $-$       &  $-$0.25  & 1 \\
NGC~3184         & SABcd & 5.3    & $-$20.0   & $-$0.18   & $-$       & $-$0.34   & $-$0.35   & 2 \\
NGC~628          & SAc   & 3.3    & $-$20.7   & $-$0.11   & $-$       & $-$0.19   & $-$0.18   & 2 \\
Milky Way        & SBbc  & 4.5    & $-$20.8   & $-$0.17   & $-$0.33   & $-$0.16   & $-$0.05   &  3  \\
M101             & SABcd & 8.6    & $-$21.0   & $-$0.24   & $-$0.33   & $-$0.23   & $-$0.18   & 4, 2 \\
M31              & SAb   & 9.4    & $-$21.2   & $-$0.13   & $-$0.86:  & $-$       & $-$0.36   & 4 \\
NGC~5194         & SAbc  & 3.6    & $-$21.3   & $-$0.07   & $-$       & $-$0.23   & $-$0.18   & 2 \\
\hline           
           \multicolumn{9}{l}{1 --  \citet{ToribioSanCipriano:2016}.} \\
           \multicolumn{9}{l}{2 --  \citet{Berg:2020}.} \\
           \multicolumn{9}{l}{3 --  This work.} \\
           \multicolumn{9}{l}{4 --  \citet{Esteban:2020}.} \\           
  
  \end{tabular} 
\end{table*}